%% file: enviar.arxiv.tex
\documentclass[10pt, apj]{emulateapj}
\submitted{Received 2017 February 14; accepted 2017 October 31}

\usepackage{graphicx} \usepackage{natbib} \usepackage{amsmath}
\usepackage{amssymb} \usepackage{fancyhdr} \usepackage{latexsym}
\usepackage{color} \usepackage{txfonts}
\usepackage[hyperfootnotes=false, colorlinks = true, linkcolor = blue,
  urlcolor = blue]{hyperref}
\usepackage{mydefs_comparison_hinode_hmi}
\graphicspath{{./}}
\usepackage{soul}

\newcommand\asdok[1]{{\color{black} #1}}
\newcommand\newcorr[1]{{\color{black} #1}}

\begin{document}

\title{A STATISTICAL COMPARISON BETWEEN PHOTOSPHERIC VECTOR MAGNETOGRAMS \\ OBTAINED BY SDO/HMI and Hinode/SP}

\author{Alberto Sainz Dalda\altaffilmark{1}$^,\,$\altaffilmark{2}}
\altaffiltext{1}{Stanford-Lockheed Institute for Space Research,
	Palo Alto, CA 94304, USA}
\altaffiltext{2}{High Altitude Observatory, Boulder, CO 80301, USA.}

\begin{abstract}
Since May 1, 2010, we have been able to study (almost) continuously the
vector magnetic field in the Sun, thanks to two  space-based
observatories: the Solar Dynamics Observatory (SDO)
and Hinode. Both are equipped with
instruments able to measure the Stokes parameters of Zeeman-induced
polarization of photospheric line radiation.
But the observation modes, the spectral lines, the spatial,
spectral and temporal sampling, and even the inversion codes used to
recover magnetic and thermodynamic information from the Stokes profiles are
different.
We compare the vector magnetic
fields derived from observations with the HMI instrument on board SDO,
with those
observed by the  SP instrument on Hinode. We have obtained
relationships between components of magnetic vectors in the umbra,
penumbra and plage observed in 14 maps of NOAA AR
11084.
Importantly, we have transformed SP data into observables comparable to
those of HMI, to explore possible influences of the different
modes of operation of the two instruments, and the inversion schemes used to infer the magnetic fields. The assumed
filling factor (fraction of each pixel containing a Zeeman signature)
produces the most significant differences in derived magnetic properties,
especially in the plage. The spectral and angular samplings
have the next largest effects.
\asdok{We suggest to treat the disambiguation in the same way in the data
provided by HMI and SP}. That would make the relationship between the
vector magnetic field recovered from these data stronger, what would
favor the simultaneous or complementary use of both instruments.
\end{abstract}

\section{Introduction}
The goal of this paper is to compare statistically the
 vector magnetic fields retrieved
from  SDO/HMI \citep{iSche12,iScho12} with
those from  Hinode-SOT/SP \citep{iTsu08,iLites13b}.
We are not interested in the absolute calibration of
the magnetograms, but in the comparison between them.  We
shall provide a way, based on a statistical analysis, to go from the
magnetic field measured by one instrument to the other in a similar
way as \cite{iBer03} did for the imaging polarimeter on SoHO (MDI, \citealp{iSche95}) and the slit ground-based Advanced Spectro Polarimeter
\citep{iElm92}. The present comparison is motivated by and follows a
similar methodology of \cite{iBer03}. We also analyze the influence of
the various observing configurations and inversion codes \newcorr{(ICs)}, henceforth called
{\it actors},   in the derivation of vector magnetic fields.

 In addition to the original data provided by HMI and SP\footnote{For
   the sake of clarity, we will refer to SDO/HMI and Hinode/SOT-SP
   instruments just as HMI and SP respectively.}, we
 have created pseudo-SP \asdok{(SP data spectrally sampled as HMI does)} maps to simulate the actors present in HMI. Our
rationale is that SP is an instrument with detailed spectral data and a higher angular resolution, which, being optically stable over the period of an observing run, and being free of significant seeing-induced crosstalk, should yield magnetic fields of a higher quality, serving as a kind of ``ground truth'' measurement for HMI.   In Section \ref{sec:data_analysis}
 we describe the instruments and data we are comparing, how we create
 the pseudo-SP maps, the methodology used to prepare and compare the
 data, and the comparisons we have carried on. We also remind the reader of
the
 meaning of the statistical parameters used to support our conclusions.

In Section \ref{sec:results}, we discuss the  vector magnetic field comparisons.   We also explore how that comparison might be
affected when the spectral sampling, the spatial sampling, the spectral
line, the inversion code and the disambiguation code are different.
All our results are based on
an statistical analysis made over 14 maps of an active region observed
simultaneously by HMI and SP. Finally, in Section
\ref{sec:conclusions}, we summarize our findings and suggest several
improvements that might benefit the simultaneous or complementary use
of HMI and SP data. \newcorr{In the Appendix, the interested reader may find detailed statistical tables. Although those tables make the length of this paper longer than expected, we believe they are necessary to support our results.}

\asdok{The main goal of this paper is to encourage the solar community to use HMI and SP data either simultaneously or alternatively if one data set is available in one instrument but not in the other. Nowadays, the synergy between the numerical experiments --including simulations and extrapolations-- and the observational data is critical for the advance of our knowledge of the Sun. In that sense, to speak the same language makes the communication between those data products easier. Apparently, the majority of the solar community uses the Cartesian system instead of the Spherical coordinate to express the vector magnetic field, and for that reason we decided to offer our results primarily in the Cartesian coordinates system. Therefore, we have conducted our investigation in the Cartesian on-CCD coordinates system, i.e. $\mathbf{B}=\{B_{x}, B_{y}, B_{z}\}$, although the most important comparisons have been also made for the spherical on-CCD coordinates system, i.e. \newcorr{$\mathbf{B}=\{|B|, \theta_B, \phi_B\}$\footnote{The spherical components of a vector $\mathbf{A}$ are usually denoted as \{$A_{r}$, $A_{\theta}$, $A_{\phi}$\}. In this paper, we follow the notation widely used in the solar community for the vector magnetic field $\mathbf{B}$.}}. As we will see, using the spherical on-CCD coordinates system has a great advantage to decouple the effect of the field strength and the disambiguation in the computation of the horizontal components of the vector magnetic field.} 

\section{Data analysis}\label{sec:data_analysis}
HMI and SP instruments make full-polarimetric observations (i.e., the
four Stokes parameters I, Q, U and V) of the solar disk and selected
areas in the Sun respectively. Both instruments observe spectral lines formed in the
photosphere. The polarization induced by
the Zeeman effect (e.g., Jefferies, Lites
\& Skumanich 1989)  allows us to infer the photospheric vector
magnetic field, given a suitable model for the radiating layers. These models
work with the data in what is collectively called an inversion scheme, or
simply an ``inversion''. 

\subsection{Data Selection}\label{sec:data_sel}
HMI provides full polarimetric filtergrams of the full solar disk
taken in just six carefully selected
spectral positions around the Fe {\footnotesize I} 6173
\AA\ line. \newcorr{Every  $\approx ~69m\AA$, the line is sampled with spectral filtergram with a bandpass of roughly $70m\AA$ \asdok{(spectral bandpass tunable over $680m\AA$)}}, and the spatial
sampling is $0.5\arcsec \times 0.5 \arcsec$. The field of view is the whole Sun.  While the total
polarimetric cycle takes 135 $s$\footnote{New observing scheme for HMI is recording full-Stokes data with a cadence of 90$s$. See \href{http://hmi.stanford.edu/hminuggets/?p=1596}{http://hmi.stanford.edu/hminuggets/?p=1596}.}, the vector magnetograms provided by
the official web page\footnote{HMI data are available at
  \href{http://jsoc.stanford.edu/}{http://jsoc.stanford.edu/}.} \newcorr{have a} cadence of 12 minutes, as many images are needed to build up
an average to
improve the signal-to-noise ratio.

The SP data are full slit spectro-polarimetric data in the Fe~{\footnotesize I} 6301 \& 6302 \AA\ lines. The data used here
have a spatial sampling \newcorr{of} $0.30 \arcsec \times 0.32 \arcsec$, and the
spectral sampling is $21m\AA/px$. 

\asdok{There are several data products available for both instruments. In this investigation, the main comparison is between the vector magnetic field recorded in the HMI {\it hmi.B\_720s} data and the one saved in the SP {\it Level2} data. Albeit, the main goal of this investigation is to provided to the solar community a comparison between these easy-access data {\it as they are}.
	
The {\it hmi.B\_720s} data are full-disk Milne-Eddigton inversion results with the magnetic field azimuthal ambiguity resolution applied \citep{Hoe14}. The {\it \_720s} in the name of the data product refers to the post-observation integration in time of the data  to improve the S/N ratio, as we mentioned above\footnote{Information about the HMI data may be found at:\\
\href{http://jsoc.stanford.edu/JsocSeries\_DataProducts\_map.html}{http://jsoc.stanford.edu/JsocSeries\_DataProducts\_map.html}}. 

The {\it Level2} data are selected-area Milne-Eddigton inversion results with the magnetic field azimuthal ambiguity resolution \textbf{not} applied\footnote{Information about the SP data may be found at:\\ \href{https://www2.hao.ucar.edu/csac/csac-data/sp-data-description}{https://www2.hao.ucar.edu/csac/csac-data/sp-data-description.}}. 

In addition to the SP \textit{Level2} data, we have also used the Stokes profiles originally observed by SP (\textit{Level1}). The SP \textit{Level2} data provide maps of physical variables -- such as the  field strength, the inclination and the azimuth of vector magnetic field, line-of-sight velocity, and so on-- after the inversion of the Stokes profiles recorded in the SP \textit{Level1} data. 

We have inverted the original Stokes profiles observed by SP considering a filling factor (FF) equal to 1. Since the data used to obtain the vector magnetic field are the original SP \textit{Level1} data, \newcorr{i.e. Fe~{\footnotesize I} 6301 \& 6302 \AA\ lines., and we use the inversion code  used for the official SP inversions,} we refer to this data as \textit{SP data inverted with FF equal to 1, \newcorr{or SPFF1}}. 

We have created a new data set from the original SP \textit{Level1} data. We have convolved the Stokes profiles observed by SP (\textit{Level1}) corresponding to  Fe {\footnotesize I} 6302\AA\ with the HMI
transmission profiles calculated for that line. The HMI pipeline
calculates these instrumental profiles for any pixel in the solar
disk, so it takes into account the Doppler shift due to the
differential rotation in the solar surface. For each SP map, we have
calculated the instrumental profiles for the pixel located at the center of the
FoV, and applied them to all pixels in that SP map. We refer to the SP
data sampled and filtered as HMI filtergrams are made as \textit{S2H} (after
SP to HMI). The S2H data have the same spatial sampling \newcorr{as} the
original SP data. They are inverted with the same inversion code as HMI, i.e.,
VFISV with FF equal to 1. Notice for creating the S2H data we only use 1 of the 2 lines observed by SP, since we use the same instrumental profiles used by HMI, which cover a spectral range equivalent to one spectral line in the SP spectropolarimetric data. We have chosen the line Fe~{\footnotesize I} 6302 \AA{} because its effective Land\'e factor, $g_{eff}$, is the same as the one of the line used by HMI, being $g_{eff} = 2.5$ for these two spectral lines. The code that calculates the instrumental transmission
profiles for any spectral line at a point of the solar disk is the
same used in the HMI pipeline at JSOC. This code was kindly provided in the
IDL version by Dr. S. Couvidat from Stanford University.

\newcorr{The SP slit does not always scan the solar surface} with a constant step in the perpendicular direction to the slit \citep{iCen09}. In some cases, an interpolation may be necessary to have an equally-spaced scan. However, since the ultimate goal in this paper is the comparison between SP maps (and S2H maps) and HMI maps, the interpolation of SP data to an equally-spaced grid as initial step is unnecessary, since the interpolation what matters is the one that matches the SP maps with HMI maps. Therefore, in this paper we have skipped the interpolation of the SP data to equally-spaced grid, and we have directly interpolated the SP magnetic field maps to the HMI ones. Same strategy has been applied to match the S2H maps with HMI maps. More details about the co-alignment procedure are given in Section \ref{sec:data_preparation}.

Although the databases of HMI and SP offer different data products, we have selected the data detailed above since they are the ones \newcorr{that} are more directly comparable. Recently, HMI has started to provide similar full-disk vector magnetic field similar to the ones analyzed in this paper, but with a post-observation integration time of the data of 135 and 90 $s$. A comparison between these new HMI data products and SP data would be interesting. That new comparison with the one presented in this paper would allow us to better understand the role played by the S/N. Doubtless, the influence of this factor in the comparison between different instruments is important. Nevertheless, it currently falls out of the scope of this paper.}


\begin{figure*}[]
\includegraphics[width = \textwidth]{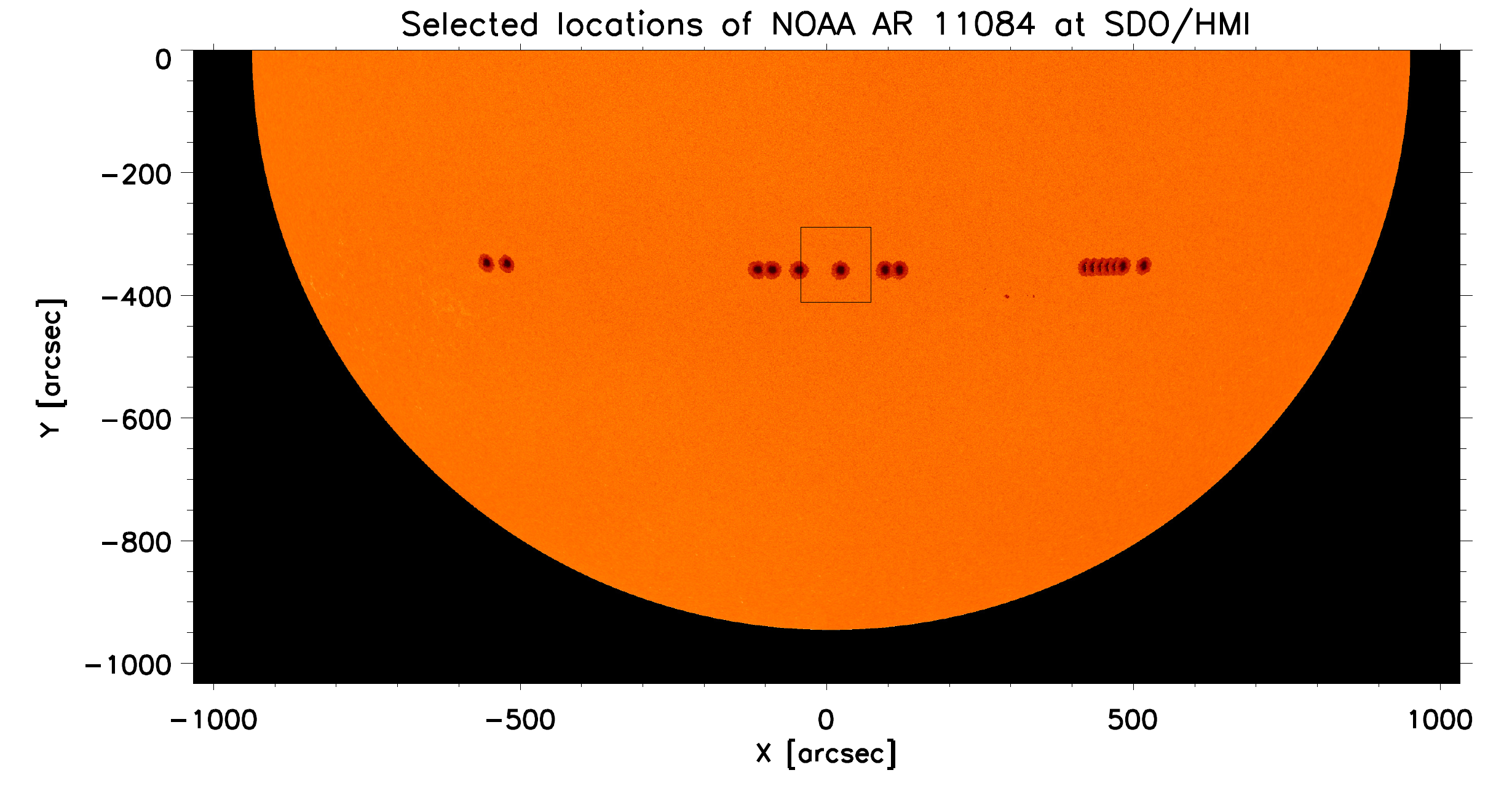}
\caption{Locations where NOAA AR 11084 were simultaneously observed by
  Hinode-SOT/SP and SDO/HMI. The square delimits the size area usually
  scanned by Hinode-SOT/SP over this AR.}\label{fig:pos_AR_HMI}
\end{figure*}

\begin{figure*}[]
\includegraphics[width=\textwidth, angle=-90, scale=0.55]{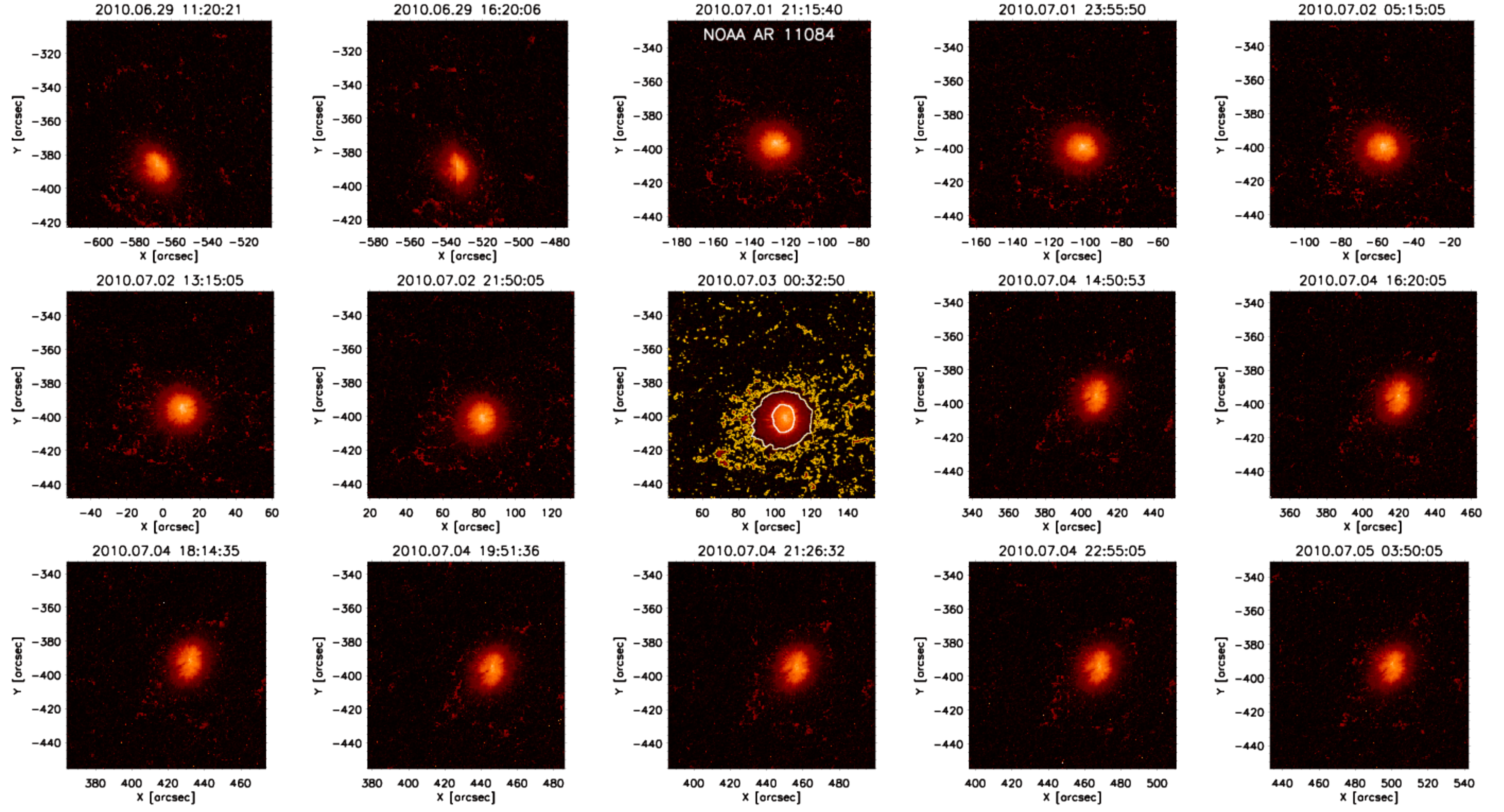}
\caption{Magnetic field strength provided by MERLIN IC on NOAA AR 11084
  by Hinode-SOT/SP. Contours overplotted on the intensity map observed at
  2010.07.03 00:32:50 delimit the umbra-penumbra and the
  penumbra-moat edges (both in white contours), and the plage (yellow
  contours). Date and time are relative to the starting time of the
  scan. Axes are given in heliocentric coordinates.}\label{fig:pos_AR_SP}
\end{figure*}

\newcorr{The SP scans required approximately} 30 minutes for this medium size
sunspot. In that time, any sunspot may have significantly evolved if
undergoing significant activity, e.g., those hosting flare activity or
during the initial (non-stationary)
emergence phase. Therefore we deliberately selected a mature active region (NOAA AR 11084) with a
well-developed sunspot, \newcorr{which} is located in a relatively isolated and quiet
region. For the
comparisons, we  selected the HMI data product which was
closest in time to the time when the slit of SP was in the middle of
its raster.  Figure \ref{fig:pos_AR_HMI} shows the location of NOAA AR 11084
and Fig. \ref{fig:pos_AR_SP} shows the magnetic field strength 
maps \newcorr{provided by the SP Level 2 data as resulting from the
{\it Milne-Eddington gRid Linear Inversion Netwrok} (hereafter MERLIN\footnote{Information about MERLIN code and the inverted data
can be found at \href{https://www2.hao.ucar.edu/csac}{https://www2.hao.ucar.edu/csac}
and
\href{http://sot.lmsal.com/data/sot/level2d}{http://sot.lmsal.com/data/sot/level2d}.}) IC on the Level 1 data.} Except for
the second map in the first row, during which the SP slit jumped,
all maps were used. Table \ref{tabla_data_sel} in Section \ref{sec:data_sel} of the
Appendix shows a list with the observational details of the selected
data.

In addition to the difference between the instruments used -- therefore
between the type of data -- we have to take into account differences
between the ICs used to retrieve the
physical information from the data. \newcorr{HMI B\_720s series
data result from the systematic inversion of observed Stokes filtergrams by
the code} {\it Very Fast Inversion of Stokes Vector}
(hereafter VFISV, \citealt{Bor11,Cen14}). SP data were inverted by MERLIN. \asdok{As mentioned above, S2H data were inverted using VFISV as well}. Both
VFISV and MERLIN are based on the nonlinear least-squares fitting
between the synthetic and the observed profiles using the
Levenberg-Marquardt algorithm.
Both ICs calculate the synthetic Stokes profiles after solving the
radiative transfer equation for polarized light under under the
Milne-Eddington approximation. Thus, the source function varies
linearly with the optical depth, while the other physical parameters
are constant through the atmosphere (non-dependence with the optical
depth).
These codes provide the vector magnetic field, magnetic filling
factor, source function and source function gradient, Doppler width,
Doppler velocity, damping parameter, and line-to-continuum absorption
coefficient.

\subsection{Data preparation}\label{sec:data_preparation}

In this section, we explain the steps we have applied to the data to
make the comparison between them possible. 
%
\asdok{We have applied a semi-automatic alignment method using the routine
\texttt{auto\_align\_images} developed by T. Metcalf, and
\texttt{poly\_2d}. Both routines and dependent ones are available in the
       {\it SolarSoftware} package. The first step is to identify
       manually several common structures in both maps. We selected 10
       of them. Then, the code matches the maps maximizing the
       correlation between them applying automatically corrections in shifts in vertical and horizontal direction, rotation and expansion or contraction. We use the
       output of the first routine as input of {\texttt poly\_2d} on
       the HMI data. All the comparisons between HMI and SP, and HMI
       and S2H have been made considering the corrections provided by this method. 

       Even when we have done our best in the alignment
       between the maps, some misalignment may be present in some
       cases. Several reasons may cause this misalignment. The hardest
       one to treat is the a non-constant spatial interpolation
       between the FoV observed by HMI and SP. That may be produced by
       a non-constant spatial sampling in the direction perpendicular
       to the slit, i.e. the size step of the raster changed during
       the scan, as it has been reported by \cite{iCen09}. Of course, other reason is the natural evolution of the solar features while the data were taken. As a visual effect, we are able to align very good
       one side of the FoV, but not so good the other side, where a ``stretching''
       effect may be observed. In the first case, a constant spatial interpolation in the SP data would help to have these data in a more appropriate spatial grid. However, we will need another interpolation to match the HMI spatial scale. We have skipped the first step, and we have to interpolate the SP data directly to HMI. For that reason, we selected 10 points homogeneously distributed in the FoV to warranty the best possible match between the maps.
       The best alignment achieved
       is under pixel sampling, i.e., under HMI spatial scale or $\sim 0\arcsec.5$."
       During the remapping process the data are interpolated using several procedures.}

For SP data, the disambiguation of the azimuth was made using AZAM
code. This code was developed by Dr. P. Seagraves and Dr. B. Lites,
and it is available in
the {\it SolarSoftware} distribution. For HMI and S2H data we have
used the disambiguation solution map provided with the original HMI
data \citep{Hoe14}.  While the main results presented in this paper
are using AZAM, we have done a detailed studied of the effect
introduced by the disambiguation. Thus, we have used three
disambiguation solutions. Two are as a result of applying AZAM in
slightly different ways. The third is AMBIG, an automatic
disambiguation code developed by \citep{Lek09}. The core of this code
is similar to the one used by HMI, although the setup parameters are
slightly different. \newcorr{All in all, there is no significant difference between the  three disambiguation results yielded in the comparison HMI $vs$ SP and the ones obtained in the comparison HMI $vs$ S2H (see Table \ref{tabla_hmi_sp_disamb} and discussion in Section \ref{sec:disambig})}. However, if we do not consider those solar features showing
opposite sign in the horizontal components -- mainly located in the
plage --, then the correlation between these data is better.  A
detailed discussion is presented in section \ref{sec:disambig} of the
Appendix.

After the co-aligment and the the disambiguation of the azimuth, we
calculate the three components of the vector magnetic field in the
observer frame, i.e., as it is observed from the Earth.

\asdok{In our study, we have selected 4 different regions of interest (RoI): umbra, penumbra, strong plage, and weak plage. The umbra and penumbra
are calculated taking into account photometric thresholds with respect to the
mean value of the continuum Intensity in the local quiet sun ($\tilde{I}_{c,lqs}$). Thus, the umbra is defined as the region where the $I_{c,lqs} < \tilde{I}_{c,lqs}\times0.55$, while the penumbra is where $  \tilde{I}_{c,lqs}\times0.55  < I_{c,lqs} < \tilde{I}_{c,lqs}\times0.95$. For the plage we have used total polarization slit-reconstructed map from the SP data. We have considered as plage as those locations outside of the umbra and penumbra where the total polarization signal is larger than 0.005. We have split the plage region into strong plage and weak plage considering the values of the components of the vector magnetic field (in Cartesian coordinates, see Section \ref{sec:comp_BSP_BHMI}) or the field strength (in spherical coordinates. The contours of the umbra, penumbra and plage are showed in the central image of fig. \ref{fig:pos_AR_SP}. Table \ref{tabla_ranges_SPcgHMI} shows the averaged (in time) values of the minimum, maximum and mean of the magnetic field strength in the RoIs considered in the comparison between \vectorb{HMI}{} and \vectorb{SP}{} in spherical coordinates.} 

\input{table_fieldstrength_SPcg_HMI_ori_dis_newcutpoly2d_ranges.ord2}


\subsection{Data treatement and Inversions Setup}\label{sec:data_insetup}

As well as re-binning and filtering SP data to match those of HMI,
we must  perform what we call {\it cross
	inversions} using the same IC with the different kinds of data or different IC with different data. Table
\ref{table:summary}  lists the various combinations we have examined. \asdok{In the following description of the comparison we have made, the expression \textit{`Original Inverted Data'} refers to use the original inversion results provided by the \newcorr{HMI and SP project official web pages}. We have re-binned and interpolated the \textit{`Original inverted data'} to make possible the spatial matching between them.} 
The comparisons made include:

\begin{itemize}
	\item{\textit{Case A: Comparison between Original Inverted Data \newcorr{(HMI $vs$ SP)}.}
		The straightforward comparison between HMI \asdok{(\textit{ hmi.B\_720s})} and SP \asdok{(\textit{Level2})} is done
		from the data accessible at the corresponding data base. In this
		case, we are comparing different
		data inverted with different inversion code in the two different
		spectral ranges \asdok{( Fe {\footnotesize I} 6173 \AA\ for HMI, and  Fe {\footnotesize I} 6301 and 6302 \AA\ for SP)}, and with
		different filling-factor conditions (FF = 1 for HMI, FF variable for
		SP).} \asdok{This is the comparison of the data \textit{as they are publicly available}}. 
	\item{\textit{Case B: Comparison between Original data with different FF \newcorr{(HMI $vs$ SPFF1)}.}
		\newcorr{In this case, we have modified MERLIN code to use FF=1 on SP data (SPFF1), instead of the data publicly offered, which have FF variable (SP). Both inversions are mode on Fe {\footnotesize I} 6301 and 6302 \AA\. This modification of MERLIN was kindly made by Dr. A. de Wijn in CSAC at HAO.}}
	\item{\textit{Case C: Comparison between data equally sampled
			spectrally \newcorr{(HMI $vs$ S2H).}} \newcorr{We use  VFISV code with FF variable to invert the S2H data. The stray light 
			profile is calculated using the same methodology used by the SP data pipeline. Then, it is convolved with the same HMI instrumental profiles we applied to the corresponding data.}
		From this comparison we infer the influence of
		the different spectral sampling between the original SP and HMI
		data.}
	\item{\textit{Case D: Comparison between data equally sampled with FF fixed to 1 \newcorr{(HMI $vs$ S2HFF1)}.} In this
		case, we use VFISV code with FF=1 to invert the S2H data \newcorr{(S2HFF1)}.  
		(By default original HMI data are inverted with the fixed FF to 1.)}
	\item{\textit{Case E: Comparison between SP data using different FF (\newcorr{(SPFF1 $vs$ SP).}).}
		We use
		the original SP Stokes profiles data \asdok{(\textit{Level1})}, and we invert them with MERLIN using
		different FF, i.e. either with FF variable \newcorr{(SP)}
		or with fixed to 1 \newcorr{(SPFF1)}. Therefore, we see the influence of the FF during the inversion on the original SP data. \asdok{In this comparison, we invert both lines Fe {\footnotesize I} 6301 \AA and 6302 \AA, as MERLIN does for the original SP data.}}
	\item{\textit{Case F: Comparison between \asdok{S2H data} inverted
			with FF fixed to 1 and the same data inverted with FF variable \newcorr{(S2HFF1 $vs$ SH2)}.
		} In this case, we investigate the effect of different FF
		during the inversion on the S2H data. The data have the same spatial
		resolution, the observed line is the same (Fe {\footnotesize I} 6302\AA), and the IC used is the
		same (VFISV).}
	\item{\textit{Case G:  Comparison between S2H and SP data using FF variable \newcorr{(S2H $vs$ SP)}.} In this case, the data have the same spatial and temporal resolution, and are perfectly co-aligned. \asdok{SP observes Fe {\footnotesize I} 6301 \&  6302 \AA\ spectral lines, while S2H data only considers Fe {\footnotesize I} 6302 \AA.} We use different spectral sampling and different ICs.}
	\item{\textit{Case H: Comparison between S2H and SP data using FF fixed to 1 \newcorr{(SH2FF1 $vs$ SPFF1)}.} Same than case G but using FF fixed to 1, } 
\end{itemize}

\input{tablecomparisons}

We have intentionally left the impact introduced by the
disambiguation solution as the last step of our comparisons. While all
the others actors play a significant role in the process of
calculation of the field strength, inclination and azimuth of the
vector magnetic field, the disambiguation solution will play a different
role in
the composition of that vector. This is the only case where the information is not contained in the data themselves, we must add information to obtain an ``optimal'' disambiguated solution \citep{Hoe14}.
As we shall show, the chosen solution to
the disambiguation problem may have a
relevant impact in the correlation between the magnetic field observed
by the two instruments. That relationship might be easily improved if
both data set solved the disambiguation problem with the same code. A
detailed discussion on this topic is given in Section
\ref{sec:disambig} of the Appendix.

\subsection{Comparison of the Data: Statistical Analysis}

\begin{figure*}[]
\includegraphics[width = 1.\textwidth]
                {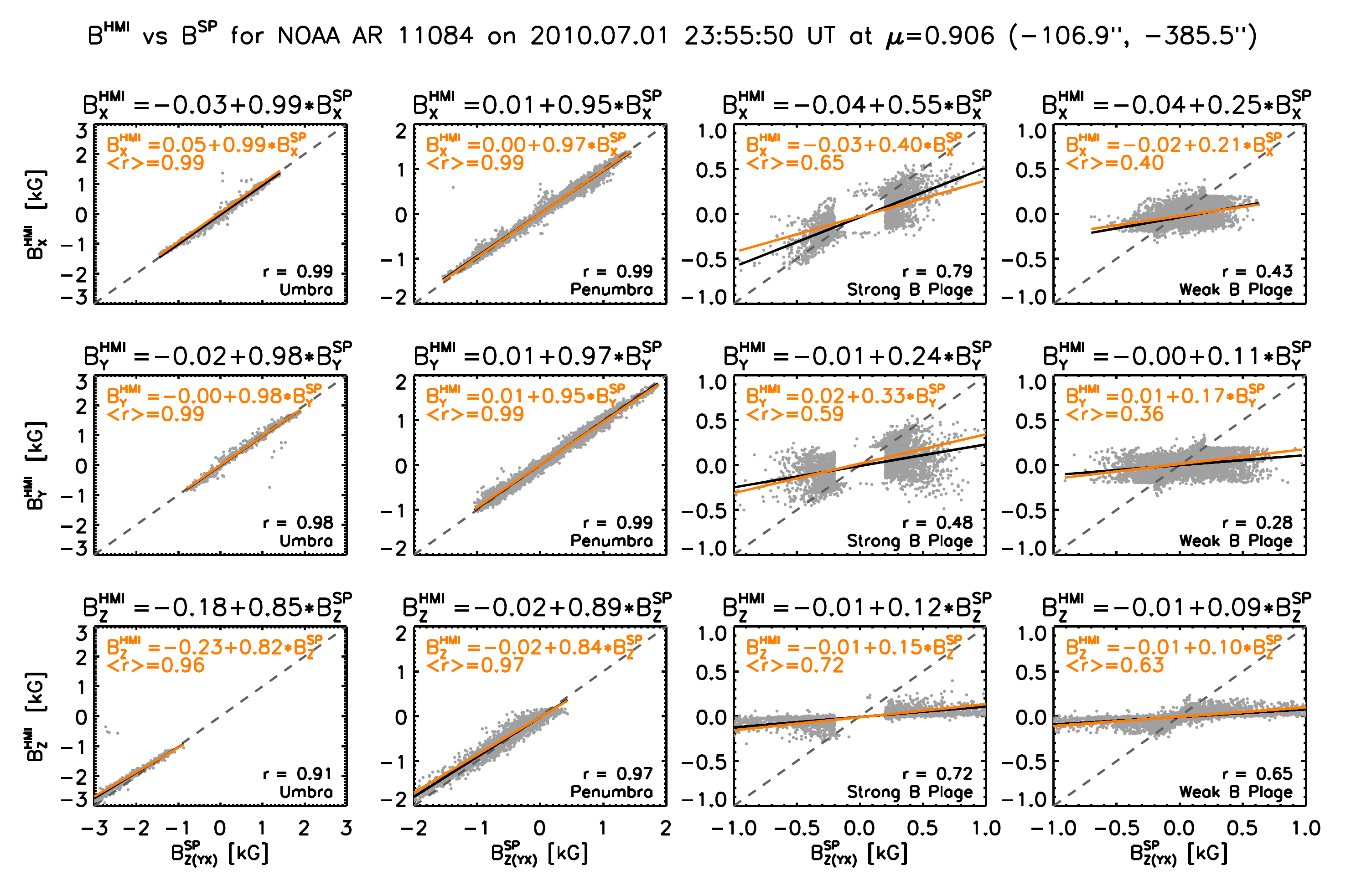}
\caption{Scatter plot between the \newcorr{Cartesian }components of the vector magnetic
  field of \vectorb{HMI}{} and \vectorb{SP}{}. The linear regression
  fits and the correlation coefficients for the scatter plots of this
  particular map are indicated in black. The averaged linear
  regression fits and correlation coefficients for all the maps are
  given in orange.}\label{fig:plot_linreg}
\end{figure*}

Figure \ref{fig:plot_linreg} shows the scatter plots for the
components of the vector magnetic field \vectorb{HMI}{} versus
\vectorb{SP}{} at the RoIs, \newcorr{i.e. the comparison between the HMI and SP data as they are provided by the respective official data centers (case A).} They show a linear dependency. Hence, we
have evaluated a linear regression fit between the components of the
vectors \vectorb{HMI}{} and \vectorb{SP}{} for the RoIs at each of the
14 selected maps. Tables \ref{tabla_linreg_all_xy} and
\ref{tabla_linreg_all_z} in Section \ref{sec:comp_HMISP_XYZ} of the
Appendix list all the parameters calculated for the comparison between
\vectorb{HMI}{} and \vectorb{SP}{}. For  these linear regression
fits we obtain the following parameters: intercept $a$, slope $b$,
(estimated) standard deviation of the residuals $s_e$, correlation
coefficient $r$, and coefficient of determination $r^2$. The
interpretation of these parameters in this paper is strictly referred
to a simple linear regression calculated using the least squares
method\footnote{A least squares line should formally be written as
  $\widehat{y} = a + b\times x$, being $\widehat{y}$ the prediction of
  $y$ resulting for a particular value of $x$. Therefore, we should
  formally denote our results as, e.g.,
  $\widehat{\mathbf{B^{\mathbf{HMI}}_{\mathbf{Z}}}} = a + b\times
  \mathbf{B^{\mathbf{SP}}_{\mathbf{Z}}}$. For the sake of clarity in
  the typography, we have denoted the prediction of the dependent
  variable in the linear regression fit without the wide tilde on it,
  e.g., $\mathbf{B^{\mathbf{HMI}}_{\mathbf{Z}}} = a + b\times
  \mathbf{B^{\mathbf{SP}}_{\mathbf{Z}}}$. For simplicity, in this
  example, we shall talk about {\it ``the vertical component of
    $\mathbf{B^{\mathbf{HMI}}_{}}$''}, when we should do as {\it ``the
    vertical component of the predicted
    $\mathbf{B^{\mathbf{HMI}}_{}}$''.}}.
Finally, we have calculated averages of  intercepts and
slopes,  averages of the estimated standard deviation \sval,
correlation coefficients \rval, and coefficients
r-squared \rsqval over the linear regression fits of the all 14 maps
analyzed in this paper. The errors of these averages are the standard
deviation to their means.

$s_e$, is a measure of the typical residual from the
least squares line. It is given in the same units of the
dependent variable of the linear regression (kG). The
correlation coefficients $r$ tell us how well the data fit into a
linear regression, i.e. $r$ is a measure of the extent to which the
independent and the dependent variable are linearly related. A linear
regression is considered showing a strong linear relationship when
$1{<}|r|{<}0.8$, a moderate linear relationship when
$0.8{<}|r|{<}0.5$, and a weak or negligble
relationship when $0.5{<}|r|{\le}0$. $r^2$
may be interpreted as how much the data
are spread with respect to the linear fit: High $r^2$ represents a small
spread, and vice versa.  A linear regression fit with a low $r^2$ implies
  a large range of values of one variable
(e.g. $B^{HMI}_{Y}$) for a given value of the other variable
(e.g. $B^{SP}_{X}$). $r^2$ also tells us how much of the variability of one
variable can be explained by its (linear) relationship with the
other. Therefore, $r^2$ is also the fraction of variation that is shared
between these variable. For instance, if $r^2$ = 0.75, it is said that
a 75\% of the total variance of the Y variable ($B^{HMI}_{X}$ in our
example) is explained by the independent variable ($B^{SP}_{X}$) with
the corresponding linear regression fit.

In summary, the average values of $s_e$, $r$ and $r^2$ analyzed in
this article tell us, in a statistical sense, about the error of the
estimate, the goodness of the linear relationship, and the percentage
of the variation explained of the compared, fitted data using a linear
regression fit.

\section{Results}\label{sec:results}
In section \ref{sec:comp_BSP_BHMI}, we give the
averaged values of the linear regression fit between the components
of \vectorb{HMI}{} and \vectorb{SP}{} {\it as they are} (case A).
That section
is based on the data released by the respective projects
to the public. \newcorr{We investigate the impact of different actors playing a role in the process of recovering the vector magnetic field from the Stokes parameters. We address that investigation} through the comparison between the HMI data
and the pseudo-SP data (cases B to D), and between SP data and
pseudo-SP data (cases F to H) in sections \ref{sec:HMI_pseudoMaps} and
\ref{sec:SP_pseudoMaps} respectively.  Separately,
in section \ref{sec:comp_Bapp},
we study the relationship between the apparent magnetic flux density
and the total flux between for the cases A to D.

\newcorr{In the main body text, we show and explain the statistical values that support our results. Figures and tables showed in the main body text are focused in the comparison between HMI and SP. In the Appendix, we include the statistical tables and other figures related with all the comparisons made in our study.}

\subsection{Comparison between \vectorb{HMI}{} and \vectorb{SP}{}\label{sec:comp_BSP_BHMI}}

We have compared the inversion results available at HMI and SP
databases. As we have mentioned before, the disambiguation is the only step
not automatically provided with the SP data. For HMI data, this step
is easily implemented, since JSOC provides automatically
disambiguated inversions. The assumptions behind the algorithm used are
summarized in \cite{Hoe14}. For the SP maps, we used AZAM code to solve the
disambiguation problem (see section \ref{sec:disambig} in the Appendix
for more details.

Figure \ref{fig:plot_linreg} shows the linear regression fits between
the components of the vector magnetic field of HMI and SP for one
individual map (lines and fonts in black). We have over-plotted the
averaged linear regression fit over the 14 selected maps (line and
fonts in orange). The averaged linear regression fits between the
components of \vectorb{HMI}{} and \vectorb{SP}{} in each of the RoIs,
and the averaged parameters described in the previous section are
given in detail in Table \ref{tabla_compar_SPHMIpoly2d}. They can
be interpreted as a {\it translator} between the vector magnetic field
observed by one instrument to the other.  Such data are listed for
the RoIs of every map of
our data set (see section \ref{sec:app_data_sel}).

\input{tablaHMIvsSPcgpoly2d}
\input{tablaHMIvsSP_fieldaziincli_dis_newcutpoly2d_soloHMISP}

On average, both in the umbra and penumbra the horizontal components
of the vector magnetic field ($B_X$ and $B_Y$, all values are given in kG units) show a slope very close
to 1 (\baver$\geq0.95$), while for the \newcorr{longitudinal component}  ($B_Z$) is
smaller ($\sim0.83 \pm 0.05$). In the next sections we will explore
this significant discrepancy.
The errors in the intercept for the umbra are larger
than for the penumbra, especially if we compare them with the errors
at the other RoIs ($\leq0.01$). The \rval and \rsqval for
these RoIs are very close to 1 for the horizontal components and
slightly smaller for the vertical component. Therefore, we can say
that the values of the components of vector magnetic field
\vectorb{HMI}{} at the umbra and penumbra can be accurately predicted
by \vectorb{SP}{} from the linear relationship showed in the first six
rows of Table \ref{tabla_compar_SPHMIpoly2d}.

In the {\it strong plage}, i.e. where 200 G $ <
(\mid$\vectorb{HMI}{XYZ}$\mid$ or $\mid$\vectorb{SP}{XYZ}$\mid) < $
1000 G,
the \sval is $160 $G  for the horizontal components
 and 70 G  for the vertical component with  correlations
 \rval $\sim 0.60$, and $0.72$ respectively.
Although the errors of \rval for the horizontal components
are larger,
what indicates a large scattering of the $r$ value of the individual comparisons of the
horizontal components. The percentage of variation shared by the
variables is \rsqval $\sim 50\%$ for the vertical component. For the
horizontal components, \rsqval slightly smaller, being $44\%$ and
$35\%$ for the X and Y component respectively. The averaged slope
\baver for the horizontal components are between $2$ and $3$ times
larger than for the vertical component.  Therefore, the components
of the vector magnetic field at HMI can be predicted from those at SP 
and vice versa, with a moderate confidence.

Finally, in  the {\it weak plage}, i.e. where $(\mid$\vectorb{HMI}{XYZ}$\mid$ or $\mid$\vectorb{SP}{XZY}$\mid) < 200
G$,  \sval for the horizontal
components is $\sim100 $G and $\sim50G$ for the vertical. The \rval for the horizontal components shows only a weak linear
relationship, while for the vertical component the linear relationship
is moderate. Only around $15\%$ of the variation of \vectorb{HMI}{XY}
may be explained from \vectorb{SP}{XY}, and $40\%$ for the vertical
component. The errors in all the averaged statistical variables for
the weak plage are small. That means, the scattering of the individual statistical variables is small: there is a trend in these variables. On average, the behavior in the weak plage, having a weak
linear correlation, is similar in all the individual maps.
\asdok{We have to be careful with the interpretation of the small values of the errors for the plage. The errors for the weak plage are smaller than for the strong plage. However, the normalized variance, i.e. the standard deviation divided by the average of the field values, is higher in the weak plage than in the strong plage, as one may expect from a poorer S/N in the weak plage \newcorr{--because of the low signal there--,} with respect to the S/N in the strong plage \newcorr{--where the signal is higher.}} 

\asdok{Table \ref{tabla_compar_SPcgHMIdiscut_soloHMISP} compares the components of the vector magnetic field in spherical coordinates. In this case, the weak and strong plage are defined as those pixels of the plage region (pixels outside of the sunspot with $P_{tot} > 0.005$) where $(|B|^{HMI}$ or $|B|^{SP}) < 200 G$ and $ 200 < (|B|^{HMI}$ or $|B|^{SP}) < 1500 G$ respectively.}

Fig. \ref{fig:maps_B_HMI_SP_SPFF1_S2H} shows, in the \first and the
\second column respectively, the components of the magnetic field of
\vectorb{HMI}{} and \vectorb{SP}{} of the maps compared through the
scatter plots in Fig. \ref{fig:plot_linreg}. In the top panels, from
top to bottom, we show $B_X, B_X, B_Z$ for the sunspot zoomed
in. Similarly, in the bottom panels we show the plage. We have masked
the sunspot to emphasize its values. A visual inspection of the maps
qualitatively shows the same quantitative results as in Table
\ref{tabla_compar_SPHMIpoly2d} does. The values in the umbra and
penumbra are very similar in the \first and \second columns. All the
components of the vector magnetic field (especially in X and Y) in the
region of the plage closer to the sunspot are rather similar between
them, but in the outer part they are very different. There is a
region, between $30<X (\arcsec)<85$ and $0<Y (\arcsec)<30$, where
there are several solar features hosting different sign between the
values observed by \vectorb{HMI}{Y} and \vectorb{SP}{Y} -- two circles
point some of them out in the $5^{th}$ row of the figure. These
differences are due to the manner in which the
disambiguation algorithms work on the diverse kinds of data explored here. In
the SP maps, the disambiguation results arise
from the noisy Q and U signals which in turn result from low values of the
horizontal field components in the plage. This divides those maps in two regions
each with a dominant sign. Section
\ref{sec:disambig} of the Appendix proves that different solutions
provided by different disambiguation methods
produce statistically similar results to the ones showed in Table
\ref{tabla_compar_SPHMIpoly2d}, that is, in the comparison between the magnetic
field for HMI and SP.
However, if we would consider those pixels sharing the same sign
-- that means, there where both disambiguation code yield the same
solution--, then the improvement of the correlation between the
horizontal components is important (e.g. the variation of
\vectorb{HMI}{XY} explained by \vectorb{SP}{XY} would increase in
about $\sim35\%$ in the strong plage and $\sim40\%$ in the weak plage,
see Table \ref{tabla_HMISP_posneg}).

\asdok{Figures \ref{fig:maps_B_HMI_SP_SPFF1_S2H_spher_1} and \ref{fig:maps_B_HMI_SP_SPFF1_S2H_spher_2} show a comparison of the 
components of the \vectorb{HMI}{} and \vectorb{SP}{} in spherical coordinates for the sunspot and plage respectively. The last row of these figures show the disambiguated azimuth obtained from different methods. A detailed comparison of the solutions for the disambiguation is given in the Appendix. However, at a glance, it is easy to see an important result: the disambiguation method used has an strong effect to determine the vector magnetic field, especially in the plage. Thus, the values of field strength ($|B|$), inclination ($\theta_B$), and azimuth ($\phi_B$) of the vector magnetic field derived from the data observed by HMI and SP are very similar in the sunspot (\first and \second column of Fig. \ref{fig:maps_B_HMI_SP_SPFF1_S2H_spher_1}). However, those values are quite different in the plage. Notice \vectorb{SP}{} is computed considering FF variable, while \vectorb{HMI}{} is done with FF equals to 1. Therefore, the components of the vector magnetic field in SP data may show values rather different than the corresponding in  HMI data. As we mentioned above, this effect is more notorious in the plage than in the umbra and penumbra, where the FF is close to 1. Thus, in the plage, there are regions where $|B|$ reaches values above 1 $kG$ in the SP data (in light green in the images of the \first row of Fig. \ref{fig:maps_B_HMI_SP_SPFF1_S2H_spher_1}, and in white in the images of the \first row of Fig. \ref{fig:maps_B_HMI_SP_SPFF1_S2H_spher_2}). While for those location the values are a few tens in HMI data (in dark violet in those images). For that reason, in the Table \ref{tabla_ranges_SPcgHMI}, the values of $|B|$ in the weak and strong plage in the analyzed SP data are showing a large range and a large dispersion, while the analyzed HMI data show a tighter range and a smaller dispersion. As a consequence of the large difference in these values, the linear relationship between $|B|^{HMI}$ and $|B|^{SP}$ in the plage is close to 0 (see Table \ref{tabla_compar_SPcgHMIdiscut_soloHMISP}), and the linear correlation between these values is very weak. However, when we consider $|B|$ with the inclination and azimuth, as we compose the vector magnetic field in the Cartesian coordinates, the relationships between the components in that reference system  (see Table \ref{tabla_compar_SPHMIpoly2d}) show a low relationship but different than 0, and a moderate and a weak linear correlation for the strong and weak plage respectively.}

\dosfigvert{maps_sunspot_ok_Bcg_HMI_SP_SPFF1_S2HFF1_20100701_235550_newcutpoly2d}
           {maps_plage_ok_Bcg_HMI_SP_SPFF1_S2HFF1_20100701_235550_newcutpoly2d_plage_circle}
           {Components of the vector magnetic field \newcorr{in Cartesian coordinates} observed by HMI
             (\first column), SP (\second column), SP with FF=1
             (\third column), S2H ($4^{th}$ column), S2H with FF=1
             ($5^{th}$ column), and, for making the visual comparison
             easier to the reader, again HMI ($6^{th}$ column) in the
             sunspot (\first to \third row) and the plage ($4^{th}$ to
             $6^{th}$).\label{fig:maps_B_HMI_SP_SPFF1_S2H}}

\unafig{width=\textwidth}{maps_sunspot_ok_fieldaziincli_cg_HMI_SP_SPFF1_S2HFF1_20100701_235550_newcutpoly2d_sinff_condis}
{Components of the vector magnetic field \newcorr{in spherical coordinates} observed by HMI 
	(\first column), SP (\second column), SP with FF=1
	(\third column), S2H ($4^{th}$ column), S2H with FF=1
	($5^{th}$ column), and, for making the visual comparison
	easier to the reader, again HMI ($6^{th}$ column) in the
	sunspot (\first to \third row) and the plage ($4^{th}$ to
	$6^{th}$).\label{fig:maps_B_HMI_SP_SPFF1_S2H_spher_1}}

\unafig{width=\textwidth}{maps_plage_ok_fieldaziincli_cg_HMI_SP_SPFF1_S2HFF1_20100701_235550_newcutpoly2d_sinff_condis_circle}
{Components of the vector magnetic field \newcorr{in spsherical coordinates} observed by HMI 
	(\first column), SP (\second column), SP with FF=1
	(\third column), S2H ($4^{th}$ column), S2H with FF=1
	($5^{th}$ column), and, for making the visual comparison
	easier to the reader, again HMI ($6^{th}$ column) in the
	sunspot (\first to \third row) and the plage ($4^{th}$ to
	$6^{th}$).\label{fig:maps_B_HMI_SP_SPFF1_S2H_spher_2}}

\subsection{Comparison between HMI and pseudo-SP maps\label{sec:HMI_pseudoMaps}}
This section is devoted to understand the role played for those actors
which differ between HMI and SP data.  They are: the spectral
sampling, the spatial sampling, the spectral line, the inversion code,
the disambiguation code, and the filling factor.

We call the modified SP maps  {\it pseudo-SP
  maps}, \newcorr{and we refer to them as S2H}. Table \ref{table:summary} documents the calculated
modifications  to the original SP data, and
Section \ref{sec:data_insetup} offers details about these
comparisons. The complete Table \ref{tabla_compar_SPHMIS2H} in the Appendix,
similar to the one used for the straightforward comparison, supports
our findings.

The $3^{rd}$ column in the panels of the
Fig. \ref{fig:maps_B_HMI_SP_SPFF1_S2H} shows the components of the
vector magnetic field as result of inverting the SP data with
FF=1. The $4^{th}$ shows the S2H data inverted with FF variable,
i.e., the SP data are filtered as HMI does, inverted with VFISV and
disambiguated as HMI does. The $5^{th}$ column of that figure shows
the S2H data inverted similarly but with FF=1.  For a better visual
comparison, HMI data are displayed again in the last ($6^{th}$)
column. Figure \ref{fig:maps_B_HMI_SP_SPFF1_S2H}, and
Fig. \ref{fig:maps_B_HMI_SP_SPFF1_S2H_plage_az2leka} in the Appendix,
visually summarize most of \newcorr{our} findings.

Again,  values of all the components of the vector magnetic field in the
umbra and penumbra are very similar in all the cases. In these RoIs,
the FF of the original SP data is variable but is very close to
1, so that the inversions assuming a FF=1 are not largely in error.
The signal of the Stokes
profiles is much larger than the noise in these regions, therefore, even with a
sample of 6 spectral samples -- as HMI does--, VFISV is able to
reproduce and fit adequately the Stokes profiles, and so to recover the same
physical information as inversions using the full spectrum. The
richness in the spatial structures of the penumbra may be
distinguished in the SPFF1, S2H and S2HFF1 maps. A visual comparison
of pseudo-SP maps with the SP original data reveals a penumbra
slightly blurrier, but sharper than the equivalent HMI images. Therefore, the
spatial information, understood as the physical information by pixel,
remains despite of the spectral sampling (S2H maps) or being inverted
with FF fixed to 1 (SPFF1 and S2HFF1). The three components of the
vector magnetic field of SPFF1 and S2HFF1 are {\it visually closer} to
the ones observed by HMI than the corresponding to SP and S2H. Therefore
{\em the treatment of FF during the inversion seems to be
more important than the spectral sampling or the inversion code
used}. Note that S2H and S2HFF1 data are inverted using VFISV and the
same disambiguation solution used by HMI, while SP and SPFF1 data are
inverted using MERLIN and a disambiguation solution obtained with AZAM.
In the moat -- an annular region around the sunspot mainly visible in the
magnetograms \citep{Vra74} as the place where the moving magnetic features \citep{Har73} run away from the penumbra as extension of the penumbral filaments \citep{Sai05} --  the sign of the horizontal components seem to match
in all the cases, so the disambiguation has not an apparent effect in
that region, as it does not have in the umbra and penumbra either.

Data for the plage are displayed in the bottom panel of
Fig. \ref{fig:maps_B_HMI_SP_SPFF1_S2H}. \newcorr{For the SPFF1 maps, we see the same effect in the horizontal components as for the SP data; half of the plage
shows a predominant sign, and the other half shows the opposite.} It is due to disambiguation solution chosen. Despite this visual effect, this has no
impact in the statistical values of the comparison between the vector
magnetic field of HMI and SP (see \ref{sec:disambig} in the Appendix).
For the S2H and SH2FF1 data, as we are using the same disambiguation solution that
HMI, the sign of the horizontal component matches with HMI in most of
the map. However, the sign in the solar feature located in the circles
in the HMI data (colored as blue and mixed blue-red) seems to mach
only with the S2H and S2HFF1 maps (blue and mixed colors), but not with
the SP and SPFF1 maps (hosting red-colored values). These kind of
regions, which were assigned with a different sign by the
disambiguation code have an important impact in the correlation of the
variables studied, as we have proven in Section \ref{sec:disambig} of
the Appendix.

\newcorr{Figure \ref{fig:maps_B_HMI_SP_SPFF1_S2H} allows us to appreciate visually the impact of the FF, the spatial and the spectral sampling, and the disambiguation in the comparison carried out in our study. Thus, HMI maps (\first column) seem to match visually better with maps on the far right: being the worse match with  
SP (\second column), the match is a bit better with SPFF1 (\third column), becoming acceptable with the S2H ($4^{th}$ column), and finally, with the S2HFF1 ( $5^{th}$ column), where the visual match seems to be the best possible.}

\asdok{In the comparison between the spherical components (Figs. \ref{fig:maps_B_HMI_SP_SPFF1_S2H_spher_1} and \ref{fig:maps_B_HMI_SP_SPFF1_S2H_spher_2}), we can appreciate that field strength and inclination of S2HFF1 ($5^{th}$ column) are very similar to the ones of HMI ($6^{th}$ column). Nevertheless, for the azimuth, the pseudo-SP maps are visually closer to SP data than to HMI. This is specially evident in the plage maps (see Fig. \ref{fig:maps_B_HMI_SP_SPFF1_S2H_spher_2}).  We can see the effect introduced by the disambiguation method to calculate the azimuth used to compose the horizontal components of \vectorb{}{} in the last column of Fig. \ref{fig:maps_B_HMI_SP_SPFF1_S2H_spher_2}. We can visually distinguish how close the disambiguated S2H azimuth data come to the disambiguated HMI data (both using the same disambiguation method denoted as 'DIS'), and how different the disambiguated SP data are from the HMI DIS.}  Tables \ref{tabla_compar_SPHMIS2H} and \ref{tabla_compar_SPcgHMIdiscut} in the Appendix support these visual impressions for the disambiguation method used in the core of this paper, and others method used to investigate this effect: AZAM2 and AMBIG, see \ref{sec:disambig} in the Appendix. 

Averaged statistical values in the umbra and penumbra are very similar in all the cases. The  slope, \baver, of the linear regression between vertical component of \vectorb{HMI}{} and the one of the pseudo-SP maps, shows lower values
($\sim0.85$ in the umbra, and $\sim0.90$ in the penumbra) than for the
horizontal components ($\sim0.99$). As noted above, a similar result
was found between \vectorb{HMI}{Z} and \vectorb{SP}{Z}. Therefore, this  behavior must be due to other factor(s) than spectral sampling, filling
factor or the combination of both. 

In the strong plage, the SPFF1 data can account for
$\sim20\%$ more of the variation of HMI than the basic
SP data do. In
weak plage, the improvement is $9\%$ for the horizontal components,
and $18\%$ for the vertical component. In the case of S2H, where the
improvement is only relevant for the X component of the vector
magnetic field, \vectorb{S2H}{X}  accounts for  $\sim30\%$ more of the
variation of \vectorb{HMI}{X} than \vectorb{SP}{X}. This
improvement in  X  alone  may be produced by
\newcorr{a more accurate} sampling of the Stokes Q and/or U profiles than the Stokes V.
Again, the averages show that
FF has a more important role than the spectral sampling: more
variation in \vectorb{HMI}{}
data can be explained with \vectorb{SPFF1}{} than with
\vectorb{S2H}{}. In plages, a
combination of both  FF and the spectral
sampling improves all the statistical correlations.
Values of \rval for X and Y  are $0.92$ and $0.84$ respectively, while
for the Y component is very close ($0.72$).  The \rval
for the X and Z components are 0.78 and 0.76 respectively. For the
Y component, the \rval is smaller, 0.42.
In the strong plage, the X and Z component of
S2HFF1 are respectively able to explain $84\%$ and $72\%$ of the
variation of HMI, i.e. $40\%$ and $18\%$ more than SP does. For the
weak plage, these values are $\sim60\%$, that means a $43\%$ more for X,
and $17\%$ more for Z than for SP. The percentages for Y are sightly
slower than the percentages of the case SPFF1, but larger than the
ones of the case S2H.

\asdok{In the Y maps of Fig. \ref{fig:maps_B_HMI_SP_SPFF1_S2H}, we have pointed out some structures with different sign in the SP data with respect to the HMI and S2H maps. A region with a
strong signal in the plage is present in all the SP and pseudo-SP
maps. However, in HMI these structures cannot be distinguished
(colored in light green, grey, and light blue.) That region is pointed
out by an ellipse in the Y map of the plage for the case
S2HFF1. There, the values are predominantly negative (blue) in the SP
and SPFF1 maps, while it is mixed (blue and red) in the S2H and S2HFF1
maps. Thus, the low, mixed values of that region in the HMI map are
correlated with values in the SP and SPFF1 maps having a same sign,
while for S2H and S2HFF1 maps, the HMI values are correlating with
mixed values, therefore, introducing a slight spread in the
correlation. That is not the case in the X and Z component, where most
of the solar structures seem to share the same sign in all the maps,
although they may be more intense in the SP and the pseudo-SP maps
than in the HMI ones (e.g., red in the former, yellow in the latter).

\asdok{Table \ref{tabla_compar_S2HcgHMIdiscut} in the Appendix shows the correlation between the spherical components of \vectorb{HMI}{} and \vectorb{S2H/S2HFF1}{}. The relationships for field strength, inclination and azimuth are very similar to the ones obtained for the comparison between HMI and SP. However, for the disambiguated azimuth the difference is important, especially in the plage. For the strong and weak plage, the linear relationship between the disambiguated data of HMI and S2H/S2HFF1 is  $\sim 0.8$ and $\sim 0.7$, with similar values for \rval. The reason for that moderately high values is the disambiguation code used with the S2H and S2HFF1, which is the same that HMI uses (denoted as DIS). Observing the \third row of Fig. \ref{fig:maps_B_HMI_SP_SPFF1_S2H_spher_2}, one may see a difference between the azimuth recovered from either SP/SPFF1 or S2H/S2HFF1 data with respect to HMI. , the azimuth is very similar between SP and SPFF1, and between S2H and S2HFF1. 

An inspection in the \newcorr{pre-disambiguation} azimuth  maps in the \third column of Fig. \ref{fig:maps_B_HMI_SP_SPFF1_S2H_spher_2}} reveals  of locations previously pointed out an interesting result: the values of the structures at these selected regions are very similar, with a little variation in the intensity of the colors in the main structures enclosed in circles or the ellipse. Thus, in the circle located in the left side, the values of the main structure seems to be red with a blue dot in the middle, both in HMI and SP, while in the structure located in the circle at right side, it is blue in the bottom and red in the top in both instruments. Similar correspondence is found in the complex structure located in the ellipse. In all the cases, the azimuth values are rather different in those pixels not related with a structure, i.e. with a location having a significant S/N. Therefore, the difference observed in the Y maps in those selected structures is due to the disambiguation, since they share similar values of the azimuth before the disambiguation is applied. In structures where the S/N of the instruments may be very different --mainly in the weak plage--, the \newcorr{pre-disambiguation} azimuth in those instruments may differ between each other. As we will see in the next section, this difference (and similarity) can be only explained in terms of the different S/N of the Stokes Q and U of one instrument with respect to the other one, and cannot be assigned to other actor in the comparison.}

\subsection{Comparison between SP and pseudo-SP data\label{sec:SP_pseudoMaps}}

Here, we compare the SP data with the pseudo-SP data, which
same spatial sampling and the same
observed spectral line. These comparisons correspond to
 cases E to H described above. The linear regression fits for all
these cases are presented in \newcorr{Table
\ref{tabla_compar_SP_S2H} (Cartesian components) and Table  \ref{tabla_compar_spher_SP_S2H} (spherical coordinates)} in the Appendix.
Cases E and F share all the actors, except the FF.
Note that some actors are different between the case E and F, 
e.g. the inversion code in case E is MERLIN, while for the case F is
VFISV.

The values of \baver, \rval, and \rsqval for the cases E and F
inform us about the role of the FF. Thus, for case E, both
in the strong and weak plage, the values of \baver for the horizontal
components are as large as twice the vertical
components. This behavior is similar for case F.
The values of \rsqval
tell us that the difference between considering the FF variable
or fixed equal to 1 in the same data, e.g. SPFF1 vs SP, is introducing a reduction in the
explained variation of the variable inverted with FF equals to 1, e.g. SPFF1,
with respect to the one considering FF variable, e.g. SP.
This reduction is $\sim20\%$ and $\sim35\%$ in the strong and weak plage respectively for the components in the Cartesian coordinates (see Table \ref{tabla_compar_SP_S2H}).

A visual comparison between the
\second and \third column (case E) or between the $4^{th}$ and the
$5^{th}$ column (case F) of Fig. \ref{fig:maps_B_HMI_SP_SPFF1_S2H}
confirms these results. The plage (botton panels) of the \third and
$5^{th}$ column seems to contain visually less solar structures than
in the \second and $4^{th}$ respectively.

The cases G and H simulate the
comparison between S2H and SP. Since
S2HFF1 data are the closest one to HMI, we compare them with the SP
data inverted with FF variable and fixed to 1. In these comparisons,
we are using data with \newcorr{sharing a common line (Fe {\footnotesize I} 6302\AA))} and the same spatial
resolution, but with different sampling, different inversion code,
different treatment of the FF in the ICs, and different disambiguation
algorithm. The statical values in the 
comparison between \vectorb{S2HFF1}{} and \vectorb{SP}{}
have rather similar results at each RoI to the
corresponding ones in the comparison between the \vectorb{HMI}{} and
\vectorb{SP}{}, except for the vertical component. In this case, \newcorr{the
values of \baver for the horizontal components in the linear regression fit between \vectorb{S2HFF1}{Z} and \vectorb{SP}{Z} are much
closer between them than in the linerar regression between 
\vectorb{HMI}{} and \vectorb{SP}{}}. That is significant in the plage,
where the ratio between the averaged slopes in the horizontal
components and the vertical becomes very close to 1, while for the
comparison between the original HMI and SP data is 2:1. In the umbra
and penuumbra, the values of \baver for the vertical component are $0.96\pm0.04$ and
$0.91\pm0.03$ respectively, i.e., the difference with respect to
comparison between HMI and SP is notable ($0.82\pm0.06$ and
$0.84\pm0.05$ respectively). Again, if we considered the SP treated
with a FF fixed to 1, all the statistical values associated to the
linear regression fit are improved, and particularly the ones
corresponding to the vertical component, which become very close to 1
in all the RoIs, including the strong and weak plage.

\asdok{As we mentioned above, the azimuth of SP/SPFF1 and S2H/S2HFF1 are very similar, even when the inversion code used for the former cases is MERLIN, and for the latter cases is VFISV. Both codes are Milne-Eddington, but they may have a different treatment of the weights of the Stokes parameters in their inversion scheme. Despite that fact, the azimuths are very similar, both in the sunspot (see Fig. \ref{fig:maps_B_HMI_SP_SPFF1_S2H_spher_1}) and in the plage (see Fig. \ref{fig:maps_B_HMI_SP_SPFF1_S2H_spher_2}). The significant difference comes in the relationship for $|B|$. The linear relationships for the spherical coordinates are shown in Table \ref{tabla_compar_spher_SP_S2H}.

The only difference between the data compared in the \first and \second sub-tables, i.e. for the case E (SPFF1 vs SP) and F (S2HFF1 vs S2H), is the FF used to recover \vectorb{}{}. The data compared share all the other actors (including same S/N) and use the same inversion code to recover \vectorb{}{} (MERLIN for case E, VFISV for case F). The corresponding sub-tables in Table \ref{tabla_compar_spher_SP_S2H} show: a strong linear correlation between all the components in the sunspot; moderate in the inclination and azimuth in the plage; and low in the strong plage and very low (\baver = 0.04) in the weak plage. All these effects are strictly produced for how the FF was considered during the inversion of the data. Neither the spatial sampling nor the spectral sampling nor the S/N nor the inversion code play a role in the differences shown in these sub-tables.

It is interesting to compare the statistical values obtained for the comparison between SPFF1 and SP, and those obtained for the comparison between S2HFF1 and S2H \newcorr{(see Table \ref{tabla_compar_SP_S2H} and \ref{tabla_compar_spher_SP_S2H})}. All the statistical values (\aaver, \baver, \sval, \rval, and \rsqval), in all the RoIs, are rather similar between these two comparisons. That means, the effect of considering a FF fixed to 1 versus considering it variable in the same data is similar, whether the data set has been sampled with a fine spectral sampling (SP/SPFF1) or with a coarse spectral sampling (S2H/S2HFF1). We observe the same behavior in the two first sub-tables of these tables, when the comparison is made between the Cartesian components of \vectorb{}{}. That means, the combination of spectral sampling and disambiguation is not introducing a strong effect in these comparisons. After these results, we may be tempted to consider to use a FF variable in data sampled with a coarse spectral sampling. We shall show that may be a wrong decision.   

For the cases G and H (sub-tables \third and $4^{th}$) the linear relationship between for $|B|$, $\theta_B$ and $\phi_B$ obtained in the umbra and penumbra by S2H and SP is very high, showing a strong correlation. In the plage the statistical values are showing a moderate relationship for the inclination and the azimuth in all the comparisons (cases E to H). However, the \baver and \rval  for the relationship of $|B|$ on all these cases are very low, except for the comparison between S2HFF1 and SPFF1. Only when we consider the FF equals to 1 in both data set (last sub-table), the statistical values improve. 

On the other hand, when we consider FF variable in both data set (\third sub-table), we obtain low statistical values, and not too different than for the previous sub-tables, even when in this case (G) we are comparing data inverted with different code. As we mentioned above, the effects on the relationship between the components of \vectorb{S2H}{} and \vectorb{S2HFF1}{} seem to be the same that for comparison between  \vectorb{SP}{} and  \vectorb{SPFF1}{}. We feel tempted to interpret this results as an evidence that calculating the vector magnetic field on data sampled coarsely considering a FF variable as we do with SP -- i.e. data with a fine spectral sampling data, and a stray light profile calculated as the averaged Intensity profile where the $P_{tot}$ is smaller than $35\%$ --is not a good approach. However, we should keep in mind that for an instrument where the S/N is not as good as for SP data, the situation may be different. We think a further investigation about the role played by the stray light, other way(s) to calculate it, and use it in the FF on data coarsely sampled, as HMI data are, may be very helpful. 
 
As a conclusion, comparing vector magnetic field recovered from inversions considering a different treatment of the FF (FF variable versus FF fixed to 1) reveals a small linear relationship for the $|B|$ in the plage. Notice the data compared in the \third and $4^{th}$ sub-tables \newcorr{of Tables \ref{tabla_compar_SP_S2H} and \ref{tabla_compar_spher_SP_S2H}} (S2H vs SP and S2HFF1 vs SPFF1 respectively) have the same temporal and spatial sampling (no co-alignment is needed), and the same S/N. The differences between them are the spectral sampling, the spectral lines considered (since S2H data only use Fe {\footnotesize I} 6302\AA), and the inversion code used to obtain \vectorb{}{}. Since these data share the same S/N, we conclude the difference in $|B|$ in the plage for these cases is due to the combination of how we consider the FF and the spectral sampling of the data.}

\subsubsection{About the  formation of the spectral lines observed by HMI and SP\label{sec:lineform}}

One question remains open from the previous sections: in the umbra and
penumbra, why are the values of \baver for the horizontal component
closer to 1 than for the vertical component? Thus, for the straight
comparison between \vectorb{HMI}{Z} and \vectorb{SP}{Z} the \baver is
$0.82$ in the umbra, and $0.84$ in the penumbra. As we mentioned
above, these values are slightly higher in the comparisons between HMI
and the pseudo-SP maps, but they are still far of 1 (the closest one
is for the comparison between \vectorb{HMI}{Z} and
\vectorb{S2HFF1}{Z}, with \baver = $0.93$). Table
\ref{tabla_compar_SP_S2H} sheds light on that question. The values of
\baver for the vertical component in the cases considered in this
section and presented in that table are very close to 1, both in the
umbra and penumbra. The comparisons between the vertical component
observed by SP and pseudo-SP show a little deviation from 1 (between
$0.09$ and $0.01$, although in average it is $0.04$). In fact, for the
comparison between the horizontal components \vectorb{HMI}{XY} and
\vectorb{SP}{XY}, we obtained a deviation of \baver from 1 in a
similar range of values (between $0.05$ and $0.01$). That means, the
statistical values for the vertical components in the linear
regression fit for the cases E to H are similar to the ones for the
horizontal components in the comparison between HMI and SP.

It is difficult to isolate the roles of the spectral sampling and
FF for the cases E to H because  \baver for these
cases are too close to one another, and to 1. Nevertheless, we
can say that in comparing SP and pseudo-SP maps,
i.e. observing the same spectral line, \baver is $\sim1$ in the umbra
and penumbra no matters what the sampling, the inversion code or the
treatment given to the FF are. However, for the plage, the case H tells
us that giving the same treatment of the FF fixed to 1 makes that
statistical value gets close to 1 for the vertical component, and increases the corresponding ones for the horizontal components, although the latter are still far from 1.

\begin{figure*}[]
\includegraphics[width = 1.\textwidth]{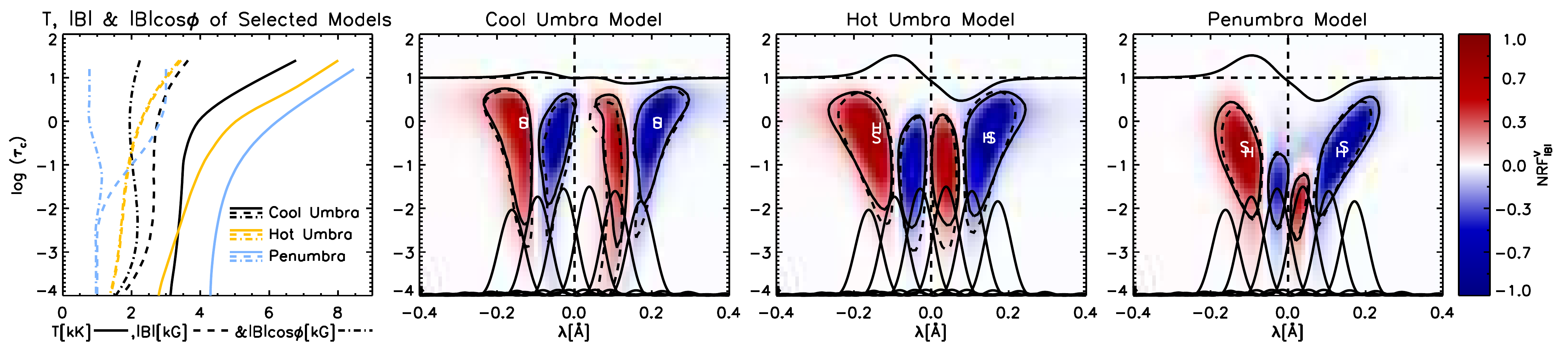}
\caption{From left to right, the $1^{st}$ panel shows the temperature (thick line), the magnetic field strength (dashed line), and the product $|B|cos\phi_B$ (dot-dashed line) through the atmosphere of the semi-empirical models used to calculate the normalized response functions of Stokes V to the magnetic field strength ($NRF^{V}_{|B|}$). The $2^{nd}, 3^{rd}$ and $4^{th}$ panels show the $NRF^{V}_{|B|}$ for Fe {\footnotesize I} 6302\AA. The thick-lined contours show the regions where the $|NRF^{V}_{|B|}|$ are larger than 0.2 for Fe {\footnotesize I} 6302\AA, and the dash-lined contours do for Fe {\footnotesize I} 6173\AA. The location of the maximum and minimum of the $NRF^{V}_{|B|}$ for those lines are marked by a "S" (SP) and a "H" (HMI).\label{fig:RFs}}
\end{figure*}

One possible scenario might explain the values of \baver in the
comparison between \vectorb{HMI}{Z} and \vectorb{SP}{Z}. In an ideal
vertical flux tube in the photosphere, the magnetic field strength
observed along the line of sight decreases as it expands in the higher
layers. Thus, if the region where the spectral line Fe~{\footnotesize
  I} 6302 \AA{} is sensitive to the magnetic field
is slightly lower than the region where the line Fe {\footnotesize I}
6173 \AA\ is, then SP would observed a slightly stronger magnetic
field than the one observed by HMI.

\cite{2011Fle} calculated the height formation of Fe {\footnotesize I}
6173 \AA\ cross-correlating Doppler velocities observed by HMI with 3D
radiation-hydrodynamic simulations. Taking into account the spatial
resolution of the HMI Dopplergrams, the authors found the formation
height of Fe {\footnotesize I} 6173 \AA\ is about $140-150\ km$. On
the other hand, several studies found the height formation for Fe
{\footnotesize I} 6302 \AA\ in higher layers of the photosphere.
\cite{2001Shch} found the height where the LTE optical depth at the
line center is $\tau_{line} = 1$ for Fe {\footnotesize I} 6302 \AA\ to
be $400\ km$ and $350\ km$ for a granule and intergranule,
respectively. \cite{2010Gre} found the line formation height located
between 138 to $201\ km$ for above the continuum formation height, and
the height where optical depth is 1 for this line to be
$244\ km$. \cite{2012Fau}, found a variation of the formation height
from $200\ km$ in the disk center to $470\ km$ close to the limb. In
all these cases, the formation height of Fe {\footnotesize I} 6302
\AA\ is higher than for Fe {\footnotesize I} 6173 \AA.

 Expressions such as {\it ``a spectral
  line is formed at a particular height in the solar atmosphere''}, or in the
best case, {\it ``a spectral line is formed within a given region''}
are an over-simplification of complex 3D physics \citep{Jud153D}. It is
 more precise to say {\it ``a spectral line is
  sensitive to a physical parameter in the region ranging from A to B''}, where
A and B are values of height given in an appropriate scale, e.g.,
geometrical or optical depth, and/or within a certain kind of magnetic structure (flux tube versus intergranular lane, for example).
Several authors have addressed such
ideas \citep{Del96, San96} by making use of {\em response functions}
(RFs, \citealt{Mei71, Bec75, Rui94}). The RF is defined as the
variation of a Stokes parameter with respect the variation of a
physical atmospheric parameter at a given optical depth and
wavelength.
\cite{Bor14} compared the results obtained from three Milne-Eddington
ICs with 3D MHD numerical simulations. The authors concluded that
using generalized RFs to determine the height at which Milne-Eddington
ICs measure physical parameters is more meaningful than comparisons of inverted parameters at a
fixed height.

We have computed the RF of Stokes V to the
magnetic field strength for those spectral lines using the SIR IC
\citep{Rui92}. That code computes the RFs under the assumption of
local thermodynamic equilibrium, what is appropriate for the spectral
lines investigated in this paper.

Figure \ref{fig:RFs} shows the behavior of temperature and the magnetic field strength
with respect to the logarithm of the continuum optical depth at 5000\AA.
The following panels show the normalized RF of Stokes V to $|B|$ ($NRF^{V}_{|B|}$)
for Fe {\footnotesize I} 6302 \AA\ calculated using these models.
The RF has been normalized to the maximum value of the RF of Stokes I to $|B|$.
The thick contours delimit where the $|NRF^{V}_{|B|}|$ for Fe {\footnotesize I}
6302 \AA\ is  greater than 0.2. The dash-lined
contours do for the $NRF^{V}_{|B|}$ calculated for Fe {\footnotesize I}
6173\AA. Figure \ref{fig:RFs} shows that the region (expressed in
$log(\tau_c)$) where those lines are sensitive to the magnetic field
strength are basically the same. The one corresponding to HMI extends
a bit farther to higher layers, but just in the tail of the
distribution of the NRF. The letters "H" and "S" pointed out the
location of the minimum and maximum of the $NRF^{V}_{|B|}$ for HMI and
SP respectively. These locations either overlap one to each other
(cool umbra) or are significantly close in $log(\tau)$ and/or in
spectral axis. The region where those lines are sensitive to $|B|$ is
basically the same. The Stokes V profile for Fe {\footnotesize I} 6302
\AA\ synthesized from the atmospheres used in this study, and the
instrumental profiles used by HMI and in the pseudo-SP maps displayed
as reference at the corresponding wavelength. One can argue about the
difference in the spectral sampling used by the instruments.  However,
the comparison between SP and S2H does not show a such a significant
difference between the \baver for the horizontal components and the
vertical one in the umbra and penumbra, being all the values very
close to 1. Therefore, we are not able to explain why \vectorb{HMI}{Z}
observes $\sim0.85$ times the apparent \asdok{longitudinal} component of the magnetic
field observed by SP, \vectorb{SP}{Z}.

We may speculate about the effects introduced by the main contributors in
the RF of the Stokes V to the magnetic field strength following the analysis
made by \cite{Cab05}. The authors calculated the expression $RF^{V}_{|B|}$
based in a phenomenological model for the weak spectral lines.
In the strong field regime, the $\sigma$ components of the Stokes V profile are represented by
(Eq. 7 in \citealt{Cab05}):

\begin{equation}
V(\lambda) = \pm A^V_0exp\left[ \frac{-(\lambda-\lambda_0 \pm \lambda_B)^2}{2(A^V_1)^2} \right]
\end{equation}

where $A^V_0$ and $A^V_1$ are the amplitude and width of the
Stokes V lobes, and the Zeeman splitting is $\lambda_B = C g_{eff} B \lambda^2_0$
, with $C \equiv 4.67 \times 10^{-13} \AA^{-1} G^{-1}$ and B the magnetic field strength (assumed
to be constant with height).
The authors found the $RF^{V}_{|B|}$ reaches the maximum sensitivity in the winds of the
$\sigma$ components of the Stokes V. Thus, for the negative sign of $A^V_1$ (Eq. 17 in \citealt{Cab05}):

\begin{equation}
R^V_{|B|} (\lambda_{max}) = e^{−1/2} C \frac{A^V_0}{A^V_1}g_{eff} \lambda ^2_0
\end{equation}\label{eq:max_rfv2b}

\newcorr{with $e$ being the Euler's number}. Therefore, for the expression for the maximum sensitivity we should know the ratio between the amplitude and the
width, and the Land\'e factor, $g_{eff}$, of the spectral line \newcorr{located at} $\lambda_0$. Any variation introduced in those factors might
change the response of the Stokes V to a variation in the magnetic field strength. Thus,
the parameters $A^V_0$ and $A^V_1$ determined from an observation taken with HMI, i.e., with a coarse
spectral sampling of $\sim 69m\AA$ in 6 points, may introduce a larger uncertainty than from an observation taken with SP, i.e., with a finer spectral sampling in more points in the spectral line.

\subsection{Comparison Between the Apparent \asdok{Longitudinal} Magnetic Flux  ($B_{app}$), Total Magnetic Flux ($\Phi$), and unsigned magnetic flux ($\widehat{\Phi}$), observed by HMI and SP. \label{sec:comp_Bapp}.}

\input{tabla_unsigFlux_HMIvsSPcgpoly2d_okmx}

\begin{figure*}[t!]
\includegraphics[width=\textwidth]{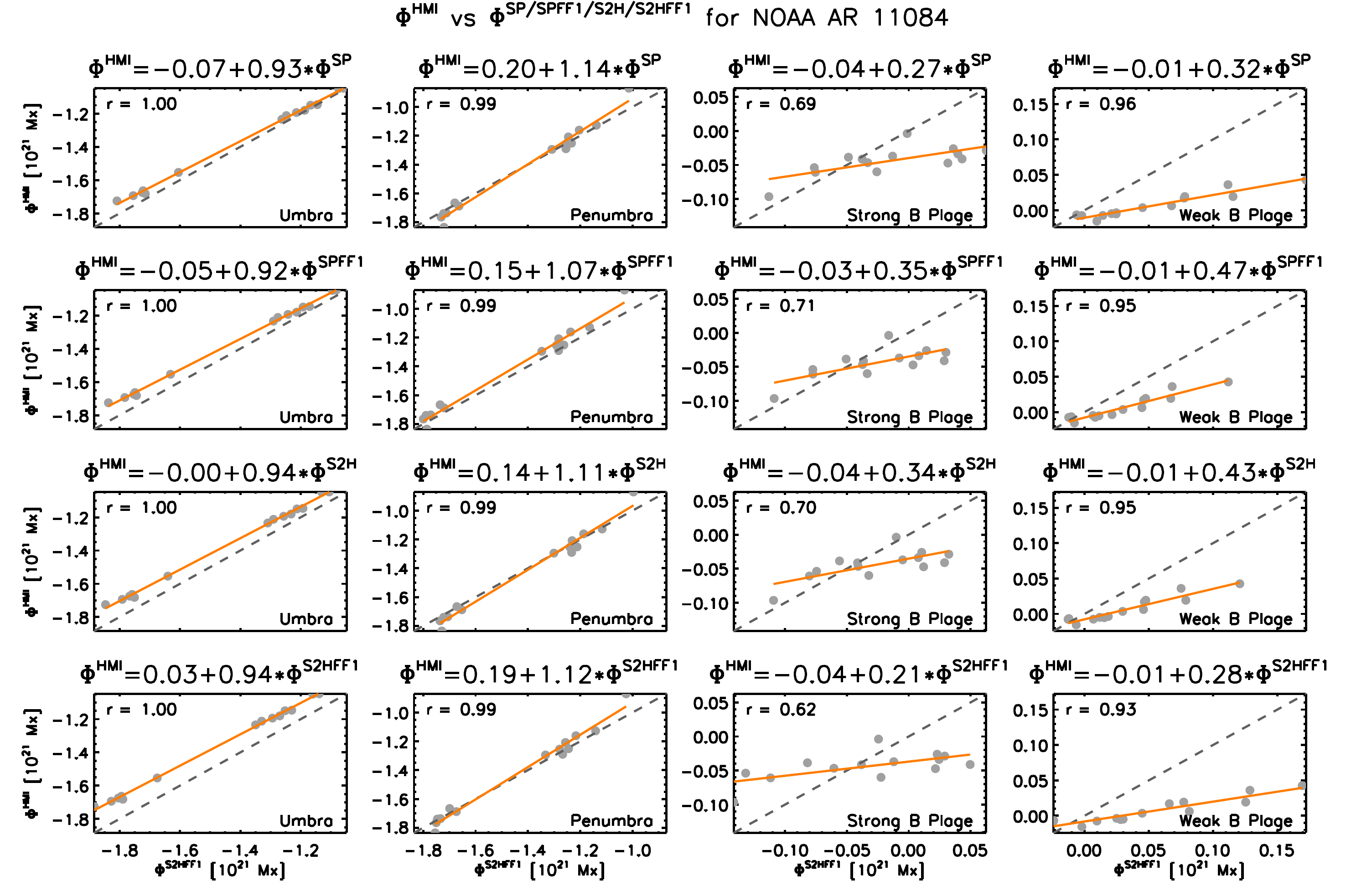}
\caption{Comparison between the total \asdok{(signed)} magnetic flux, $\Phi$, observed by HMI and SP,
  SPFF1, S2H, and S2HFF1. \asdok{Values are given in $10^{21} Mx$.}\label{fig:totflux_HMIvsSP}}
\end{figure*}

We have studied the apparent \asdok{longitudinal} magnetic flux density, $\mathit{B_{app}}$\footnote{The interested reader may find a good explanation about the magnetic field strength, \newcorr{magnetic flux and magnetic flux density} in \cite{iKel94}. In this paper, for the sake of clarity, we use $\mathit{B_{app}}$ to refer the apparent \asdok{longitudinal}  magnetic flux density, instead of $\mathit{B_{app}^{L}}$, \newcorr{which has been} used by other authors.},
 observed by HMI and SP, being $\mathit{B_{app}} =
FF\times |\mathbf{B}|\times cos\phi_B$, and expressed in $Mx/cm^2$. We
follow the notation and the conceptual interpretation given by
\cite{iBer03}, i.e.: {\it ``we denote calibrated magnetic flux density
  quantities as ‘B app’, the ‘apparent’ magnetic flux density measured
  by instrument ‘X’, in order to emphasize the inherent instrumental
  and observational limitations of the measurement.''}

Table \ref{tabla_compar_flux_SPvsHMI} shows the averaged statistical
values of the linear regression between the apparent \asdok{longitudinal} magnetic flux
measured by HMI and the one measured by SP in the RoIs. In addition, we present a
similar comparison between HMI and the pseudo-SP maps in Table \ref{tabla_compar_flux_SPHMIS2H}
in the Appendix, i.e. for the cases A to D in Table \ref{table:summary}.
It is obvious that in those cases where the FF is equal to one, the
relationship between the $B_{app}^{HMI}$ and $B_{app}^{SPFF1}$ or
$B_{app}^{S2HFF1}$, is the same as the one between the
\vectorb{HMI}{Z} and \vectorb{SPFF1}{Z} or \vectorb{S2HFF1}{Z} respectively
(see Table \ref{tabla_compar_SPHMIS2H}). 

In the umbra and penumbra, the
statistical values for the cases A to D are very similar, except for the averaged slope \baver in the
comparison between $B_{app}^{HMI}$ and $B_{app}^{SPFF1}$. The other
statistical values show a very strong linear relationship between the
apparent \asdok{longitudinal} magnetic flux density of the data compared, without a clear
impact because of considering a different spectral line, spectral
sampling, spatial sampling, filling factor or the other actors
evaluated in the comparison. The apparent \asdok{longitudinal} magnetic flux observed by
HMI in the umbra and penumbra is $84\%$ and $95\%$ respectively of the
one observed by SP.
The influence of  FF in the calculation of
$B_{app}$ is evident in the penumbra, where it is close to 1, but
mostly different than 1, for the SP, while it is virtually equal to 1
everywhere in the umbra for the SP. Therefore, the relation
$B_{app}^{HMI}$ and $B_{app}^{SP}$ in the umbra is similar to the
relationship between \vectorb{HMI}{Z} and \vectorb{SP}{Z} in that
RoI. The \sval in both RoIs is $\sim90 Mx/cm^2$.

In the plage, the situation is slightly different. Again, the maps
which consider a FF variable show lower averaged slope and lower
averaged statistical values than the ones with a FF equal to
1. Although, that different behavior in the relationship of the
$B_{app}$ observed by the maps with a different FF is little (e.g.,
less than $10\%$ in \baver). Therefore, the linear relationship
between the $B_{app}$ observed in the plage by HMI, both the strong
and the weak plage, and the one observed by SP is moderate. In the
strong and weak plage, HMI observes $32\%$ of the apparent \asdok{longitudinal} magnetic flux observed by SP, following a high-moderate linear relationship. The FF
has a little impact in that relationship in the plage, while the other
factors do not seem to play an important role in it. The \sval in the
strong plage is $\sim90 Mx/cm^2$, and $\sim50 Mx/cm^2$ in the weak
plage.

Since the errors in the averaged slope and the statistical values in
all the RoIs are small, we can conclude that the behavior in the
relationship between $B_{app}^{HMI}$ and $B_{app}^{SP}$ is consistent
through the maps analyzed in this study. That means, between $0.7 <
\mu < 1$, the relationship between $B_{app}^{HMI}$ and $B_{app}^{SP}$
is strong in the umbra and penumbra, being in the plage moderate in
the high range.

The sub-table in the middle of Table \ref{tabla_compar_flux_SPvsHMI}
shows the relationship between the total magnetic flux $\Phi$ observed
by HMI and SP. Similarly, we have done that comparison between HMI and
the pseudo-SP maps (see sub-table in the bottom of Table
\ref{tabla_compar_flux_SPHMIS2H} in the Appendix). In these
comparisons, one linear regression fit is calculated for each RoIs of
the 14 analyzed maps. As in the rest of this paper, the RoIs are the
areas of the umbra, penumbra, plage with a strong field, and plage
with a weak field. As all the maps, after being corrected to match to
each other, they share the same spatial scale. Therefore, the area
used to calculate the total flux in the (compared) maps is the
same. Of course, the value of the magnetic field inside these areas is
what differs from one map to the other. Figure
\ref{fig:totflux_HMIvsSP} shows the values for the total magnetic flux
for the RoIs observed by HMI, SP and the pseudo-SP maps and the linear
regression fit between them. The values of $\Phi$ in the panels of
Fig. \ref{fig:totflux_HMIvsSP} and in Table
\ref{tabla_compar_flux_SPHMIS2H} are expressed in $10^{21}Mx$. Note that in
this case, we are doing the linear regression fit over 14 values
corresponding to the 14 studied maps. Therefore, the intercept, slope
and statistical values ($s_e, r, r^2$) are not average values since
they correspond to only one linear fit. The errors in the intercept
and the slope are the standard errors defined for a linear regression
fit. The errors of $s_e, r, r^2$ are the rounding error in these
values.

HMI observes $93\pm1\%$ of the total magnetic flux observed by SP
($\Phi^{SP}$) in the umbra. However, for the penumbra, HMI
overestimates in $14\pm5\%$ the $\Phi^{SP}$. The statistical values
show a very strong linear relationship between $\Phi^{HMI}$ and
$\Phi^{SP}$ both in the umbra and penumbra. Similar behavior is observed in the comparison between  $\Phi^{HMI}$ and the total magnetic flux of the pseudo-SP maps (see Table \ref{tabla_compar_flux_SPHMIS2H}). \asdok{In the strong and weak plage,  HMI observes $\sim 30\%$ of the total magentic flux observed by SP. However, in the strong plage the correlation is moderate, while in the weak plage the correlation is very strong (\rval =  $0.96 \pm 0.01$). Similar behavior is observed in the comparison between $\Phi^{HMI}$ and 
the pseudo-SP maps.}



In the bottom of Table \ref{tabla_compar_flux_SPvsHMI}, we
show the comparison between the unsigned total magnetic flux measured by HMI,
$\widehat{\Phi}^{HMI}$ and the one measured by SP, $\widehat{\Phi}^{SP}$,
in the different RoIs.  The values for the umbra and penumbra are very similar to
the ones presented for the total magnetic flux. In these RoIs, as we have
proven in Section \ref{sec:comp_BSP_BHMI}, the magnetic field vector
observed by HMI and SP are very similar (see top panels
in Fig. \ref{fig:maps_B_HMI_SP_SPFF1_S2H}). The most striking value in the
comparison between $\widehat{\Phi}^{HMI}$ and $\widehat{\Phi}^{SP}$ are the
ones corresponding to the strong plage. There, the slope, \baver, is
$0.08\pm0.05$, and the variation of $\widehat{\Phi}^{HMI}$ explained by that
relationship is around just the $20\%$. The values for the weak plage are slightly
larger. The sub-table in the bottom of Table \ref{tabla_compar_flux_SPHMIS2H}
in the Appendix shows similar comparisons between the unsigned total magnetic
flux measured by HMI and the one measured by SPFF1, S2H and S2HFF1. Those comparisons
may help us to understand what happens in the plage for the comparison between
$\widehat{\Phi}^{HMI}$ and $\widehat{\Phi}^{SP}$. The statistical values for the comparison between $\widehat{\Phi}^{HMI}$ and $\widehat{\Phi}^{SPFF1}$  and $\widehat{\Phi}^{S2HFF1}$ in the strong plage are larger than those considering FF variable, but they are still
considerably smaller than the corresponding ones in the weak plage. If we look in detail the
vertical component of the map showed in Fig. \ref{fig:maps_B_HMI_SP_SPFF1_S2H}, we can many pixels showing strong values (dark blue and dark red) in the SP and S2H maps that they are associated with smaller values in HMI (and in SPFF1 and S2HFF1). \newcorr{While the
slopes in the relationships between \vectorb{HMI}{Z} and \vectorb{SP}{Z} for the
strong and weak plage are similar (\baver $\sim 0.15$), the total sum of the unsigned flux in the locations showing strong values in SP  shall be significantly larger than for the same locations at HMI, and a bit smaller if we consider FF fixed to 1. On the other hand, those locations showing weak values in HMI, i.e. the weak plage, have also associated weak values in SP. Therefore, the correlation between the total flux is better for the weak values, i.e. the weak plage, than for the strong values, i.e. the strong plage. Thus, the \rval values are again, as for $\Phi$, larger for the weak plage than for the strong plage.  In addition, the significant difference between the \rval values in the strong and the weak plage may be due to the difference in the area defined as strong plage in the HMI maps and the one defined in the SP and S2H maps.}

\asdok{
\subsection{Comparison Between the Apparent Longitudinal Magnetic Flux  ($B_{app}$), Total Magnetic Flux ($\Phi$), and unsigned magnetic flux ($\widehat{\Phi}$), observed by SP and pseudo-SP data. \label{sec:comp_Bapp}}
 
In this section, we discuss the comparison of $B_{app}$, $\Phi$, and $\widehat{\Phi}$ obtained for SP, SPFF1, S2H and S2H data, i.e. corresponding
to the comparisons E to H of Table \ref{table:summary}. 
The comparison of those magnitudes for the case E (SP vs SPFF1) is particularly interesting, and they are shown in Table \ref{tabla_compar_flux_SP_SPFF1_only}. Notice that the only difference between SP and SPFF1 is the way the FF was considered during the inversion od the Stokes profiles, being variable for the SP data, and fixed to 1 for SPFF1.
All other actors are exactly the same. 

The apparent longitudinal magnetic flux in the umbra is similar either considering FF fixed to one or variable. As we have mentioned, in the umbra the values of FF, when it is considered variable,  are virtually equal to 1 in most of the pixels, and close, but different, to 1 in the penumbra. In this RoI, the  $B_{app}^{SPFF1}$ shows an overestimation of $6\%$ with respect to $B_{app}^{SP}$. On the plage, things are different. Considering the FF fixed to 1 produce an underestimation in the apparent longitudinal magnetic flux $B_{app}^{SPFF1}$ of $\sim18\%$ and $\sim26\%$ in the strong and weak plage with respect to the case considering a FF variable, $B_{app}^{SP}$. 

For the total magnetic flux and the total unsigned magnetic flux, we observe a similar behavior, i.e. a small overestimation ($\sim 5\%$) for the relationship of those magnitudes in the penumbra, and an important underestimation in the strong ($\sim 60\%$ for both magnitudes) and weak ($\sim 60\%$ for $\Phi$, and $\sim 45\%$ for $\widehat{\Phi}$) plage when considering a FF fixed to 1 instead of being variable. 	 
That means, the underestimation in these magnitudes in the plage appears (in this comparison) only because of the way the inversion is made, i.e. how the FF is considered. Therefore, we have to keep in mind this effect \newcorr{when we compare those magnitudes for HMI (FF fixed to 1) and SP (FF variable) data in Table  \ref{tabla_compar_flux_SP_SPFF1_only}.} 

\input{tabla_unsigFlux_SP_SPFF1_only_ori_sincg_okmx}

To evaluate the effect of other actors, we have made the comparison of these magnitudes between SP and pseudo-SP data, as are described to the case E to H in Section \ref{sec:data_insetup}. The statistical values of these comparison are showed in the Table \ref{tabla_compar_flux_SP_S2H} of the Appendix. We can compare the values corresponding to the case G (S2HFF1 vs SP) with the values of the case A (HMI vs SP, Table \ref{tabla_compar_flux_SPvsHMI}), since the S2HFF1 are the closest data to HMI. Thus, if we take the pair of values of \baver ($\%$) for the relationship of those magnitudes at the strong and weak plage in the case A as \baver(strong/weak plage) [$B_{app} = 32/32, \phi = 27/32, \widehat{\phi}$ = 8/15 ], while for the case G is \baver(strong/weak plage) [$B_{app} = 79/72, \phi = 54/64, \widehat{\phi}$ = 55/50]. The difference between these values may be explained in terms of the different actors not included either in the comparison A or G, that means: the spatial sampling, the spectral line and the S/N. In other words, the largest difference comes from considering the FF variable or fixed to 1, then the spectral sampling, finally the other actors. As we have showed in the Section \ref{sec:lineform}, the lines observed by HMI and SP shows a similar response variation to the magnetic field, so we should expect a similar magnetic flux observed by those lines. We have to concluded that the spatial resolution and the S/N are the last actors (in importance) in the list of contributors to the difference in the comparison of $B_{app}$, $\phi$, and  $\widehat{\phi}$.


}

\section{Conclusions}\label{sec:conclusions}

We have compared the vector magnetic field observed by HMI with the
one observed by SP. In addition, we have used pseudo-SP data to know
about the role played for the actors what differ between the original
HMI and SP data sets. \asdok{These actors are summarized in Table \ref{table:summary}. We have to admit that other actors may introduce an effect in an eventual comparison between HMI and SP data, e.g.: the temporal evolution of solar structures during the scan of SP. Thus, while we have done our best to analyze, explain and quantify the effects of the actors mentioned in the Table \ref{table:summary}, we cannot guarantee that other actors may introduce an
	effect in our results. However, in a comparisons between HMI and SP data similar to the ones used in this study (temporal sampling, solar structures in the field of view, S/N, etc.), we expect similar results to the ones found in this study.} 

Our results are divided in two
parts. On one side, the results concerning to the data available to
the public {\it as they are}. On the other side, the results that
provide a better understanding of the instrumental and treatment
effects on the original data. Since, we have applied the same
methodology -- a linear regression fit -- to the analyzed data set (14
quasi-simultaneous maps of NOAA AR 11084), our results must be
understood in a statistical sense. \asdok{Our study has been carried out comparing the Cartesian and the \newcorr{spherical} components of the vector magnetic field \newcorr{obtained} from one
instrument to the other. Through the comparison of the \newcorr{spherical components}, we earn a better understanding of the effect introduced by the disambiguation of the azimuth. All our results are referred to both system of coordinates, since they are complementary and consistent between them, but here we summarize them referring to the Cartesian
coordinates.}

Table \ref{tabla_compar_SPHMIpoly2d} and \ref{tabla_compar_SPcgHMIdiscut_soloHMISP} show the linear relationship between
the components -- Cartesian and spherical respectively-- of the vector magnetic field \vectorb{HMI}{} and
\vectorb{SP}{} in four different regions: umbra, penumbra, plage with
a strong magnetic field, and plage with a weak magnetic field. The
values of the straightforward comparison, the averaged statistical
values and the errors found for those relationships. Similarly, Table \ref{tabla_compar_flux_SPvsHMI} shows the relation between the apparent
 \asdok{longitudinal} magnetic flux, the signed and unsigned total magnetic flux measured by HMI and SP.
Here, in the sake of
simplicity, we present our findings as an approximation of
those values. While Tables \ref{tabla_compar_SPHMIpoly2d}, \ref{tabla_compar_SPcgHMIdiscut_soloHMISP} and \ref{tabla_compar_flux_SPvsHMI} represent the main findings
of this investigation --visually supported by Figs. \ref{fig:maps_B_HMI_SP_SPFF1_S2H}, \ref{fig:maps_B_HMI_SP_SPFF1_S2H_spher_1} and \ref{fig:maps_B_HMI_SP_SPFF1_S2H_spher_2} --, we summarize here the physical meaning of our results. 

\asdok{Table \ref{tabla_ranges_SPcgHMI} shows the ranges of the values in the RoIs compared in this study. We have to consider our results valid in those intervals, and different behavior may happen in regions with different ranges of values.} \newcorr{For instance, larger cool umbrae with higher magnetic field
strengths than the one used in this study may present molecular spectral bands that likely influence MERLIN
or VFISV magnetic field inversions.} 
	
In the umbra and penumbra, the components of the vector
magnetic field observed by HMI and SP are very similar. They are showing a
strong linear correlation, having an averaged slope very close to 1 for the horizontal
components and $\sim 0.83$ for the vertical component. After a detailed study,
we can only speculate about the origin of that difference in the averaged slope.
We feel inclined to believe that difference comes from the spectral sampling
of HMI and the uncertainty introduced by sampling the Stokes V profile, therefore the
region where the atmosphere is sensitive to the magnetic field.  The averaged standard deviation about the least squares line for the components of the predicted \vectorb{HMI}{}, \sval, is $\sim 100 G$ both in the umbra and penumbra. In the umbra, the apparent \asdok{longitudinal}  magnetic flux observed by HMI, $B_{app}^{HMI}$, is $84\%$ of the one observed by SP, $B_{app}^{SP}$. In the penumbra, $B_{app}^{HMI}$ observes $95\%$ of $B_{app}^{SP}$. In both RoIs, the relationship between these variables is very strong ($\sim 0.98$). The signed and unsigned total flux measured by HMI is underestimating the one measured by SP in the umbra in about $\sim 7\%$, and overestimating it in the penumbra about $\sim 14\%$.

For the plage hosting strong magnetic field, i.e., in those pixels in
the plage where $200 G < (\mid$\vectorb{HMI}{XYZ}$\mid$ or
$\mid$\vectorb{SP}{XYZ}$\mid) < 1000 G $, we have found a moderate linear
relationship between \vectorb{HMI}{XYZ} and \vectorb{SP}{XYZ}. The averaged standard deviation about the least squares line for the predicted \vectorb{HMI}{} is $\sim160 G$ for the horizontal components, and $70 G$ for the vertical component. The linear relationships found
are able to explain about $40\%$ of the variation of the predicted \vectorb{HMI}{XY} and $50\%$ for the \vectorb{HMI}{Z}. The errors in the averaged slope and the averaged statistical
  parameters of the horizontal components are significantly larger
  than the errors for the vertical component. That means, the
  distribution of those parameters for the 14 maps is more scattered
  for the horizontal components than for the vertical one. In the plage, HMI measures $32\%$ of the apparent longitudinal magnetic flux measured by SP, with an averaged standard deviation about the least squares line for
the predicted ${B^{HMI}_{app}}$ is $90 Mx/cm^2$. Both the apparent magnetic flux and the
(signed) total magnetic flux measured by HMI and SP show a strong linear correlation, while
 the unsigned total magnetic flux shows a weak correlation with a very small slope $\sim 0.08$. In other words, in the strong plage HMI measures just $8\%$ of the unsigned total magnetic flux observed by SP in that RoI.

For the plage where the magnetic field is weak, i.e., in those pixels
in the plage where $(\mid$\vectorb{HMI}{XYZ}$\mid$ or
$\mid$\vectorb{SP}{XYZ}$\mid) < 200 G$, the relationship between
\vectorb{HMI}{XY} and \vectorb{SP}{XY} is weak, while between
\vectorb{HMI}{Z} and \vectorb{SP}{Z} is moderate. The averaged standard
deviation about the least squares line for
  the predicted \vectorb{HMI}{} is $100 G$ for the horizontal
  components, and $50 G$ for the vertical component. Note the large
  values of \sval for the horizontal components, meaning that there is
  a large scatter around the least squares line. The errors in the
  averaged slope and the averaged
  statistical parameters of the horizontal components are
  significantly larger than the errors of the vertical
  component. That is, the distribution of those parameters for the
  14 maps is more scattered for the horizontal components than for the
  vertical one. In the weak plage, the linear relationship between  $B_{app}^{HMI}$ and $B_{app}^{SP}$
  is high-moderate, while the relationships between $\Phi^{HMI}$ and $\Phi^{SP}$, and between
  $\widehat{\Phi}^{HMI}$ and $\widehat{\Phi}^{SP}$ are very strong.
  The averaged standard deviation about the least squares line for
  ${B^{HMI}_{app}}$ is $50 Mx/cm^2$.

From the comparison between HMI and the psesudo-SP maps, we conclude
that the actor having a major impact in the statistical values
analyzed is the filling factor, followed by the spectral sampling, and
finally the spatial sampling. The \newcorr{effect introduced by} the FF is important especially in
the plage. Thus, if the magnetic field observed by HMI and the
corresponding for SP where both calculated with an inversion
considering FF equal to 1, the improvement in the explained variation
of HMI by SP would be of $\sim20\%$ in the strong plage for all the
components of \vectorb{HMI}{}, and $9\%$ and $19\%$ for the horizontal
and vertical components respectively in the weak plage. The
improvements in the explained variation of corresponding predicted
variables are larger between HMI and the maps of SP sampled as HMI
does (both S2H and S2HFF1) than between HMI and SPFF1. However, the
slope for the horizontal component are worse than for the comparison
between HMI and SP with FF=1. The biggest improvements in all the
statistical values comes from considering both FF equal to 1 and same
spectral sampling between HMI and SP, i.e. when the SP data become
closest to HMI data.

\asdok{\newcorr{Considering the FF fixed to 1 or variable on the same data (SPFF1 $vs$ SP) has an important impact in the apparent longitudinal magnetic flux, the total flux and the total unsigned flux.} The apparent longitudinal magnetic flux ($B_{app}$) is overestimated by $5\%$ in the penumbra,  underestimated by $18\%$ in the strong plage, and underestimated by $26\%$ in the weak plage  when the magnetic field has been obtained considering a FF fixed to 1 ($B_{app}^{SPFF1}$), compared with when the FF is variable ($B_{app}^{SP}$). The total magnetic flux and the unsigned total magnetic flux have a similar behavior in those regions.}

The error introduced by different disambiguation methods is negligible
in the umbra and penumbra. While for the plage, the scatter introduced
by the different resolved azimuth is smaller than the scatter
introduced by the other actors. Nevertheless, if we compare only those
pixels sharing the same sign after the disambiguation, the positive impact in
all the variables might be important. For the comparison between
the original HMI and SP data, the improvement in the explained
variation of the horizontal components of the predicted
\vectorb{HMI}{} with respect considering all the pixels is $\sim40\%$
and $\sim[20-35]\%$ in the strong and weak plage. If we simply
inverted the SP data with a FF fixed to 1, as HMI does, the
correlation coefficients for between the horizontal components in
those pixels sharing the same sign in both maps become \rval
$\sim0.92$ and $\sim83$ for the strong and the weak plage. 
comparison between HMI and S2HFF1, the explained variation of the
horizontal components of the predicted \vectorb{HMI}{} becomes
$\sim85\%$ and $67\%$ in the strong and the weak plage respectively.

\newcorr{Using different disambiguation methods, we have found that about
$\sim25\%$ and $\sim35\%$ of the pixels in the plage observed and
disambiguated by HMI have an opposite sign with respect to the ones
observed and disambiguated by SP. Similar values are obtained even when  
the same data (SP) are disambiguated using different methods (AMBIG $vs$ AZAM)}. \asdok{The analysis of the disambiguation of the azimuth has proved that the opposite sign in the horizontal components of those structures in the plage is due to use different disambiguation methods, and not to opposite values of the pre-disambiguation azimuth. We suggest to investigate the option to  
use the same disambiguation code to solve the azimuth observed by HMI
and SP data. That would improve the results of a comparison between them, especially in the plage, but, more important, it would favor the quasi-simultaneous and complementary use of both data.}

In summary, the correlation between the components of the magnetic
field observed by HMI and the ones observed by SP are very high in the
umbra and penumbra, while it is moderate and weak for the strong and
weak magnetic field of the plage respectively. Similar behavior
happens for the apparent magnetic field and the total flux. The
uncertainty in predicting \vectorb{HMI}{}, $B_{app}^{HMI}$,
$\Phi^{HMI}$ and $\widehat{\Phi}^{HMI}$ from the respective values
observed by SP becomes larger as we move from the regions with
strong magnetic field and a filling factor closer to 1, to the
regions where the magnetic field is weaker
and the filling factor either is closer to 0 or more difficult to
determine, i.e. as the observation goes from the umbra to the
penumbra, then from there to strong plage, and finally from there to
the weak plage. The only exception to this behavior happens in the
statistical values of the comparison between $\Phi^{HMI}$ vs $\Phi^{SP}$ and $\widehat{\Phi}^{HMI}$ vs $\widehat{\Phi}^{SP}$ in the strong plage, which are  worse than the ones corresponding to the weak plage.

Considering inversions with a similar treatment for the filling factor
and the disambiguation of the azimuth would improve significantly the
comparison between the vector magnetic field observed by HMI and
SP. \asdok{However, to consider a FF variable for HMI data may be not reliable due to S/N constrain (more than for the fact of using 6 spectral points). A careful study would be necessary to implement or discard this option.} For the apparent \asdok{longitudinal} magnetic flux and the total flux, the impact of
the filling factor, the spectral sampling and their combination is
very little, being again the filling factor the most important actor.

\acknowledgments
\asdok{
The author thanks to the anonymous reviewer for the critical analysis of this paper. His/her corrections, comments and requests have helped to improve the final version of it.

The present paper is the result of an investigation --definitively longer than it was initially planned-- started when I was working at LMSAL and at Stanford University, and continued at HAO (UCAR/NCAR). During that time, many useful technical clarifications, comments, ideas and inspiration were given by T. Tarbell, P. Scherrer, T. Hoeksema, M. Bobra, K. Hayashi, S. Couvidat, X. Sun and R. Centeno Elliott. 
I thank P. Judge for his support, and for carefully reading this paper, providing useful comments and improvements. R. Casini and T. del Pino Alem\'an gave a useful, critical point of view on the results. 

I have used codes developed by B. Lites, P. Seagraves, J.M. Borrero, R. Centeno Elliott, S. Couvidat, K. D. Leka, G. Barnes, A. Crouch, and A. de Wijn. I thank their generosity in sharing their codes and answering my questions about how to use them properly. 

This work was supported by NASA Contract NAS5-02139
(HMI) to Stanford University and by NASA contract NNM07AA01C (Hinode) at LMSAL. Hinode is a Japanese mission developed and launched by ISAS/JAXA, collaborating with NAOJ as a domestic partner, NASA and STFC (UK) as international partners. Scientific operation of the Hinode mission is conducted by the Hinode science team organized at ISAS/JAXA. This team mainly consists of scientists from institutes in the partner countries. Support for the post-launch operation is provided by JAXA and NAOJ(Japan), STFC (U.K.), NASA, ESA, and NSC (Norway).
The National Center for Atmospheric Research is sponsored by the National Science Foundation.
}

\bibliography{aamnemonic,mib.mispapers2015,mib.instrum,mib.dea,mib.mmfs,mib.feiform,mib.newinv}
\bibliographystyle{apj}

\newpage

\appendix In this appendix we show information about the data,
detailed statistical values for the straightforward comparison between
HMI and SP, and the averaged statistical values for the comparison
between HMI and the pseudo-SP data. Despite the tediousness that all this
information may represent for the reader, we consider it valuable for
supporting our study and results.
\\
\section{Data Selection}\label{sec:app_data_sel}

Table \ref{tabla_data_sel} shows the observation date (\first column),
the starting time and the middle time for the scan in the SP
instrument (\second and \third columns), and the closest in time HMI
filtergram to the latter one ($4^{th}$ column). The positions of the
selected maps are given in heliocentric coordinates ($5^{th}$
column), and by $\mu = cos(\theta)$, being $\theta$ the heliocentric
angle ($6^{th}$ column). The size of the SP maps is given in $\arcsec\times\arcsec$ ($7^{th}$ column).

\tabladatasel

\section{Individual Linear Regression Fits and Statistical Values for the Comparison HMI vs SP}\label{sec:comp_HMISP_XYZ}
Tables \ref{tabla_linreg_all_xy} and \ref{tabla_linreg_all_z} show
several statistical values about the linear regression fit for the
regions of interest analyzed in the individual maps presented in Table
\ref{tabla_data_sel}. The statistical values \baver, \sval, \rval and
\rsqval showed in the Table \ref{tabla_compar_SPHMIpoly2d} are the
mean (averaged) values calculated from the columns $2^{nd}$, $4^{th}$,
$5^{th}$ and $6^{th}$ respectively. The errors presented in Table \ref{tabla_compar_SPHMIpoly2d} are the standard error of the mean values of the 
the quantities showed in Tables \ref{tabla_linreg_all_xy} and \ref{tabla_linreg_all_z}.
\\

\input{tablaSPcgHMInewcutXY}
\input{tablaSPcgHMInewcutz}

\section{Linear Regression and Statistical Values for the Comparison between HMI vs SP, HMI vs pseudo-SP Maps and SP vs pseudo-SP Map }
In this section, we show the averaged statistical values for the
linear regression fit between the components of the vector magnetic
field observed by HMI and SP (cases A to D, Table \ref{tabla_compar_SPHMIS2H}), and between pseudo-SP and SP (cases E to H, Table \ref{tabla_compar_SP_S2H}).  
maps. The first sub-table corresponds to the case A (HMI vs SP), and it  is the same than the Table \ref{tabla_compar_SPHMIpoly2d}. We keep it here for a better
comparison of its results with those corresponding to the cases B, C
and D of Table \ref{table:summary}. That means, for the comparison
between HMI and the SP maps inverted with FF=1 ($2^{nd}$ sub-table),
SP data sampled as HMI does ($3^{rd}$ sub-table), and SP data sampled
as HMI does and inverted with FF=1 ($4^{th}$ sub-table). The meaning
of the variables is the same as in Table
\ref{tabla_compar_SPHMIpoly2d}. See Section \ref{sec:HMI_pseudoMaps}
for details about the interpretation on the values presented here.

\input{tablaHMIvsSPcgSPFF1S2HS2HFF1poly2d}

Similarly, Tables \ref{tabla_compar_SP_S2H} and \ref{tabla_compar_spher_SP_S2H} show the averaged
statistical values for the comparison between the SP and the pseudo-SP
maps. They correspond to the cases E, F, G, and H described in Table
\ref{table:summary}. Section \ref{sec:SP_pseudoMaps} discusses the
physical meaning of the values of these comparisons.
\input{tablasSPcgSPFF1S2HS2HFF1poly2d_v2}
\input{tablas_fieldincliazi_SPcgSPFF1S2HS2HFF1poly2d_sincg}

Table \ref{tabla_compar_flux_SPHMIS2H} is an extension of the Table
\ref{tabla_compar_flux_SPvsHMI}. Here, in addition to the comparison of
the apparent magnetic flux ($B_{app}$, top sub-table) the total magnetic flux
($\Phi$, middle sub-table), and total unsigned magnetic flux ($\widehat{\Phi}$, bottom sub-table) calculated by HMI and SP, we show these relationships
between HMI and the pseudo-SP maps. A discussion about the values of
Table \ref{tabla_compar_flux_SPvsHMI} may be found in Section
\ref{sec:comp_Bapp}. Table \ref{tabla_compar_flux_SP_S2H} is equivalent to Table \ref{tabla_compar_flux_SPHMIS2H} for the comparisons between SP, SPFF1, S2H and S2HFF1, i.e. for the cases E to H.

\input{tabla_unsigFlux_HMIvsSPcgSPFF1S2HS2HFF1poly2d_okmx}
\input{tabla_unsigFlux_SP_S2H_ori_sincg_okmx}

\section{Study of the Impact of the Disambiguation in the Comparison between HMI and SP}\label{sec:disambig}

\asdok{Since our study has been mainly devoted to the comparison of the vector magnetic field in Cartesian coordinates, i.e. with the ambiguity of the azimuth resolved, we have found the need to explore in detail the impact of the procedures used to solve that problem, \newcorr{which} in the solar community is known as \textit{disambiguation} of the azimuth of the vector magnetic field. 

In this section, we show different solutions found for three codes: the one used by HMI project (denoted in this paper as `DIS'); AZAM is a code available at \textit{SolarSoft} for Hinode/SP data; and AMBIG, an automatic code ready to be used on Hinode/SP data and very similar to the one used with HMI data\footnote{Reference of the codes are properly given in Section \ref{sec:data_preparation}.}. The \textit{disambiguation} of the azimuth is based in several assumptions concerning to the magnetic field, for instance: azimuth which best matches the potential field, the minimum $J_{z}$, \textit{center of azimuths} and etc. This paper is not an exhaustive study of the disambiguation problem. \cite{iLek09}, \cite{iGeo12}, and \cite{iLek12} provide an interesting discussion on this topic, and the limitations that different methods have. 
Being that said, the results presented in the Section, about the impact using different solution of the ambiguity of the azimuth when comparing \vectorb{}{} obtained from different instruments, should be considered constrained to the type of data compared in this study: regular size sunspot, with a plage homogeneously distributed around the sunspot, in a small field of view around the sunspot, and for the specific instrumental features of each data set compared. Thus,
this investigation represents an example of the important role played by the disambiguation when one wants to recover the vector magnetic field from real observations.}

We have used three different disambiguation methods to calculate the
vector magnetic field observed by SP: i) using AZAM with the option
{\it ``center''} in the sunspot, then we apply the option {\it
  ``smooth''} in the whole map, ii) using AZAM with the manual method
option {\it ``center''} in the sunspot, then we use the manual {\it
  "wads"} action on the plage and small areas of the sunspot, finally
the whole map is smoothed, referred as \vectorb{SP\ AZAM2}{}; and iii)
using the automatic disambiguation method AMBIG, developed by
\citealt{Lek09}\footnote{AZAM is available through the SolarSoft
  package at
  \href{http://www.lmsal.com/solarsoft/}{http://www.lmsal.com/solarsoft/}. AMBIG
  is available at
  \href{https://www.cora.nwra.com/AMBIG/}{https://www.cora.nwra.com/AMBIG/}.},
referred as \vectorb{SP\ AMBIG}{}. All the results presented in the
main body of this paper have been calculated using the disambiguation
method i). In this case, we simply referred to is as
$\mathbf{B^{\mathbf{SP}}}$ when we refer to the Cartesian coordinates. To distinguish the azimuth for the disambiguated azimuth, we have explicitly denoted them by $\phi_B^{SP}$ and $\phi_B^{SP\ AZAM}$ respectively.

We first study the effect of the disambiguation on the azimuth (i.e. in spherical coordinates). Then, we give a detailed discussion of the effect of the disambiguation on the Cartesian coordinates. 

\unafig{width=\textwidth}{maps_sunspot_ok_fieldaziincli_cg_HMI_SP_SPFF1_S2HFF1_20100701_235550_dis_newcutpoly2d_sinff_allazi_2filas}{Comparison between the azimuth disambiguated on HMI (explicitly denoted as HMI DIS) and SP using AZAM and AMBIG (\first row), and SP using AZAM2 and S2H using DIS (\second row), for inversions considering FF variable (SP an S2H) and FF fixed to 1 (SPFF1 and S2HFF1).\label{fig:azi_dis_all}}

\input{tablaHMIvsSP_fieldaziincli_dis_ff1_newcutpoly2d}
\input{tablaHMIvsS2H_fieldaziincli_dis_ff1_newcutpoly2d}

\asdok{Figure \ref{fig:azi_dis_all} shows the azimuth disambiguated for HMI in the \first and $6^{th}$ column in both rows. We have explicitly denoted the disambiguated azimuth observed by HMI as `HMI DIS'. In the \second and \third images of the \first row, we show the disambiguated azimuth for SP and SFFF1. The same azimuth values disambiguated with AMBIG are shown in the $4^{th}$ and $5^{th}$ images. In the \second row, the \second and \third maps are showing the disambiguated SP and SPFF1 azimuth with AZAM2, and the $4^{th}$ and $5^{th}$ images show S2H and S2HFF1 azimuth disambiguated as HMI does, therefore, they are also denoted with `DIS'.
	At a glance, it is clear that different disambiguation codes may give slightly different solutions to the ambiguty of the azimuth (SP AZAM vs SP AZAM2), or quite different solutions (SP AZAM or AZAM2 vs SP AMBIG), even when the codes are working on the same data (azimuth of SP data). \newcorr{This difference is most noticeable in the plage.} The solution found by the same code (DIS) applied to similar data (S2H and HMI) gives a close solution between those data set. The statistical values of this visual comparison are given in Table \ref{tabla_compar_SPcgHMIdiscut} and \ref{tabla_compar_S2HcgHMIdiscut}. The \baver for the relationship between the disambiguated azimuths $\phi_B^{HMI DIS}$ and $\phi_B^{S2H DIS}$ in the weak and strong plage  are as high as $\sim0.8$ and $0.6$ respectively, showing a moderately strong correlation (\rval $\sim0.85$ and $0.75$ respectively). However, it would be a mistake to considerer a solution to be the \textit{`right solution'} because of these high values in the correlation between of the azimuths. Since, what these statistical values tell us is that the solution for the ambiguity in the azimuth for the data compared is very close, if not the same, in both data set.  We  conclude that using different methods for the same data set may provide different solutions (e.g. AZAM vs AMBIG on SP data), but using similar methods with similar data set provides similar solutions (SP AZAM vs SP AZAM2 or HMI DIS vs S2H DIS).}

Table \ref{tabla_hmi_sp_disamb} shows the results for the
straightforward comparison between HMI and SP, being SP disambiguated
with the methods mentioned above. Again, for a better comparison we
keep the results showed in Table \ref{tabla_compar_SPHMIpoly2d} as the
\first sub-table. The \second and \third sub-tables show the same
comparison between the components of the magnetic field observed by
HMI and the ones observed by SP when they have been disambiguated with
the methods ii) and iii) respectively. The results for the three
disambiguation methods are statistically similar in all the RoIs. That
means, to apply these other disambiguation methods to SP data does not
have an impact in the comparison with HMI disambiguated data.
Figure \ref{fig:maps_B_HMI_SP_SPFF1_S2H_plage_az2leka} shows the same maps
that Fig. \ref{fig:maps_B_HMI_SP_SPFF1_S2H}, but here they are disambiguated with method AZAM2 (top panels)
and AMBIG (bottom panels). The maps in Fig. \ref{fig:maps_B_HMI_SP_SPFF1_S2H_plage_az2leka}
 show a more coherent spatial distribution of the values than for the maps in Fig. \ref{fig:maps_B_HMI_SP_SPFF1_S2H}. However, this fact has not an impact in the
 statistical values for the comparison of the components of the vector magnetic field,
 since most of the pixels showing that different solution are not considered in the comparison, not even as part of the weak plage, because their polarization signal is below the criteria used in this paper (see Section \ref{sec:data_preparation}).

\dosfigvert{maps_plage_ok_Bcg_HMI_SP_SPFF1_S2HFF1_20100701_235550_newdis_newcutpoly2d}
           {maps_plage_ok_Bcg_HMI_SP_SPFF1_S2HFF1_20100701_235550_lekadis_newcutpoly2d}
           {Components of the \newcorr{Cartesian components of the } vector magnetic field observed by HMI
             (\first column), SP (\second column), SP with FF=1
             (\third column), S2H ($4^{th}$ column), S2H with FF=1
             ($5^{th}$ column), and, for making the visual comparison
             easier to the reader, again HMI ($6^{th}$ column) in the
             sunspot (\first to \third row) and the plage ($4^{th}$ to
             $6^{th}$. Top: maps disambiguated using AZAM2. Bottom: maps
             disambguated using AMBIG.
             \label{fig:maps_B_HMI_SP_SPFF1_S2H_plage_az2leka}}

\input{tablaHMIvsSP_SPAZAM2_Leka_poly2d}

However, as we pointed out in Section \ref{sec:comp_BSP_BHMI}, there
are regions where the horizontal components of the vector magnetic
field observed by HMI show opposite (or mixed) polarity with respect
to the ones observed by SP. We have studied in how many pixels in the
comparison between HMI and SP that happens. Since the three
disambiguations \newcorr{methods} provide statistically the same results for the umbra
and the penumbra (see Table \ref{tabla_hmi_sp_disamb}), we only offer
the results for the strong and weak plage in the following tables.
Tables \ref{tabla_HMISP_posneg},
\ref{tabla_compar_SPFF1cgHMInewcutposneg}, and
\ref{tabla_SPSPLeka_posneg} are divided as follows. In the top of
these tables, there is the comparison between the HMI and SP/SPFF1
without considering any relationship between the sign of the
components, as it has been done in the main body of the paper (e.g.,
the sub-table in the top of \ref{tabla_HMISP_posneg} is the same that
\ref{tabla_compar_SPHMIpoly2d}). In the middle and the bottom of these
tables, we have done the same analysis, i.e., linear regression fit
and averaged statistical values for the components of the magnetic
field observed by two data set at the pixels showing a positive
relationship (middle sub-table) and a negative relationship (bottom
sub-table) between their horizontal components. In addition, in these
tables we show the percentage of the pixels evaluated in the
comparison. For the sub-table in the top, the percentages refer to the
total size of the map, while for the sub-tables in the middle and the
bottom they refer to the size of the RoI considered. Table
\ref{tabla_HMISP_posneg} refers to the straightforward comparison
between \vectorb{HMI}{} and \vectorb{SP}{}, i.e., as it is considered
in Section \ref{sec:comp_BSP_BHMI}.  We have done a similar analysis
for the comparison between HMI and SP data disambiguated with AZAM
method ii) and with AMBIG code. For the three disambiguation methods
used in this paper, the percentages of pixels having the same sign for
the horizontal components of the magnetic field are very similar in
the three tables, either for the strong plage ($73-80\%$) or the weak
plage ($61-67\%$). The number of pixels with opposite sign in the
comparison is, although calculated independently, the complementary to
$100\%$ of those values. That happens even when we consider the same
data, from the same instrument (same spatial and spectral sampling,
same spectral line) and solved with the same inversion code (IC) as it is
showed in Table \ref{tabla_SPSPLeka_posneg}, where we compare
\vectorb{SP\ AMBIG}{} with \vectorb{SP}{}. Because of all the results
are similar in the three cases, we conclude that the different
disambiguation solutions used in this study - AZAM methods i) and ii)
and AMBIG - introduce a change of sign in the solar structures with
respect to the solution found by HMI observed about $\sim25\%$ in the
strong plage, and about $\sim35\%$ in the weak plage.

It is obvious that disentangling the mixed sings favor a better linear
relationship between the horizontal components. If we would only
consider those pixels where the relationship of the horizontal
component is positive -- i.e., the solutions of disambiguation
methods used here match with the one found by HMI--, all the averaged
statistical values \sval, \rval, and \rsqval will have a significant \newcorr{improvement},
both in the strong and the weak plage. Table
\ref{tabla_compar_SPFF1cgHMInewcutposneg} compares the HMI data with
the SP data inverted considering a FF fixed to 1. The correlation for
between the horizontal components in those pixels sharing the same
sign in both maps become \rval $\sim0.92$ and $\sim83$ for the strong
and the weak plage.

The values associate to the Y component of the vector magnetic field
is the one showing the biggest improvement. As we explained in Section
\ref{sec:HMI_pseudoMaps}, there are several solar structures in the
\vectorb{HMI}{Y} maps of the plage showing opposite or mixed sign with
respect the SP or pseudo-SP maps (see Figure 
\ref{fig:maps_B_HMI_SP_SPFF1_S2H}). In summary, if we do not consider
those structures the linear relationship between HMI and SP will
become stronger. Applying a reliable disambiguation code able to solve
the azimuth ambiguity to both HMI and SP data, would benefit the
comparison between them. Of course, the solution found for such a code will
rely on the azimuth found by the inversions results, but using the
same disambiguation code might favor to find similar disambiguation
solutions.

\input{tablaHMIvsSPnewcutposneg}
\input{tablaHMIvsSPFF1newcutposneg}
\input{tablaSPcgvsSPcglekadisnewcutposneg}

\end{document}

%% file: table_fieldstrength_SPcg_HMI_ori_dis_newcutpoly2d_ranges.ord2.tex
 
{
\begin{table*}[]
\scriptsize
\begin{center}
\caption{Averaged (in time) values of the minimum, the spatial mean, and the  maximum of the magetic field strength in the RoIs of NOAA AR 11084 observed by HMI and SP. Values are given in $G$.\label{tabla_ranges_SPcgHMI}}
\begin{tabular}{lcccccccccc} 
 $\mathbf{Solar Feature}$ & & $\mathbf{<|B|^{HMI}_{min}>}$ & $\mathbf{<|B|^{SP}_{min}>}$ & & $\mathbf{<|B|^{HMI}_{mean}>}$ & $\mathbf{<|B|^{SP}_{mean}>}$ & & $\mathbf{<|B|^{HMI}_{max}>}$ & $\mathbf{<|B|^{SP}_{max}>}$  \\
\hline
Umbra
 & & $ 1723\pm 234$ & $ 1881\pm 155$ & & $ 2384\pm 75 $ & $ 2528\pm 63$  & & $ 3006\pm 141$ & $ 3225\pm 123 $ \\
Penumbra
 & & $ 162\pm 90$ & $ 151\pm 81$ & & $ 1106\pm 34 $ & $ 1177\pm 24$  & & $ 2391\pm 61$ & $ 2484\pm 45 $ \\
Strong B Plage
 & & $ 33\pm 8$ & $ 47\pm 19$ & & $ 225\pm 14 $ & $ 534\pm 22$  & & $ 864\pm 67$ & $ 1773\pm 60 $ \\
Weak B Plage
 & & $ 31\pm 8$ & $ 21\pm 9$ & & $ 133\pm 6 $ & $ 423\pm 46$  & & $ 433\pm 80$ & $ 2208\pm 854 $ \\
\hline
\end{tabular}
\end{center}
\end{table*}
}

%% file: tablecomparisons.tex
\begin{table*}[t] 
\begin{center}  
\caption{Cross inversions setups. The checkmark means coincidence in the value used 
 for both data set. The cross indicates different value used between both instruments. We use `$1/2$' to explicitly mean the number spectral lines used by the inversion code for S2H (Fe {\footnotesize I} 6302\AA) and SP (Fe {\footnotesize I} 6301 \&  6302\AA) data.   
The bullet ($\bullet$) means a filling factor fixed to 1 used during the inversion of data A ($1^{st}$ symbol) or B ($2^{nd}$ symbol) - while the dot inside 
 circle ($\odot$) means a filling factor variable.  See section \ref{sec:data_insetup} for 
 a detailed description.\label{table:summary}} 
\begin{tabular}{cccccccc} 
\hline
Case & Data A vs Data B  & Spectral Sampling & Spatial Sampling & Spectral Line & Inversion Code & Disambiguation Code & Filling Factor  \\
\hline
A & HMI/SP   & $\times$ & $\times$ & $\times$ & $\times$ & $\times$ & $\bullet\odot$ \\
B & HMI/SP   & $\times$ & $\times$ & $\times$ & $\times$ & $\times$ & $\bullet\bullet$ \\
C & HMI/S2H  & $\checkmark$ & $\times $ & $\times$ & $\checkmark$ & $\checkmark$ & $\bullet\odot$ \\
D & HMI/S2H  & $\checkmark$ & $\times $ & $\times$ & $\checkmark$ & $\checkmark$ & $\bullet\bullet$ \\
E & SP/SP    & $\checkmark$ & $\checkmark$ & $\checkmark$ & $\checkmark$ & $\checkmark$ & $\bullet\odot$ \\
F & S2H/S2H  & $\checkmark$ & $\checkmark$ & $\checkmark$ & $\checkmark$ & $\checkmark$ & $ \bullet\odot$ \\
G & S2H/SP   & $\times$ & $\checkmark$ & $1/2$ & $\times$ & $\times$ & $ \bullet\odot$ \\
H & S2H/SP   & $\times$ & $\checkmark$ & $1/2$ & $\times$ & $\times$ &  $ \bullet\bullet$ \\
\hline 
\end{tabular} 
\end{center} 
\end{table*}

%% file: tablaHMIvsSPcgpoly2d.tex
 
{
\begin{table*}[]
\scriptsize
\begin{center}
\caption{Comparison between the Cartesian components of the vector magnetic field observed by HMI and SP. The meaning of the columns is as follows: 
1) solar feature or region of interest (RoI); 2) averaged linear regression fit between the 
 variables compared; 3) averaged estimated standard error about the least squares 
 line, ${<}s_e{>}$; 4) averaged correlation coefficient, ${<}r{>}$; 5) averaged coefficient of determination, ${<}r^2{>}$.
The errors in the averaged variables are the standard deviation of the corresponding mean. Values for the magnetic field components are given in $kG$.
\label{tabla_compar_SPHMIpoly2d}}
\begin{tabular}{llccc} 
 $\mathbf{Solar\ Feature}$ &  \vectorb{HMI}{}  vs \vectorb{SP}{} & $\mathbf{<s_{e}>}$ & $\mathbf{<r>}$ & $\mathbf{<r^{2}>}$ \\
\hline
Umbra
 & $ B^{HMI}_{X} = (+0.05\pm 0.10) + (+0.99\pm 0.02)\times B^{SP}_{X} $  &  
$ 0.09\pm 0.02 $ & $ 0.99\pm 0.00 $ & $ 0.99\pm 0.01 $ \\
 & $ B^{HMI}_{Y} = (-0.00\pm 0.04) + (+0.98\pm 0.02)\times B^{SP}_{Y} $  &  
$ 0.09\pm 0.02 $ & $ 0.99\pm 0.00 $ & $ 0.98\pm 0.01 $ \\
 & $ B^{HMI}_{Z} = (-0.23\pm 0.10) + (+0.82\pm 0.06)\times B^{SP}_{Z} $  &  
$ 0.10\pm 0.03 $ & $ 0.96\pm 0.02 $ & $ 0.93\pm 0.04 $ \\
 & & & \\
Penumbra
 & $ B^{HMI}_{X} = (+0.00\pm 0.01) + (+0.97\pm 0.02)\times B^{SP}_{X} $  &  
$ 0.10\pm 0.01 $ & $ 0.99\pm 0.00 $ & $ 0.98\pm 0.01 $ \\
 & $ B^{HMI}_{Y} = (+0.01\pm 0.01) + (+0.95\pm 0.02)\times B^{SP}_{Y} $  &  
$ 0.11\pm 0.02 $ & $ 0.99\pm 0.00 $ & $ 0.97\pm 0.01 $ \\
 & $ B^{HMI}_{Z} = (-0.02\pm 0.01) + (+0.84\pm 0.05)\times B^{SP}_{Z} $  &  
$ 0.11\pm 0.01 $ & $ 0.97\pm 0.01 $ & $ 0.94\pm 0.01 $ \\
 & & & \\
Strong B Plage
 & $ B^{HMI}_{X} = (-0.03\pm 0.01) + (+0.40\pm 0.11)\times B^{SP}_{X} $  &  
$ 0.16\pm 0.02 $ & $ 0.65\pm 0.12 $ & $ 0.44\pm 0.14 $ \\
 & $ B^{HMI}_{Y} = (+0.02\pm 0.01) + (+0.33\pm 0.08)\times B^{SP}_{Y} $  &  
$ 0.16\pm 0.01 $ & $ 0.59\pm 0.10 $ & $ 0.35\pm 0.12 $ \\
 & $ B^{HMI}_{Z} = (-0.01\pm 0.01) + (+0.15\pm 0.01)\times B^{SP}_{Z} $  &  
$ 0.07\pm 0.01 $ & $ 0.72\pm 0.02 $ & $ 0.52\pm 0.04 $ \\
 & & & \\
Weak B Plage
 & $ B^{HMI}_{X} = (-0.02\pm 0.01) + (+0.21\pm 0.04)\times B^{SP}_{X} $  &  
$ 0.10\pm 0.01 $ & $ 0.40\pm 0.05 $ & $ 0.17\pm 0.04 $ \\
 & $ B^{HMI}_{Y} = (+0.01\pm 0.01) + (+0.17\pm 0.04)\times B^{SP}_{Y} $  &  
$ 0.10\pm 0.00 $ & $ 0.36\pm 0.05 $ & $ 0.14\pm 0.04 $ \\
 & $ B^{HMI}_{Z} = (-0.01\pm 0.00) + (+0.10\pm 0.02)\times B^{SP}_{Z} $  &  
$ 0.05\pm 0.00 $ & $ 0.63\pm 0.03 $ & $ 0.40\pm 0.04 $ \\
\hline
\end{tabular}
\end{center}
\end{table*}
}

%% file: tablaHMIvsSP_fieldaziincli_dis_newcutpoly2d_soloHMISP.tex
 
{
\begin{table*}[]
\scriptsize
\begin{center}
\caption{Comparison between the spherical components of the vector magnetic field observed by HMI and SP. Magnetic field strength ($|B|$) is given in $kG$. Inclination ($\theta_{B}$) and the azimuth ($\phi_{B}$) of the magnetic field  are given in $degree$. \label{tabla_compar_SPcgHMIdiscut_soloHMISP}}
\begin{tabular}{llccc} 
 $\mathbf{Solar Feature}$ &  \vectorb{HMI}{}  vs \vectorb{SP}{} & $\mathbf{<s_{e}>}$ & $\mathbf{<r>}$ & $\mathbf{<r^{2}>}$ \\
\hline
Umbra
 & $ |B|^{HMI} = (-0.11\pm 0.31) + (0.99\pm 0.12)\times |B|^{SP} $  &  
$ 0.10\pm 0.03 $ & $ 0.92\pm 0.05 $ & $ 0.84\pm 0.09 $ \\
 & $ \theta_{B}^{HMI} = (12.60\pm 1.89) + (0.91\pm 0.01)\times \theta_{B}^{SP} $  &  
$ 1.89\pm 0.57 $ & $ 0.99\pm 0.01 $ & $ 0.98\pm 0.02 $ \\
 & $ \phi_{B}^{HMI} = (14.77\pm 6.60) + (0.82\pm 0.08)\times \phi_{B}^{SP} $  &  
$ 29.04\pm 5.67 $ & $ 0.77\pm 0.12 $ & $ 0.61\pm 0.18 $ \\
 & $ \phi_{B}^{HMI\ DIS} = (5.48\pm 10.49) + (0.97\pm 0.05)\times \phi_{B}^{SP\ AZAM} $  &  
$ 19.11\pm 3.85 $ & $ 0.91\pm 0.04 $ & $ 0.83\pm 0.08 $ \\
 & & & \\
Penumbra
 & $ |B|^{HMI} = (0.01\pm 0.01) + (0.93\pm 0.02)\times |B|^{SP} $  &  
$ 0.13\pm 0.02 $ & $ 0.96\pm 0.01 $ & $ 0.91\pm 0.02 $ \\
 & $ \theta_{B}^{HMI} = (14.37\pm 2.01) + (0.86\pm 0.02)\times \theta_{B}^{SP} $  &  
$ 4.98\pm 0.40 $ & $ 0.97\pm 0.02 $ & $ 0.94\pm 0.03 $ \\
 & $ \phi_{B}^{HMI} = (13.00\pm 4.82) + (0.85\pm 0.04)\times \phi_{B}^{SP} $  &  
$ 24.95\pm 2.49 $ & $ 0.87\pm 0.03 $ & $ 0.75\pm 0.05 $ \\
 & $ \phi_{B}^{HMI\ DIS} = (13.11\pm 6.93) + (0.92\pm 0.03)\times \phi_{B}^{SP\ AZAM} $  &  
$ 36.28\pm 6.33 $ & $ 0.92\pm 0.03 $ & $ 0.85\pm 0.05 $ \\
 & & & \\
Strong B Plage
 & $ |B|^{HMI} = (0.19\pm 0.01) + (0.07\pm 0.03)\times |B|^{SP} $  &  
$ 0.12\pm 0.01 $ & $ 0.16\pm 0.08 $ & $ 0.03\pm 0.03 $ \\
 & $ \theta_{B}^{HMI} = (56.19\pm 1.83) + (0.39\pm 0.02)\times \theta_{B}^{SP} $  &  
$ 11.06\pm 0.77 $ & $ 0.83\pm 0.01 $ & $ 0.68\pm 0.02 $ \\
 & $ \phi_{B}^{HMI} = (52.10\pm 7.52) + (0.43\pm 0.09)\times \phi_{B}^{SP} $  &  
$ 42.11\pm 2.15 $ & $ 0.46\pm 0.06 $ & $ 0.21\pm 0.06 $ \\
 & $ \phi_{B}^{HMI\ DIS} = (122.92\pm 7.75) + (0.45\pm 0.03)\times \phi_{B}^{SP\ AZAM} $  &  
$ 82.86\pm 5.05 $ & $ 0.47\pm 0.04 $ & $ 0.22\pm 0.04 $ \\
 & & & \\
Weak B Plage
 & $ |B|^{HMI} = (0.13\pm 0.01) + (0.00\pm 0.01)\times |B|^{SP} $  &  
$ 0.04\pm 0.00 $ & $ 0.01\pm 0.04 $ & $ 0.00\pm 0.00 $ \\
 & $ \theta_{B}^{HMI} = (62.72\pm 2.41) + (0.30\pm 0.02)\times \theta_{B}^{SP} $  &  
$ 11.17\pm 0.97 $ & $ 0.78\pm 0.02 $ & $ 0.61\pm 0.03 $ \\
 & $ \phi_{B}^{HMI} = (72.95\pm 4.70) + (0.19\pm 0.03)\times \phi_{B}^{SP} $  &  
$ 44.44\pm 0.88 $ & $ 0.20\pm 0.02 $ & $ 0.04\pm 0.01 $ \\
 & $ \phi_{B}^{HMI\ DIS} = (174.89\pm 9.38) + (0.23\pm 0.02)\times \phi_{B}^{SP\ AZAM} $  &  
$ 85.83\pm 2.88 $ & $ 0.26\pm 0.02 $ & $ 0.07\pm 0.01 $ \\
\hline
\end{tabular}
\end{center}
\end{table*}
}

%% file: tabla_unsigFlux_HMIvsSPcgpoly2d_okmx.tex
 
{
\begin{table*}[]
\scriptsize
\begin{center}
\caption{Comparison between the apparent magnetic flux density ($B_{app}$), the total magnetic flux ($\Phi$), and the total unsigned magnetic flux ($\widehat{\Phi}$) for HMI and SP. $B_{app}$ is given in $Mx/cm^2$. $\Phi$ and $\widehat{\Phi}$ are given in $10^{21} Mx$.\label{tabla_compar_flux_SPvsHMI}}
\begin{tabular}{llccc} 
 $\mathbf{Solar\ Feature}$ &  $B_{app}^{HMI}$  vs $B_{app}^{SP}$ & $\mathbf{<s_{e}>}$ & $\mathbf{<r>}$ & $\mathbf{<r^{2}>}$ \\
\hline
 & & & \\
Umbra  & $ B^{HMI}_{app} = (-0.26\pm 0.07) + (0.84\pm 0.04)\times B^{SP}_{app} $  &  
$ 0.09\pm 0.03 $ & $ 0.97\pm 0.02 $ & $ 0.93\pm 0.04 $ \\
Penumbra  & $ B^{HMI}_{app} = (-0.02\pm 0.01) + (0.95\pm 0.02)\times B^{SP}_{app} $  &  
$ 0.09\pm 0.01 $ & $ 0.98\pm 0.01 $ & $ 0.96\pm 0.01 $ \\
Strong B Plage  & $ B^{HMI}_{app} = (-0.02\pm 0.01) + (0.32\pm 0.06)\times B^{SP}_{app} $  &  
$ 0.09\pm 0.01 $ & $ 0.82\pm 0.02 $ & $ 0.67\pm 0.03 $ \\
Weak B Plage  & $ B^{HMI}_{app} = (-0.00\pm 0.00) + (0.32\pm 0.08)\times B^{SP}_{app} $  &  
$ 0.05\pm 0.00 $ & $ 0.72\pm 0.03 $ & $ 0.51\pm 0.04 $ \\
 & & & \\
 $\mathbf{Solar\ Feature}$ &  $\Phi^{HMI}$  vs $\Phi^{SP}$ & $\mathbf{s_{e}}$ & $\mathbf{r}$ & $\mathbf{r^{2}}$ \\
\hline
 & & & \\
Umbra  & $ \Phi^{HMI} = (-0.07\pm 0.02) + (0.93\pm 0.01)\times \Phi^{SP} $  &  
$ 0.01\pm 0.01 $ & $ 1.00\pm 0.01 $ & $ 1.00\pm 0.01 $ \\
Penumbra  & $ \Phi^{HMI} = (0.20\pm 0.07) + (1.14\pm 0.05)\times \Phi^{SP} $  &  
$ 0.04\pm 0.01 $ & $ 0.99\pm 0.01 $ & $ 0.98\pm 0.01 $ \\
Strong B Plage  & $ \Phi^{HMI} = (-0.04\pm 0.00) + (0.27\pm 0.08)\times \Phi^{SP} $  &  
$ 0.02\pm 0.01 $ & $ 0.69\pm 0.01 $ & $ 0.47\pm 0.01 $ \\
Weak B Plage  & $ \Phi^{HMI} = (-0.01\pm 0.00) + (0.32\pm 0.03)\times \Phi^{SP} $  &  
$ 0.01\pm 0.01 $ & $ 0.96\pm 0.01 $ & $ 0.91\pm 0.01 $ \\
 & & & \\
 $\mathbf{Solar\ Feature}$ &  $\widehat{\Phi}^{HMI}$  vs $\widehat{\Phi}^{SP}$ & $\mathbf{s_{e}}$ & $\mathbf{r}$ & $\mathbf{r^{2}}$ \\
\hline
 & & & \\
Umbra  & $ \widehat{\Phi}^{HMI} = (0.07\pm 0.02) + (0.93\pm 0.01)\times \widehat{\Phi}^{SP} $  &  
$ 0.01\pm 0.01 $ & $ 1.00\pm 0.01 $ & $ 1.00\pm 0.01 $ \\
Penumbra  & $ \widehat{\Phi}^{HMI} = (-0.23\pm 0.08) + (1.13\pm 0.05)\times \widehat{\Phi}^{SP} $  &  
$ 0.03\pm 0.01 $ & $ 0.99\pm 0.01 $ & $ 0.98\pm 0.01 $ \\
Strong B Plage  & $ \widehat{\Phi}^{HMI} = (0.32\pm 0.04) + (0.08\pm 0.05)\times \widehat{\Phi}^{SP} $  &  
$ 0.02\pm 0.01 $ & $ 0.45\pm 0.01 $ & $ 0.20\pm 0.01 $ \\
Weak B Plage  & $ \widehat{\Phi}^{HMI} = (0.08\pm 0.01) + (0.15\pm 0.02)\times \widehat{\Phi}^{SP} $  &  
$ 0.01\pm 0.01 $ & $ 0.89\pm 0.01 $ & $ 0.79\pm 0.01 $ \\
 & & & \\
\hline
\end{tabular}
\end{center}
\end{table*}
}

%% file: tabla_unsigFlux_SP_SPFF1_only_ori_sincg_okmx.tex
 
{
\begin{table*}[]
\scriptsize
\begin{center}
\caption{Comparison between the apparent longitudinal magnetic flux density ($B_{app}$), the total magnetic flux ($\Phi$), and the total unsigned magnetic flux ($\widehat{\Phi}$) for SP and SP with FF=1. $B_{app}$ is given in $Mx/cm^2$. $\Phi$ and $\widehat{\Phi}$ are given in $10^{21} Mx$.\label{tabla_compar_flux_SP_SPFF1_only}}
\begin{tabular}{llccc} 
 $\mathbf{Solar\ Feature}$ &  $B_{app}^{SPFF1}$  vs $B_{app}^{SP}$ & $\mathbf{<s_{e}>}$ & $\mathbf{<r>}$ & $\mathbf{<r^{2}>}$ \\
\hline
 & & & \\
Umbra  & $ B^{SPFF1}_{app} = (-0.07\pm 0.05) + (0.98\pm 0.02)\times B^{SP}_{app} $  &  
$ 0.03\pm 0.01 $ & $ 1.00\pm 0.00 $ & $ 0.99\pm 0.00 $ \\
Penumbra  & $ B^{SPFF1}_{app} = (0.01\pm 0.00) + (1.05\pm 0.01)\times B^{SP}_{app} $  &  
$ 0.03\pm 0.01 $ & $ 1.00\pm 0.00 $ & $ 1.00\pm 0.00 $ \\
Strong B Plage  & $ B^{SPFF1}_{app} = (-0.01\pm 0.00) + (0.82\pm 0.04)\times B^{SP}_{app} $  &  
$ 0.07\pm 0.01 $ & $ 0.98\pm 0.00 $ & $ 0.96\pm 0.00 $ \\
Weak B Plage  & $ B^{SPFF1}_{app} = (-0.00\pm 0.00) + (0.74\pm 0.07)\times B^{SP}_{app} $  &  
$ 0.03\pm 0.00 $ & $ 0.95\pm 0.01 $ & $ 0.90\pm 0.01 $ \\
 & & & \\
 $\mathbf{Solar\ Feature}$ &  $\Phi^{SPFF1}$  vs $\Phi^{SP}$ & $\mathbf{<s_{e}>}$ & $\mathbf{<r>}$ & $\mathbf{<r^{2}>}$ \\
\hline
 & & & \\
Umbra  & $ \Phi^{SPFF1} = (-0.02\pm 0.00) + (1.00\pm 0.00)\times \Phi^{SP} $  &  
$ 0.00\pm 0.01 $ & $ 1.00\pm 0.01 $ & $ 1.00\pm 0.01 $ \\
Penumbra  & $ \Phi^{SPFF1} = (0.05\pm 0.01) + (1.06\pm 0.00)\times \Phi^{SP} $  &  
$ 0.00\pm 0.01 $ & $ 1.00\pm 0.01 $ & $ 1.00\pm 0.01 $ \\
Strong B Plage  & $ \Phi^{SPFF1} = (-0.02\pm 0.00) + (0.66\pm 0.04)\times \Phi^{SP} $  &  
$ 0.01\pm 0.01 $ & $ 0.98\pm 0.01 $ & $ 0.96\pm 0.01 $ \\
Weak B Plage  & $ \Phi^{SPFF1} = (-0.00\pm 0.00) + (0.61\pm 0.04)\times \Phi^{SP} $  &  
$ 0.00\pm 0.01 $ & $ 0.98\pm 0.01 $ & $ 0.96\pm 0.01 $ \\
 & & & \\
 $\mathbf{Solar\ Feature}$ &  $\widehat{\Phi}^{SPFF1}$  vs $\widehat{\Phi}^{SP}$ & $\mathbf{<s_{e}>}$ & $\mathbf{<r>}$ & $\mathbf{<r^{2}>}$ \\
\hline
 & & & \\
Umbra  & $ \widehat{\Phi}^{SPFF1} = (0.02\pm 0.00) + (1.00\pm 0.00)\times \widehat{\Phi}^{SP} $  &  
$ 0.00\pm 0.01 $ & $ 1.00\pm 0.01 $ & $ 1.00\pm 0.01 $ \\
Penumbra  & $ \widehat{\Phi}^{SPFF1} = (0.01\pm 0.01) + (1.03\pm 0.00)\times \widehat{\Phi}^{SP} $  &  
$ 0.00\pm 0.01 $ & $ 1.00\pm 0.01 $ & $ 1.00\pm 0.01 $ \\
Strong B Plage  & $ \widehat{\Phi}^{SPFF1} = (0.18\pm 0.02) + (0.60\pm 0.02)\times \widehat{\Phi}^{SP} $  &  
$ 0.01\pm 0.01 $ & $ 0.99\pm 0.01 $ & $ 0.99\pm 0.01 $ \\
Weak B Plage  & $ \widehat{\Phi}^{SPFF1} = (0.10\pm 0.01) + (0.44\pm 0.04)\times \widehat{\Phi}^{SP} $  &  
$ 0.01\pm 0.01 $ & $ 0.95\pm 0.01 $ & $ 0.89\pm 0.01 $ \\
 & & & \\
\hline
\end{tabular}
\end{center}
\end{table*}
}

%% file: tablaSPcgHMInewcutXY.tex
 
{
\begin{table*}[t] 
\scriptsize
\begin{center}
\caption{Linear regression fit between the horizontal components of the vector magentic field observed by HMI and SP. Colums are as follows: 1) location of the
enter of map given as $\mu = cos(\theta)$, being $\theta$ the heliocentric angle; 2) intercept ($a$) and slope ($b$) of the linear regression fit; 
3) confidence interval at a $95\%$ confidence level for $a$ and $b$; 4) estimated standard error about the least squares line, $s_e$; 
5) correlation coefficient, $r$; 6) coefficient of determination, $r^2$, and 7) Spearman's rank correlation coefficient $r_s$.
\label{tabla_linreg_all_xy}} 
\begin{tabular}{ccccccccccccccc}
 & & & & & & & & & & & & & & \\
 \multicolumn{15}{c}{$B_X$} \\
 \cline{1-15}\\
 & \multicolumn{7}{c}{Umbra}\hspace{.3cm} & \multicolumn{7}{c}{Penumbra}  \\
   $\mu$ & & (a,b) & (ci a,ci b) & $s_e$ & r & $r^{2}$ & $r_{S}$ &  & (a,b) & (ci a,ci b) & $s_e$ & r & $r^{2}$ & $r_{S}$ \\ 
0.708 & & 
$(-0.18,+1.02)$  &  $(\pm0.03,\pm0.02)$  &  $+0.13$  &  $+0.98$  &  $+0.97$  &  $+0.98$
 & & 
$(+0.03,+0.93)$  &  $(\pm0.00,\pm0.01)$  &  $+0.11$  &  $+0.98$  &  $+0.97$  &  $+0.99$
 \\
0.903 & & 
$(+0.03,+0.96)$  &  $(\pm0.01,\pm0.01)$  &  $+0.09$  &  $+0.99$  &  $+0.99$  &  $+0.99$
 & & 
$(+0.02,+0.96)$  &  $(\pm0.00,\pm0.00)$  &  $+0.10$  &  $+0.99$  &  $+0.98$  &  $+0.99$
 \\
0.906 & & 
$(-0.03,+0.99)$  &  $(\pm0.01,\pm0.01)$  &  $+0.12$  &  $+0.99$  &  $+0.98$  &  $+0.98$
 & & 
$(+0.01,+0.95)$  &  $(\pm0.00,\pm0.00)$  &  $+0.09$  &  $+0.99$  &  $+0.98$  &  $+0.99$
 \\
0.910 & & 
$(-0.04,+0.99)$  &  $(\pm0.01,\pm0.01)$  &  $+0.08$  &  $+1.00$  &  $+0.99$  &  $+1.00$
 & & 
$(+0.00,+0.96)$  &  $(\pm0.00,\pm0.00)$  &  $+0.10$  &  $+0.99$  &  $+0.98$  &  $+0.99$
 \\
0.912 & & 
$(-0.03,+0.97)$  &  $(\pm0.01,\pm0.01)$  &  $+0.08$  &  $+1.00$  &  $+0.99$  &  $+1.00$
 & & 
$(+0.01,+0.95)$  &  $(\pm0.00,\pm0.00)$  &  $+0.09$  &  $+0.99$  &  $+0.98$  &  $+0.99$
 \\
0.909 & & 
$(-0.00,+0.96)$  &  $(\pm0.01,\pm0.01)$  &  $+0.09$  &  $+0.99$  &  $+0.99$  &  $+1.00$
 & & 
$(-0.00,+0.96)$  &  $(\pm0.00,\pm0.00)$  &  $+0.08$  &  $+0.99$  &  $+0.99$  &  $+0.99$
 \\
0.907 & & 
$(+0.01,+0.98)$  &  $(\pm0.01,\pm0.01)$  &  $+0.08$  &  $+1.00$  &  $+0.99$  &  $+1.00$
 & & 
$(-0.00,+0.97)$  &  $(\pm0.00,\pm0.00)$  &  $+0.08$  &  $+0.99$  &  $+0.99$  &  $+0.99$
 \\
0.807 & & 
$(+0.07,+0.98)$  &  $(\pm0.01,\pm0.01)$  &  $+0.07$  &  $+1.00$  &  $+0.99$  &  $+1.00$
 & & 
$(-0.01,+0.97)$  &  $(\pm0.00,\pm0.01)$  &  $+0.10$  &  $+0.99$  &  $+0.98$  &  $+0.98$
 \\
0.801 & & 
$(+0.11,+0.99)$  &  $(\pm0.01,\pm0.01)$  &  $+0.08$  &  $+0.99$  &  $+0.99$  &  $+0.99$
 & & 
$(-0.00,+0.98)$  &  $(\pm0.00,\pm0.01)$  &  $+0.11$  &  $+0.99$  &  $+0.98$  &  $+0.98$
 \\
0.793 & & 
$(+0.19,+1.01)$  &  $(\pm0.01,\pm0.01)$  &  $+0.09$  &  $+0.99$  &  $+0.99$  &  $+0.99$
 & & 
$(+0.01,+0.99)$  &  $(\pm0.01,\pm0.01)$  &  $+0.12$  &  $+0.99$  &  $+0.97$  &  $+0.98$
 \\
0.788 & & 
$(+0.14,+0.98)$  &  $(\pm0.01,\pm0.01)$  &  $+0.08$  &  $+0.99$  &  $+0.99$  &  $+0.99$
 & & 
$(+0.00,+0.99)$  &  $(\pm0.01,\pm0.01)$  &  $+0.12$  &  $+0.99$  &  $+0.97$  &  $+0.98$
 \\
0.780 & & 
$(+0.14,+1.00)$  &  $(\pm0.01,\pm0.01)$  &  $+0.07$  &  $+0.99$  &  $+0.99$  &  $+1.00$
 & & 
$(-0.01,+0.99)$  &  $(\pm0.01,\pm0.01)$  &  $+0.12$  &  $+0.99$  &  $+0.97$  &  $+0.98$
 \\
0.773 & & 
$(+0.13,+1.00)$  &  $(\pm0.01,\pm0.01)$  &  $+0.08$  &  $+0.99$  &  $+0.99$  &  $+0.99$
 & & 
$(-0.00,+0.98)$  &  $(\pm0.00,\pm0.01)$  &  $+0.11$  &  $+0.99$  &  $+0.97$  &  $+0.98$
 \\
0.749 & & 
$(+0.13,+1.01)$  &  $(\pm0.01,\pm0.01)$  &  $+0.08$  &  $+0.99$  &  $+0.99$  &  $+0.99$
 & & 
$(+0.01,+0.97)$  &  $(\pm0.00,\pm0.01)$  &  $+0.10$  &  $+0.99$  &  $+0.98$  &  $+0.98$
 \\
   & & & & & & & & & & & & & & \\
 & \multicolumn{7}{c}{Strong Plage}\hspace{.3cm} & \multicolumn{7}{c}{Weak Plage}  \\  
   $\mu$ & & (a,b) & (ci a,ci b) & $s_e$ & r & $r^{2}$ & $r_{S}$ &  & (a,b) & (ci a,ci b) & $s_e$ & r & $r^{2}$ & $r_{S}$ \\ 
0.708 & & 
$(-0.01,+0.16)$  &  $(\pm0.01,\pm0.02)$  &  $+0.20$  &  $+0.34$  &  $+0.11$  &  $+0.30$
 & & 
$(-0.02,+0.11)$  &  $(\pm0.00,\pm0.01)$  &  $+0.11$  &  $+0.26$  &  $+0.07$  &  $+0.33$
 \\
0.903 & & 
$(-0.03,+0.52)$  &  $(\pm0.01,\pm0.02)$  &  $+0.16$  &  $+0.73$  &  $+0.53$  &  $+0.71$
 & & 
$(-0.03,+0.18)$  &  $(\pm0.00,\pm0.01)$  &  $+0.10$  &  $+0.32$  &  $+0.10$  &  $+0.32$
 \\
0.906 & & 
$(-0.04,+0.55)$  &  $(\pm0.01,\pm0.02)$  &  $+0.14$  &  $+0.79$  &  $+0.62$  &  $+0.76$
 & & 
$(-0.04,+0.25)$  &  $(\pm0.00,\pm0.01)$  &  $+0.09$  &  $+0.43$  &  $+0.18$  &  $+0.41$
 \\
0.910 & & 
$(-0.02,+0.52)$  &  $(\pm0.01,\pm0.02)$  &  $+0.14$  &  $+0.77$  &  $+0.59$  &  $+0.75$
 & & 
$(-0.03,+0.24)$  &  $(\pm0.00,\pm0.01)$  &  $+0.09$  &  $+0.42$  &  $+0.18$  &  $+0.41$
 \\
0.912 & & 
$(-0.02,+0.51)$  &  $(\pm0.01,\pm0.02)$  &  $+0.14$  &  $+0.76$  &  $+0.58$  &  $+0.74$
 & & 
$(-0.03,+0.24)$  &  $(\pm0.00,\pm0.01)$  &  $+0.09$  &  $+0.44$  &  $+0.19$  &  $+0.42$
 \\
0.909 & & 
$(-0.01,+0.50)$  &  $(\pm0.01,\pm0.02)$  &  $+0.15$  &  $+0.74$  &  $+0.55$  &  $+0.72$
 & & 
$(-0.01,+0.22)$  &  $(\pm0.00,\pm0.01)$  &  $+0.10$  &  $+0.40$  &  $+0.16$  &  $+0.41$
 \\
0.907 & & 
$(-0.02,+0.49)$  &  $(\pm0.01,\pm0.02)$  &  $+0.14$  &  $+0.75$  &  $+0.56$  &  $+0.70$
 & & 
$(-0.02,+0.28)$  &  $(\pm0.00,\pm0.01)$  &  $+0.09$  &  $+0.48$  &  $+0.23$  &  $+0.49$
 \\
0.807 & & 
$(-0.04,+0.34)$  &  $(\pm0.01,\pm0.02)$  &  $+0.17$  &  $+0.62$  &  $+0.38$  &  $+0.59$
 & & 
$(-0.02,+0.19)$  &  $(\pm0.00,\pm0.01)$  &  $+0.10$  &  $+0.41$  &  $+0.17$  &  $+0.46$
 \\
0.801 & & 
$(-0.04,+0.33)$  &  $(\pm0.01,\pm0.02)$  &  $+0.17$  &  $+0.60$  &  $+0.36$  &  $+0.58$
 & & 
$(-0.03,+0.20)$  &  $(\pm0.00,\pm0.01)$  &  $+0.10$  &  $+0.42$  &  $+0.18$  &  $+0.48$
 \\
0.793 & & 
$(-0.04,+0.34)$  &  $(\pm0.01,\pm0.02)$  &  $+0.17$  &  $+0.63$  &  $+0.39$  &  $+0.61$
 & & 
$(-0.02,+0.20)$  &  $(\pm0.00,\pm0.01)$  &  $+0.10$  &  $+0.44$  &  $+0.20$  &  $+0.50$
 \\
0.788 & & 
$(-0.03,+0.31)$  &  $(\pm0.01,\pm0.02)$  &  $+0.17$  &  $+0.57$  &  $+0.33$  &  $+0.56$
 & & 
$(-0.02,+0.18)$  &  $(\pm0.00,\pm0.01)$  &  $+0.11$  &  $+0.40$  &  $+0.16$  &  $+0.46$
 \\
0.780 & & 
$(-0.03,+0.36)$  &  $(\pm0.01,\pm0.02)$  &  $+0.18$  &  $+0.61$  &  $+0.38$  &  $+0.58$
 & & 
$(-0.02,+0.19)$  &  $(\pm0.00,\pm0.01)$  &  $+0.11$  &  $+0.41$  &  $+0.17$  &  $+0.46$
 \\
0.773 & & 
$(-0.04,+0.33)$  &  $(\pm0.01,\pm0.02)$  &  $+0.17$  &  $+0.61$  &  $+0.37$  &  $+0.58$
 & & 
$(-0.02,+0.20)$  &  $(\pm0.00,\pm0.01)$  &  $+0.10$  &  $+0.41$  &  $+0.17$  &  $+0.48$
 \\
0.749 & & 
$(-0.04,+0.33)$  &  $(\pm0.01,\pm0.02)$  &  $+0.17$  &  $+0.62$  &  $+0.39$  &  $+0.58$
 & & 
$(-0.02,+0.18)$  &  $(\pm0.00,\pm0.01)$  &  $+0.10$  &  $+0.39$  &  $+0.15$  &  $+0.44$
 \\
 & & & & & & & & & & & & & & \\
 \multicolumn{15}{c}{$B_Y$} \\
 \cline{1-15}\\
 & \multicolumn{7}{c}{Umbra}\hspace{.3cm} & \multicolumn{7}{c}{Penumbra}  \\
   $\mu$ & & (a,b) & (ci a,ci b) & $s_e$ & r & $r^{2}$ & $r_{S}$ &  & (a,b) & (ci a,ci b) & $s_e$ & r & $r^{2}$ & $r_{S}$ \\ 
0.708 & & 
$(-0.10,+1.02)$  &  $(\pm0.01,\pm0.01)$  &  $+0.10$  &  $+0.99$  &  $+0.98$  &  $+0.99$
 & & 
$(-0.00,+0.95)$  &  $(\pm0.00,\pm0.01)$  &  $+0.09$  &  $+0.99$  &  $+0.98$  &  $+0.99$
 \\
0.903 & & 
$(-0.05,+1.01)$  &  $(\pm0.01,\pm0.01)$  &  $+0.08$  &  $+0.99$  &  $+0.99$  &  $+0.99$
 & & 
$(+0.00,+1.00)$  &  $(\pm0.00,\pm0.01)$  &  $+0.10$  &  $+0.99$  &  $+0.98$  &  $+0.99$
 \\
0.906 & & 
$(-0.02,+0.98)$  &  $(\pm0.02,\pm0.02)$  &  $+0.14$  &  $+0.98$  &  $+0.96$  &  $+0.98$
 & & 
$(+0.01,+0.97)$  &  $(\pm0.00,\pm0.00)$  &  $+0.09$  &  $+0.99$  &  $+0.98$  &  $+0.99$
 \\
0.910 & & 
$(-0.01,+0.99)$  &  $(\pm0.01,\pm0.01)$  &  $+0.08$  &  $+0.99$  &  $+0.99$  &  $+0.99$
 & & 
$(-0.00,+0.95)$  &  $(\pm0.00,\pm0.01)$  &  $+0.13$  &  $+0.98$  &  $+0.96$  &  $+0.98$
 \\
0.912 & & 
$(-0.02,+0.99)$  &  $(\pm0.01,\pm0.01)$  &  $+0.09$  &  $+0.99$  &  $+0.99$  &  $+0.99$
 & & 
$(+0.00,+0.95)$  &  $(\pm0.00,\pm0.01)$  &  $+0.10$  &  $+0.99$  &  $+0.98$  &  $+0.99$
 \\
0.909 & & 
$(-0.03,+0.98)$  &  $(\pm0.01,\pm0.01)$  &  $+0.10$  &  $+0.99$  &  $+0.98$  &  $+0.99$
 & & 
$(+0.00,+0.98)$  &  $(\pm0.00,\pm0.00)$  &  $+0.09$  &  $+0.99$  &  $+0.98$  &  $+0.99$
 \\
0.907 & & 
$(-0.01,+0.99)$  &  $(\pm0.01,\pm0.01)$  &  $+0.08$  &  $+0.99$  &  $+0.99$  &  $+0.99$
 & & 
$(+0.01,+0.96)$  &  $(\pm0.00,\pm0.00)$  &  $+0.09$  &  $+0.99$  &  $+0.98$  &  $+0.99$
 \\
0.807 & & 
$(+0.02,+0.96)$  &  $(\pm0.01,\pm0.01)$  &  $+0.07$  &  $+0.99$  &  $+0.99$  &  $+0.99$
 & & 
$(+0.01,+0.95)$  &  $(\pm0.00,\pm0.01)$  &  $+0.11$  &  $+0.99$  &  $+0.97$  &  $+0.99$
 \\
0.801 & & 
$(+0.05,+0.95)$  &  $(\pm0.01,\pm0.01)$  &  $+0.07$  &  $+0.99$  &  $+0.98$  &  $+0.99$
 & & 
$(+0.02,+0.94)$  &  $(\pm0.00,\pm0.01)$  &  $+0.10$  &  $+0.99$  &  $+0.98$  &  $+0.99$
 \\
0.793 & & 
$(+0.02,+0.95)$  &  $(\pm0.01,\pm0.01)$  &  $+0.09$  &  $+0.99$  &  $+0.98$  &  $+0.99$
 & & 
$(+0.02,+0.93)$  &  $(\pm0.01,\pm0.01)$  &  $+0.13$  &  $+0.98$  &  $+0.96$  &  $+0.98$
 \\
0.788 & & 
$(+0.00,+0.96)$  &  $(\pm0.01,\pm0.01)$  &  $+0.08$  &  $+0.99$  &  $+0.98$  &  $+0.99$
 & & 
$(+0.03,+0.94)$  &  $(\pm0.01,\pm0.01)$  &  $+0.13$  &  $+0.98$  &  $+0.96$  &  $+0.98$
 \\
0.780 & & 
$(+0.02,+0.96)$  &  $(\pm0.01,\pm0.01)$  &  $+0.07$  &  $+0.99$  &  $+0.99$  &  $+0.99$
 & & 
$(+0.02,+0.94)$  &  $(\pm0.01,\pm0.01)$  &  $+0.13$  &  $+0.98$  &  $+0.96$  &  $+0.98$
 \\
0.773 & & 
$(+0.04,+0.96)$  &  $(\pm0.01,\pm0.01)$  &  $+0.07$  &  $+0.99$  &  $+0.99$  &  $+0.99$
 & & 
$(+0.02,+0.95)$  &  $(\pm0.00,\pm0.01)$  &  $+0.11$  &  $+0.99$  &  $+0.97$  &  $+0.98$
 \\
0.749 & & 
$(+0.02,+0.96)$  &  $(\pm0.01,\pm0.01)$  &  $+0.07$  &  $+0.99$  &  $+0.99$  &  $+0.99$
 & & 
$(+0.00,+0.95)$  &  $(\pm0.00,\pm0.01)$  &  $+0.11$  &  $+0.99$  &  $+0.97$  &  $+0.99$
 \\
   & & & & & & & & & & & & & & \\
 & \multicolumn{7}{c}{Strong Plage}\hspace{.3cm} & \multicolumn{7}{c}{Weak Plage}  \\  
   $\mu$ & & (a,b) & (ci a,ci b) & $s_e$ & r & $r^{2}$ & $r_{S}$ &  & (a,b) & (ci a,ci b) & $s_e$ & r & $r^{2}$ & $r_{S}$ \\ 
0.708 & & 
$(+0.01,+0.26)$  &  $(\pm0.01,\pm0.02)$  &  $+0.18$  &  $+0.46$  &  $+0.21$  &  $+0.42$
 & & 
$(+0.01,+0.14)$  &  $(\pm0.00,\pm0.01)$  &  $+0.10$  &  $+0.28$  &  $+0.08$  &  $+0.32$
 \\
0.903 & & 
$(-0.00,+0.23)$  &  $(\pm0.01,\pm0.02)$  &  $+0.17$  &  $+0.43$  &  $+0.19$  &  $+0.39$
 & & 
$(-0.00,+0.11)$  &  $(\pm0.00,\pm0.01)$  &  $+0.10$  &  $+0.27$  &  $+0.07$  &  $+0.30$
 \\
0.906 & & 
$(-0.01,+0.24)$  &  $(\pm0.01,\pm0.02)$  &  $+0.16$  &  $+0.48$  &  $+0.23$  &  $+0.40$
 & & 
$(-0.00,+0.11)$  &  $(\pm0.00,\pm0.01)$  &  $+0.09$  &  $+0.28$  &  $+0.08$  &  $+0.31$
 \\
0.910 & & 
$(+0.00,+0.27)$  &  $(\pm0.01,\pm0.02)$  &  $+0.16$  &  $+0.53$  &  $+0.28$  &  $+0.48$
 & & 
$(-0.00,+0.14)$  &  $(\pm0.00,\pm0.01)$  &  $+0.10$  &  $+0.34$  &  $+0.11$  &  $+0.37$
 \\
0.912 & & 
$(+0.01,+0.26)$  &  $(\pm0.01,\pm0.02)$  &  $+0.16$  &  $+0.50$  &  $+0.25$  &  $+0.47$
 & & 
$(-0.00,+0.14)$  &  $(\pm0.00,\pm0.01)$  &  $+0.09$  &  $+0.34$  &  $+0.12$  &  $+0.37$
 \\
0.909 & & 
$(+0.01,+0.25)$  &  $(\pm0.01,\pm0.02)$  &  $+0.16$  &  $+0.49$  &  $+0.24$  &  $+0.46$
 & & 
$(+0.00,+0.15)$  &  $(\pm0.00,\pm0.01)$  &  $+0.09$  &  $+0.37$  &  $+0.14$  &  $+0.40$
 \\
0.907 & & 
$(+0.01,+0.29)$  &  $(\pm0.01,\pm0.02)$  &  $+0.17$  &  $+0.54$  &  $+0.30$  &  $+0.52$
 & & 
$(+0.00,+0.17)$  &  $(\pm0.00,\pm0.01)$  &  $+0.10$  &  $+0.39$  &  $+0.15$  &  $+0.42$
 \\
0.807 & & 
$(+0.03,+0.38)$  &  $(\pm0.01,\pm0.02)$  &  $+0.15$  &  $+0.67$  &  $+0.45$  &  $+0.64$
 & & 
$(+0.02,+0.19)$  &  $(\pm0.00,\pm0.01)$  &  $+0.09$  &  $+0.41$  &  $+0.17$  &  $+0.44$
 \\
0.801 & & 
$(+0.03,+0.39)$  &  $(\pm0.01,\pm0.02)$  &  $+0.16$  &  $+0.67$  &  $+0.45$  &  $+0.63$
 & & 
$(+0.03,+0.20)$  &  $(\pm0.00,\pm0.01)$  &  $+0.09$  &  $+0.41$  &  $+0.17$  &  $+0.45$
 \\
0.793 & & 
$(+0.02,+0.41)$  &  $(\pm0.01,\pm0.02)$  &  $+0.15$  &  $+0.70$  &  $+0.49$  &  $+0.68$
 & & 
$(+0.03,+0.17)$  &  $(\pm0.00,\pm0.01)$  &  $+0.10$  &  $+0.38$  &  $+0.15$  &  $+0.43$
 \\
0.788 & & 
$(+0.04,+0.37)$  &  $(\pm0.01,\pm0.02)$  &  $+0.16$  &  $+0.63$  &  $+0.40$  &  $+0.61$
 & & 
$(+0.02,+0.19)$  &  $(\pm0.00,\pm0.01)$  &  $+0.10$  &  $+0.41$  &  $+0.16$  &  $+0.45$
 \\
0.780 & & 
$(+0.03,+0.41)$  &  $(\pm0.01,\pm0.02)$  &  $+0.16$  &  $+0.69$  &  $+0.47$  &  $+0.65$
 & & 
$(+0.03,+0.20)$  &  $(\pm0.00,\pm0.01)$  &  $+0.09$  &  $+0.42$  &  $+0.17$  &  $+0.46$
 \\
0.773 & & 
$(+0.02,+0.42)$  &  $(\pm0.01,\pm0.02)$  &  $+0.16$  &  $+0.70$  &  $+0.49$  &  $+0.66$
 & & 
$(+0.02,+0.21)$  &  $(\pm0.00,\pm0.01)$  &  $+0.10$  &  $+0.40$  &  $+0.16$  &  $+0.44$
 \\
0.749 & & 
$(+0.02,+0.43)$  &  $(\pm0.01,\pm0.02)$  &  $+0.16$  &  $+0.70$  &  $+0.48$  &  $+0.67$
 & & 
$(+0.01,+0.22)$  &  $(\pm0.00,\pm0.02)$  &  $+0.10$  &  $+0.41$  &  $+0.17$  &  $+0.43$
 \\
\hline
\end{tabular}
\end{center}
\end{table*}
}

%% file: tablaSPcgHMInewcutz.tex
 
{
\begin{table*}[t] 
\scriptsize
\begin{center}
\caption{Linear regression fit between the vertical component of the vector magentic field observed by HMI and SP. The meaning of the columns is the same as in Table \ref{tabla_linreg_all_xy}.\label{tabla_linreg_all_z}}
\begin{tabular}{ccccccccccccccc}
 & & & & & & & & & & & & & & \\
 \multicolumn{15}{c}{$B_Z$} \\
 \cline{1-15}\\
 & \multicolumn{7}{c}{Umbra}\hspace{.3cm} & \multicolumn{7}{c}{Penumbra}  \\
   $\mu$ & & (a,b) & (ci a,ci b) & $\sigma$ & r & $r^{2}$ & $r_{S}$ &  & (a,b) & (ci a,ci b) & $\sigma$ & r & $r^{2}$ & $r_{S}$ \\ 
0.708 & & 
$(-0.30,+0.76)$  &  $(\pm0.03,\pm0.02)$  &  $+0.12$  &  $+0.97$  &  $+0.94$  &  $+0.97$
 & & 
$(+0.00,+0.75)$  &  $(\pm0.01,\pm0.01)$  &  $+0.12$  &  $+0.96$  &  $+0.92$  &  $+0.96$
 \\
0.903 & & 
$(-0.17,+0.86)$  &  $(\pm0.03,\pm0.01)$  &  $+0.07$  &  $+0.98$  &  $+0.97$  &  $+0.98$
 & & 
$(-0.03,+0.90)$  &  $(\pm0.01,\pm0.01)$  &  $+0.13$  &  $+0.96$  &  $+0.92$  &  $+0.95$
 \\
0.906 & & 
$(-0.18,+0.85)$  &  $(\pm0.07,\pm0.03)$  &  $+0.17$  &  $+0.91$  &  $+0.84$  &  $+0.96$
 & & 
$(-0.02,+0.89)$  &  $(\pm0.01,\pm0.01)$  &  $+0.11$  &  $+0.97$  &  $+0.95$  &  $+0.96$
 \\
0.910 & & 
$(-0.08,+0.90)$  &  $(\pm0.02,\pm0.01)$  &  $+0.06$  &  $+0.99$  &  $+0.98$  &  $+0.99$
 & & 
$(-0.01,+0.89)$  &  $(\pm0.01,\pm0.01)$  &  $+0.11$  &  $+0.97$  &  $+0.94$  &  $+0.96$
 \\
0.912 & & 
$(-0.17,+0.84)$  &  $(\pm0.02,\pm0.01)$  &  $+0.06$  &  $+0.99$  &  $+0.98$  &  $+0.99$
 & & 
$(-0.01,+0.87)$  &  $(\pm0.01,\pm0.01)$  &  $+0.13$  &  $+0.96$  &  $+0.92$  &  $+0.95$
 \\
0.909 & & 
$(-0.18,+0.87)$  &  $(\pm0.06,\pm0.03)$  &  $+0.13$  &  $+0.94$  &  $+0.88$  &  $+0.97$
 & & 
$(-0.02,+0.88)$  &  $(\pm0.01,\pm0.01)$  &  $+0.11$  &  $+0.97$  &  $+0.94$  &  $+0.97$
 \\
0.907 & & 
$(-0.05,+0.92)$  &  $(\pm0.03,\pm0.01)$  &  $+0.06$  &  $+0.99$  &  $+0.98$  &  $+0.99$
 & & 
$(-0.02,+0.89)$  &  $(\pm0.01,\pm0.01)$  &  $+0.10$  &  $+0.98$  &  $+0.95$  &  $+0.97$
 \\
0.807 & & 
$(-0.22,+0.82)$  &  $(\pm0.03,\pm0.02)$  &  $+0.09$  &  $+0.97$  &  $+0.94$  &  $+0.96$
 & & 
$(-0.01,+0.85)$  &  $(\pm0.01,\pm0.01)$  &  $+0.11$  &  $+0.98$  &  $+0.95$  &  $+0.98$
 \\
0.801 & & 
$(-0.26,+0.79)$  &  $(\pm0.03,\pm0.02)$  &  $+0.09$  &  $+0.96$  &  $+0.93$  &  $+0.95$
 & & 
$(-0.01,+0.84)$  &  $(\pm0.01,\pm0.01)$  &  $+0.10$  &  $+0.98$  &  $+0.96$  &  $+0.98$
 \\
0.793 & & 
$(-0.35,+0.75)$  &  $(\pm0.04,\pm0.02)$  &  $+0.11$  &  $+0.95$  &  $+0.90$  &  $+0.92$
 & & 
$(-0.02,+0.81)$  &  $(\pm0.01,\pm0.01)$  &  $+0.13$  &  $+0.97$  &  $+0.93$  &  $+0.97$
 \\
0.788 & & 
$(-0.39,+0.73)$  &  $(\pm0.05,\pm0.03)$  &  $+0.12$  &  $+0.93$  &  $+0.87$  &  $+0.89$
 & & 
$(-0.03,+0.81)$  &  $(\pm0.01,\pm0.01)$  &  $+0.11$  &  $+0.97$  &  $+0.95$  &  $+0.98$
 \\
0.780 & & 
$(-0.32,+0.77)$  &  $(\pm0.04,\pm0.02)$  &  $+0.10$  &  $+0.96$  &  $+0.92$  &  $+0.93$
 & & 
$(-0.03,+0.81)$  &  $(\pm0.01,\pm0.01)$  &  $+0.11$  &  $+0.97$  &  $+0.95$  &  $+0.98$
 \\
0.773 & & 
$(-0.29,+0.80)$  &  $(\pm0.04,\pm0.02)$  &  $+0.11$  &  $+0.96$  &  $+0.92$  &  $+0.93$
 & & 
$(-0.03,+0.82)$  &  $(\pm0.01,\pm0.01)$  &  $+0.10$  &  $+0.98$  &  $+0.95$  &  $+0.98$
 \\
0.749 & & 
$(-0.25,+0.80)$  &  $(\pm0.03,\pm0.02)$  &  $+0.09$  &  $+0.97$  &  $+0.95$  &  $+0.95$
 & & 
$(-0.02,+0.79)$  &  $(\pm0.01,\pm0.01)$  &  $+0.11$  &  $+0.98$  &  $+0.95$  &  $+0.98$
 \\
   & & & & & & & & & & & & & & \\
 & \multicolumn{7}{c}{Strong Plage}\hspace{.3cm} & \multicolumn{7}{c}{Weak Plage}  \\  
   $\mu$ & & (a,b) & (ci a,ci b) & $\sigma$ & r & $r^{2}$ & $r_{S}$ &  & (a,b) & (ci a,ci b) & $\sigma$ & r & $r^{2}$ & $r_{S}$ \\ 
0.708 & & 
$(-0.00,+0.15)$  &  $(\pm0.00,\pm0.01)$  &  $+0.07$  &  $+0.73$  &  $+0.54$  &  $+0.78$
 & & 
$(+0.00,+0.13)$  &  $(\pm0.00,\pm0.00)$  &  $+0.05$  &  $+0.67$  &  $+0.44$  &  $+0.79$
 \\
0.903 & & 
$(-0.01,+0.13)$  &  $(\pm0.00,\pm0.00)$  &  $+0.07$  &  $+0.72$  &  $+0.52$  &  $+0.78$
 & & 
$(-0.01,+0.09)$  &  $(\pm0.00,\pm0.00)$  &  $+0.05$  &  $+0.65$  &  $+0.42$  &  $+0.77$
 \\
0.906 & & 
$(-0.01,+0.12)$  &  $(\pm0.00,\pm0.00)$  &  $+0.06$  &  $+0.72$  &  $+0.52$  &  $+0.79$
 & & 
$(-0.01,+0.09)$  &  $(\pm0.00,\pm0.00)$  &  $+0.05$  &  $+0.65$  &  $+0.43$  &  $+0.78$
 \\
0.910 & & 
$(-0.01,+0.12)$  &  $(\pm0.00,\pm0.00)$  &  $+0.06$  &  $+0.70$  &  $+0.49$  &  $+0.78$
 & & 
$(-0.01,+0.08)$  &  $(\pm0.00,\pm0.00)$  &  $+0.05$  &  $+0.63$  &  $+0.40$  &  $+0.77$
 \\
0.912 & & 
$(-0.01,+0.14)$  &  $(\pm0.00,\pm0.00)$  &  $+0.07$  &  $+0.74$  &  $+0.55$  &  $+0.80$
 & & 
$(-0.01,+0.09)$  &  $(\pm0.00,\pm0.00)$  &  $+0.05$  &  $+0.65$  &  $+0.42$  &  $+0.77$
 \\
0.909 & & 
$(-0.01,+0.15)$  &  $(\pm0.00,\pm0.00)$  &  $+0.07$  &  $+0.77$  &  $+0.60$  &  $+0.81$
 & & 
$(-0.01,+0.09)$  &  $(\pm0.00,\pm0.00)$  &  $+0.05$  &  $+0.66$  &  $+0.44$  &  $+0.77$
 \\
0.907 & & 
$(-0.00,+0.16)$  &  $(\pm0.00,\pm0.01)$  &  $+0.08$  &  $+0.76$  &  $+0.57$  &  $+0.80$
 & & 
$(-0.01,+0.09)$  &  $(\pm0.00,\pm0.00)$  &  $+0.05$  &  $+0.67$  &  $+0.45$  &  $+0.78$
 \\
0.807 & & 
$(-0.01,+0.15)$  &  $(\pm0.00,\pm0.01)$  &  $+0.07$  &  $+0.72$  &  $+0.52$  &  $+0.78$
 & & 
$(-0.00,+0.11)$  &  $(\pm0.00,\pm0.00)$  &  $+0.05$  &  $+0.60$  &  $+0.36$  &  $+0.77$
 \\
0.801 & & 
$(-0.02,+0.15)$  &  $(\pm0.00,\pm0.01)$  &  $+0.08$  &  $+0.71$  &  $+0.50$  &  $+0.78$
 & & 
$(-0.01,+0.11)$  &  $(\pm0.00,\pm0.00)$  &  $+0.05$  &  $+0.60$  &  $+0.37$  &  $+0.77$
 \\
0.793 & & 
$(-0.02,+0.14)$  &  $(\pm0.00,\pm0.01)$  &  $+0.08$  &  $+0.69$  &  $+0.48$  &  $+0.75$
 & & 
$(-0.01,+0.10)$  &  $(\pm0.00,\pm0.00)$  &  $+0.06$  &  $+0.58$  &  $+0.33$  &  $+0.74$
 \\
0.788 & & 
$(-0.01,+0.16)$  &  $(\pm0.00,\pm0.01)$  &  $+0.08$  &  $+0.72$  &  $+0.52$  &  $+0.78$
 & & 
$(-0.00,+0.11)$  &  $(\pm0.00,\pm0.00)$  &  $+0.06$  &  $+0.61$  &  $+0.38$  &  $+0.76$
 \\
0.780 & & 
$(-0.01,+0.15)$  &  $(\pm0.00,\pm0.01)$  &  $+0.08$  &  $+0.69$  &  $+0.47$  &  $+0.77$
 & & 
$(-0.00,+0.11)$  &  $(\pm0.00,\pm0.00)$  &  $+0.06$  &  $+0.60$  &  $+0.36$  &  $+0.76$
 \\
0.773 & & 
$(-0.01,+0.15)$  &  $(\pm0.00,\pm0.01)$  &  $+0.07$  &  $+0.70$  &  $+0.49$  &  $+0.76$
 & & 
$(-0.00,+0.12)$  &  $(\pm0.00,\pm0.00)$  &  $+0.06$  &  $+0.62$  &  $+0.39$  &  $+0.77$
 \\
0.749 & & 
$(-0.02,+0.16)$  &  $(\pm0.00,\pm0.01)$  &  $+0.07$  &  $+0.72$  &  $+0.52$  &  $+0.78$
 & & 
$(-0.00,+0.13)$  &  $(\pm0.00,\pm0.00)$  &  $+0.06$  &  $+0.64$  &  $+0.41$  &  $+0.79$
 \\
\hline
\end{tabular}
\end{center}
\end{table*}
}

%% file: tablaHMIvsSPcgSPFF1S2HS2HFF1poly2d.tex
 
{
\begin{table*}[]
\scriptsize
\begin{center}
\caption{Comparison between the \newcorr{Cartesian components of the} vector magnetic field observed by HMI and SP, SP with FF=1, S2H, and S2H with FF=1.\label{tabla_compar_SPHMIS2H}}
\begin{tabular}{llccc} 
 $\mathbf{Solar Feature}$ &  \vectorb{HMI}{}  vs \vectorb{SP}{} & $\mathbf{<s_{e}>}$ & $\mathbf{<r>}$ & $\mathbf{<r^{2}>}$ \\
\hline
Umbra
 & $ B^{HMI}_{X} = (+0.05\pm 0.10) + (+0.99\pm 0.02)\times B^{SP}_{X} $  &  
$ 0.09\pm 0.02 $ & $ 0.99\pm 0.00 $ & $ 0.99\pm 0.01 $ \\
 & $ B^{HMI}_{Y} = (-0.00\pm 0.04) + (+0.98\pm 0.02)\times B^{SP}_{Y} $  &  
$ 0.09\pm 0.02 $ & $ 0.99\pm 0.00 $ & $ 0.98\pm 0.01 $ \\
 & $ B^{HMI}_{Z} = (-0.23\pm 0.10) + (+0.82\pm 0.06)\times B^{SP}_{Z} $  &  
$ 0.10\pm 0.03 $ & $ 0.96\pm 0.02 $ & $ 0.93\pm 0.04 $ \\
 & & & \\
Penumbra
 & $ B^{HMI}_{X} = (+0.00\pm 0.01) + (+0.97\pm 0.02)\times B^{SP}_{X} $  &  
$ 0.10\pm 0.01 $ & $ 0.99\pm 0.00 $ & $ 0.98\pm 0.01 $ \\
 & $ B^{HMI}_{Y} = (+0.01\pm 0.01) + (+0.95\pm 0.02)\times B^{SP}_{Y} $  &  
$ 0.11\pm 0.02 $ & $ 0.99\pm 0.00 $ & $ 0.97\pm 0.01 $ \\
 & $ B^{HMI}_{Z} = (-0.02\pm 0.01) + (+0.84\pm 0.05)\times B^{SP}_{Z} $  &  
$ 0.11\pm 0.01 $ & $ 0.97\pm 0.01 $ & $ 0.94\pm 0.01 $ \\
 & & & \\
Strong B Plage
 & $ B^{HMI}_{X} = (-0.03\pm 0.01) + (+0.40\pm 0.11)\times B^{SP}_{X} $  &  
$ 0.16\pm 0.02 $ & $ 0.65\pm 0.12 $ & $ 0.44\pm 0.14 $ \\
 & $ B^{HMI}_{Y} = (+0.02\pm 0.01) + (+0.33\pm 0.08)\times B^{SP}_{Y} $  &  
$ 0.16\pm 0.01 $ & $ 0.59\pm 0.10 $ & $ 0.35\pm 0.12 $ \\
 & $ B^{HMI}_{Z} = (-0.01\pm 0.01) + (+0.15\pm 0.01)\times B^{SP}_{Z} $  &  
$ 0.07\pm 0.01 $ & $ 0.72\pm 0.02 $ & $ 0.52\pm 0.04 $ \\
 & & & \\
Weak B Plage
 & $ B^{HMI}_{X} = (-0.02\pm 0.01) + (+0.21\pm 0.04)\times B^{SP}_{X} $  &  
$ 0.10\pm 0.01 $ & $ 0.40\pm 0.05 $ & $ 0.17\pm 0.04 $ \\
 & $ B^{HMI}_{Y} = (+0.01\pm 0.01) + (+0.17\pm 0.04)\times B^{SP}_{Y} $  &  
$ 0.10\pm 0.00 $ & $ 0.36\pm 0.05 $ & $ 0.14\pm 0.04 $ \\
 & $ B^{HMI}_{Z} = (-0.01\pm 0.00) + (+0.10\pm 0.02)\times B^{SP}_{Z} $  &  
$ 0.05\pm 0.00 $ & $ 0.63\pm 0.03 $ & $ 0.40\pm 0.04 $ \\
 & & & \\
 $\mathbf{Solar Feature}$ &  \vectorb{HMI}{}  vs \vectorb{SPFF1}{} & $\mathbf{<s_{e}>}$ & $\mathbf{<r>}$ & $\mathbf{<r^{2}>}$ \\
\hline
Umbra
 & $ B^{HMI}_{X} = (+0.04\pm 0.08) + (+0.98\pm 0.02)\times B^{SPFF1}_{X} $  &  
$ 0.11\pm 0.04 $ & $ 0.99\pm 0.01 $ & $ 0.97\pm 0.02 $ \\
 & $ B^{HMI}_{Y} = (-0.01\pm 0.04) + (+0.98\pm 0.02)\times B^{SPFF1}_{Y} $  &  
$ 0.09\pm 0.02 $ & $ 0.99\pm 0.00 $ & $ 0.98\pm 0.01 $ \\
 & $ B^{HMI}_{Z} = (-0.20\pm 0.11) + (+0.85\pm 0.06)\times B^{SPFF1}_{Z} $  &  
$ 0.10\pm 0.03 $ & $ 0.96\pm 0.02 $ & $ 0.93\pm 0.04 $ \\
 & & & \\
Penumbra
 & $ B^{HMI}_{X} = (+0.01\pm 0.01) + (+1.01\pm 0.02)\times B^{SPFF1}_{X} $  &  
$ 0.12\pm 0.02 $ & $ 0.99\pm 0.00 $ & $ 0.97\pm 0.01 $ \\
 & $ B^{HMI}_{Y} = (+0.01\pm 0.01) + (+1.00\pm 0.02)\times B^{SPFF1}_{Y} $  &  
$ 0.10\pm 0.02 $ & $ 0.99\pm 0.00 $ & $ 0.98\pm 0.01 $ \\
 & $ B^{HMI}_{Z} = (-0.03\pm 0.01) + (+0.90\pm 0.02)\times B^{SPFF1}_{Z} $  &  
$ 0.09\pm 0.01 $ & $ 0.98\pm 0.01 $ & $ 0.96\pm 0.01 $ \\
 & & & \\
Strong B Plage
 & $ B^{HMI}_{X} = (-0.01\pm 0.01) + (+0.74\pm 0.17)\times B^{SPFF1}_{X} $  &  
$ 0.16\pm 0.04 $ & $ 0.78\pm 0.14 $ & $ 0.62\pm 0.19 $ \\
 & $ B^{HMI}_{Y} = (+0.01\pm 0.02) + (+0.66\pm 0.12)\times B^{SPFF1}_{Y} $  &  
$ 0.17\pm 0.02 $ & $ 0.74\pm 0.10 $ & $ 0.55\pm 0.14 $ \\
 & $ B^{HMI}_{Z} = (-0.02\pm 0.01) + (+0.39\pm 0.05)\times B^{SPFF1}_{Z} $  &  
$ 0.09\pm 0.01 $ & $ 0.84\pm 0.02 $ & $ 0.71\pm 0.04 $ \\
 & & & \\
Weak B Plage
 & $ B^{HMI}_{X} = (-0.02\pm 0.01) + (+0.50\pm 0.08)\times B^{SPFF1}_{X} $  &  
$ 0.10\pm 0.01 $ & $ 0.50\pm 0.06 $ & $ 0.26\pm 0.06 $ \\
 & $ B^{HMI}_{Y} = (+0.01\pm 0.01) + (+0.42\pm 0.09)\times B^{SPFF1}_{Y} $  &  
$ 0.10\pm 0.00 $ & $ 0.47\pm 0.06 $ & $ 0.23\pm 0.06 $ \\
 & $ B^{HMI}_{Z} = (-0.00\pm 0.00) + (+0.41\pm 0.08)\times B^{SPFF1}_{Z} $  &  
$ 0.05\pm 0.00 $ & $ 0.76\pm 0.03 $ & $ 0.58\pm 0.05 $ \\
 & & & \\
 $\mathbf{Solar Feature}$ &  \vectorb{HMI}{}  vs \vectorb{S2H}{} & $\mathbf{<s_{e}>}$ & $\mathbf{<r>}$ & $\mathbf{<r^{2}>}$ \\
\hline
Umbra
 & $ B^{HMI}_{X} = (+0.03\pm 0.05) + (+0.99\pm 0.04)\times B^{S2H}_{X} $  &  
$ 0.10\pm 0.01 $ & $ 0.99\pm 0.00 $ & $ 0.98\pm 0.01 $ \\
 & $ B^{HMI}_{Y} = (+0.02\pm 0.05) + (+0.99\pm 0.03)\times B^{S2H}_{Y} $  &  
$ 0.10\pm 0.03 $ & $ 0.99\pm 0.00 $ & $ 0.98\pm 0.01 $ \\
 & $ B^{HMI}_{Z} = (-0.22\pm 0.06) + (+0.80\pm 0.04)\times B^{S2H}_{Z} $  &  
$ 0.10\pm 0.03 $ & $ 0.96\pm 0.02 $ & $ 0.93\pm 0.04 $ \\
 & & & \\
Penumbra
 & $ B^{HMI}_{X} = (+0.01\pm 0.01) + (+0.99\pm 0.01)\times B^{S2H}_{X} $  &  
$ 0.09\pm 0.01 $ & $ 0.99\pm 0.00 $ & $ 0.98\pm 0.00 $ \\
 & $ B^{HMI}_{Y} = (+0.01\pm 0.02) + (+0.99\pm 0.01)\times B^{S2H}_{Y} $  &  
$ 0.08\pm 0.01 $ & $ 0.99\pm 0.00 $ & $ 0.99\pm 0.00 $ \\
 & $ B^{HMI}_{Z} = (-0.03\pm 0.01) + (+0.92\pm 0.03)\times B^{S2H}_{Z} $  &  
$ 0.10\pm 0.01 $ & $ 0.98\pm 0.01 $ & $ 0.95\pm 0.02 $ \\
 & & & \\
Strong B Plage
 & $ B^{HMI}_{X} = (-0.00\pm 0.01) + (+0.58\pm 0.06)\times B^{S2H}_{X} $  &  
$ 0.13\pm 0.01 $ & $ 0.85\pm 0.03 $ & $ 0.73\pm 0.05 $ \\
 & $ B^{HMI}_{Y} = (+0.02\pm 0.01) + (+0.41\pm 0.06)\times B^{S2H}_{Y} $  &  
$ 0.17\pm 0.01 $ & $ 0.63\pm 0.06 $ & $ 0.40\pm 0.07 $ \\
 & $ B^{HMI}_{Z} = (-0.01\pm 0.01) + (+0.19\pm 0.01)\times B^{S2H}_{Z} $  &  
$ 0.09\pm 0.01 $ & $ 0.74\pm 0.02 $ & $ 0.54\pm 0.03 $ \\
 & & & \\
Weak B Plage
 & $ B^{HMI}_{X} = (-0.01\pm 0.00) + (+0.46\pm 0.04)\times B^{S2H}_{X} $  &  
$ 0.08\pm 0.01 $ & $ 0.70\pm 0.02 $ & $ 0.49\pm 0.03 $ \\
 & $ B^{HMI}_{Y} = (+0.01\pm 0.01) + (+0.23\pm 0.04)\times B^{S2H}_{Y} $  &  
$ 0.10\pm 0.00 $ & $ 0.38\pm 0.04 $ & $ 0.15\pm 0.04 $ \\
 & $ B^{HMI}_{Z} = (-0.00\pm 0.00) + (+0.14\pm 0.02)\times B^{S2H}_{Z} $  &  
$ 0.06\pm 0.00 $ & $ 0.58\pm 0.02 $ & $ 0.33\pm 0.03 $ \\
 & & & \\
 $\mathbf{Solar Feature}$ &  \vectorb{HMI}{}  vs \vectorb{S2HFF1}{} & $\mathbf{<s_{e}>}$ & $\mathbf{<r>}$ & $\mathbf{<r^{2}>}$ \\
\hline
Umbra
 & $ B^{HMI}_{X} = (+0.02\pm 0.05) + (+0.99\pm 0.04)\times B^{S2HFF1}_{X} $  &  
$ 0.10\pm 0.02 $ & $ 0.99\pm 0.00 $ & $ 0.98\pm 0.01 $ \\
 & $ B^{HMI}_{Y} = (+0.00\pm 0.04) + (+1.00\pm 0.02)\times B^{S2HFF1}_{Y} $  &  
$ 0.09\pm 0.03 $ & $ 0.99\pm 0.00 $ & $ 0.98\pm 0.01 $ \\
 & $ B^{HMI}_{Z} = (-0.18\pm 0.06) + (+0.84\pm 0.03)\times B^{S2HFF1}_{Z} $  &  
$ 0.09\pm 0.03 $ & $ 0.97\pm 0.02 $ & $ 0.94\pm 0.04 $ \\
 & & & \\
Penumbra
 & $ B^{HMI}_{X} = (+0.01\pm 0.01) + (+1.00\pm 0.01)\times B^{S2HFF1}_{X} $  &  
$ 0.08\pm 0.01 $ & $ 0.99\pm 0.00 $ & $ 0.99\pm 0.00 $ \\
 & $ B^{HMI}_{Y} = (+0.01\pm 0.01) + (+1.00\pm 0.01)\times B^{S2HFF1}_{Y} $  &  
$ 0.07\pm 0.01 $ & $ 0.99\pm 0.00 $ & $ 0.99\pm 0.00 $ \\
 & $ B^{HMI}_{Z} = (-0.03\pm 0.01) + (+0.93\pm 0.02)\times B^{S2HFF1}_{Z} $  &  
$ 0.09\pm 0.01 $ & $ 0.98\pm 0.01 $ & $ 0.96\pm 0.01 $ \\
 & & & \\
Strong B Plage
 & $ B^{HMI}_{X} = (-0.01\pm 0.01) + (+0.86\pm 0.08)\times B^{S2HFF1}_{X} $  &  
$ 0.11\pm 0.01 $ & $ 0.92\pm 0.02 $ & $ 0.84\pm 0.04 $ \\
 & $ B^{HMI}_{Y} = (+0.03\pm 0.01) + (+0.64\pm 0.09)\times B^{S2HFF1}_{Y} $  &  
$ 0.18\pm 0.01 $ & $ 0.72\pm 0.07 $ & $ 0.52\pm 0.09 $ \\
 & $ B^{HMI}_{Z} = (-0.02\pm 0.01) + (+0.40\pm 0.04)\times B^{S2HFF1}_{Z} $  &  
$ 0.10\pm 0.01 $ & $ 0.84\pm 0.02 $ & $ 0.70\pm 0.04 $ \\
 & & & \\
Weak B Plage
 & $ B^{HMI}_{X} = (-0.01\pm 0.00) + (+0.73\pm 0.05)\times B^{S2HFF1}_{X} $  &  
$ 0.07\pm 0.01 $ & $ 0.78\pm 0.01 $ & $ 0.60\pm 0.02 $ \\
 & $ B^{HMI}_{Y} = (+0.01\pm 0.01) + (+0.40\pm 0.07)\times B^{S2HFF1}_{Y} $  &  
$ 0.10\pm 0.00 $ & $ 0.45\pm 0.05 $ & $ 0.21\pm 0.05 $ \\
 & $ B^{HMI}_{Z} = (-0.00\pm 0.00) + (+0.43\pm 0.07)\times B^{S2HFF1}_{Z} $  &  
$ 0.05\pm 0.00 $ & $ 0.76\pm 0.03 $ & $ 0.57\pm 0.05 $ \\
 & & & \\
 \hline
\end{tabular}
\end{center}
\end{table*}
}

%% file: tablasSPcgSPFF1S2HS2HFF1poly2d_v2.tex
 
{
\begin{table*}[]
\scriptsize
\begin{center}
\caption{Comparison between the \newcorr{Cartesian components of the} vector magnetic field observed by SP and SP with FF=1, S2H and S2H with FF=1, S2H and SP, and S2H and SP both with FF=1.\label{tabla_compar_SP_S2H}}
\begin{tabular}{llccc} 
 $\mathbf{Solar\ Feature}$ &  \vectorb{SPFF1}{}  vs \vectorb{SP}{} & $\mathbf{<s_e>}$ & $\mathbf{<r>}$ & $\mathbf{<r^{2}>}$ \\
\hline
Umbra
 & $ B^{SPFF1}_{X} = (0.00\pm 0.00) + (0.99\pm 0.00)\times B^{SP}_{X} $  &  
$ 0.06\pm 0.06 $ & $ 0.99\pm 0.01 $ & $ 0.99\pm 0.02 $ \\
 & $ B^{SPFF1}_{Y} = (0.00\pm 0.00) + (0.99\pm 0.00)\times B^{SP}_{Y} $  &  
$ 0.01\pm 0.00 $ & $ 1.00\pm 0.00 $ & $ 1.00\pm 0.00 $ \\
 & $ B^{SPFF1}_{Z} = (-0.04\pm 0.02) + (0.97\pm 0.01)\times B^{SP}_{Z} $  &  
$ 0.02\pm 0.01 $ & $ 1.00\pm 0.00 $ & $ 1.00\pm 0.00 $ \\
 & & & \\
Penumbra
 & $ B^{SPFF1}_{X} = (-0.01\pm 0.01) + (0.95\pm 0.01)\times B^{SP}_{X} $  &  
$ 0.09\pm 0.02 $ & $ 0.99\pm 0.00 $ & $ 0.98\pm 0.01 $ \\
 & $ B^{SPFF1}_{Y} = (0.00\pm 0.00) + (0.95\pm 0.00)\times B^{SP}_{Y} $  &  
$ 0.05\pm 0.00 $ & $ 1.00\pm 0.00 $ & $ 0.99\pm 0.00 $ \\
 & $ B^{SPFF1}_{Z} = (0.01\pm 0.00) + (0.94\pm 0.03)\times B^{SP}_{Z} $  &  
$ 0.06\pm 0.01 $ & $ 0.99\pm 0.00 $ & $ 0.99\pm 0.00 $ \\
 & & & \\
Strong B Plage
 & $ B^{SPFF1}_{X} = (-0.01\pm 0.01) + (0.57\pm 0.03)\times B^{SP}_{X} $  &  
$ 0.11\pm 0.00 $ & $ 0.89\pm 0.01 $ & $ 0.79\pm 0.02 $ \\
 & $ B^{SPFF1}_{Y} = (0.00\pm 0.00) + (0.58\pm 0.02)\times B^{SP}_{Y} $  &  
$ 0.09\pm 0.00 $ & $ 0.92\pm 0.01 $ & $ 0.85\pm 0.02 $ \\
 & $ B^{SPFF1}_{Z} = (-0.01\pm 0.00) + (0.31\pm 0.01)\times B^{SP}_{Z} $  &  
$ 0.10\pm 0.01 $ & $ 0.88\pm 0.01 $ & $ 0.77\pm 0.03 $ \\
 & & & \\
Weak B Plage
 & $ B^{SPFF1}_{X} = (-0.00\pm 0.00) + (0.40\pm 0.02)\times B^{SP}_{X} $  &  
$ 0.06\pm 0.00 $ & $ 0.80\pm 0.02 $ & $ 0.65\pm 0.03 $ \\
 & $ B^{SPFF1}_{Y} = (0.00\pm 0.00) + (0.43\pm 0.03)\times B^{SP}_{Y} $  &  
$ 0.05\pm 0.00 $ & $ 0.88\pm 0.02 $ & $ 0.77\pm 0.03 $ \\
 & $ B^{SPFF1}_{Z} = (-0.00\pm 0.00) + (0.19\pm 0.01)\times B^{SP}_{Z} $  &  
$ 0.05\pm 0.00 $ & $ 0.79\pm 0.01 $ & $ 0.62\pm 0.02 $ \\
 & & & \\
 $\mathbf{Solar\ Feature}$ &  \vectorb{S2HFF1}{}  vs \vectorb{S2H}{} & $\mathbf{<s_e>}$ & $\mathbf{<r>}$ & $\mathbf{<r^{2}>}$ \\
\hline
Umbra
 & $ B^{S2HFF1}_{X} = (0.01\pm 0.01) + (1.00\pm 0.00)\times B^{S2H}_{X} $  &  
$ 0.04\pm 0.01 $ & $ 1.00\pm 0.00 $ & $ 1.00\pm 0.00 $ \\
 & $ B^{S2HFF1}_{Y} = (0.01\pm 0.01) + (0.99\pm 0.01)\times B^{S2H}_{Y} $  &  
$ 0.04\pm 0.01 $ & $ 1.00\pm 0.00 $ & $ 1.00\pm 0.00 $ \\
 & $ B^{S2HFF1}_{Z} = (-0.04\pm 0.02) + (0.96\pm 0.02)\times B^{S2H}_{Z} $  &  
$ 0.04\pm 0.01 $ & $ 1.00\pm 0.00 $ & $ 0.99\pm 0.00 $ \\
 & & & \\
Penumbra
 & $ B^{S2HFF1}_{X} = (-0.00\pm 0.00) + (0.99\pm 0.00)\times B^{S2H}_{X} $  &  
$ 0.03\pm 0.00 $ & $ 1.00\pm 0.00 $ & $ 1.00\pm 0.00 $ \\
 & $ B^{S2HFF1}_{Y} = (0.00\pm 0.00) + (0.99\pm 0.00)\times B^{S2H}_{Y} $  &  
$ 0.03\pm 0.00 $ & $ 1.00\pm 0.00 $ & $ 1.00\pm 0.00 $ \\
 & $ B^{S2HFF1}_{Z} = (0.00\pm 0.00) + (0.99\pm 0.01)\times B^{S2H}_{Z} $  &  
$ 0.04\pm 0.00 $ & $ 1.00\pm 0.00 $ & $ 1.00\pm 0.00 $ \\
 & & & \\
Strong B Plage
 & $ B^{S2HFF1}_{X} = (0.01\pm 0.01) + (0.69\pm 0.02)\times B^{S2H}_{X} $  &  
$ 0.09\pm 0.00 $ & $ 0.94\pm 0.00 $ & $ 0.89\pm 0.01 $ \\
 & $ B^{S2HFF1}_{Y} = (0.01\pm 0.00) + (0.63\pm 0.04)\times B^{S2H}_{Y} $  &  
$ 0.13\pm 0.02 $ & $ 0.86\pm 0.06 $ & $ 0.75\pm 0.10 $ \\
 & $ B^{S2HFF1}_{Z} = (-0.01\pm 0.01) + (0.41\pm 0.03)\times B^{S2H}_{Z} $  &  
$ 0.13\pm 0.01 $ & $ 0.86\pm 0.01 $ & $ 0.74\pm 0.01 $ \\
 & & & \\
Weak B Plage
 & $ B^{S2HFF1}_{X} = (-0.00\pm 0.00) + (0.61\pm 0.02)\times B^{S2H}_{X} $  &  
$ 0.05\pm 0.00 $ & $ 0.88\pm 0.01 $ & $ 0.78\pm 0.02 $ \\
 & $ B^{S2HFF1}_{Y} = (0.00\pm 0.00) + (0.55\pm 0.05)\times B^{S2H}_{Y} $  &  
$ 0.06\pm 0.01 $ & $ 0.81\pm 0.04 $ & $ 0.67\pm 0.07 $ \\
 & $ B^{S2HFF1}_{Z} = (-0.00\pm 0.00) + (0.23\pm 0.02)\times B^{S2H}_{Z} $  &  
$ 0.06\pm 0.00 $ & $ 0.69\pm 0.01 $ & $ 0.48\pm 0.02 $ \\
 & & & \\
 $\mathbf{Solar\ Feature}$ &  \vectorb{S2H}{}  vs \vectorb{SP}{} & $\mathbf{<s_e>}$ & $\mathbf{<r>}$ & $\mathbf{<r^{2}>}$ \\
\hline
Umbra
 & $ B^{S2H}_{X} = (0.01\pm 0.05) + (0.99\pm 0.05)\times B^{SP}_{X} $  &  
$ 0.07\pm 0.02 $ & $ 0.99\pm 0.00 $ & $ 0.99\pm 0.01 $ \\
 & $ B^{S2H}_{Y} = (-0.01\pm 0.02) + (0.98\pm 0.02)\times B^{SP}_{Y} $  &  
$ 0.06\pm 0.01 $ & $ 1.00\pm 0.00 $ & $ 0.99\pm 0.00 $ \\
 & $ B^{S2H}_{Z} = (-0.04\pm 0.06) + (1.00\pm 0.02)\times B^{SP}_{Z} $  &  
$ 0.08\pm 0.01 $ & $ 0.99\pm 0.00 $ & $ 0.97\pm 0.01 $ \\
 & & & \\
Penumbra
 & $ B^{S2H}_{X} = (-0.01\pm 0.00) + (0.98\pm 0.01)\times B^{SP}_{X} $  &  
$ 0.06\pm 0.00 $ & $ 1.00\pm 0.00 $ & $ 0.99\pm 0.00 $ \\
 & $ B^{S2H}_{Y} = (-0.01\pm 0.01) + (0.97\pm 0.01)\times B^{SP}_{Y} $  &  
$ 0.06\pm 0.00 $ & $ 1.00\pm 0.00 $ & $ 0.99\pm 0.00 $ \\
 & $ B^{S2H}_{Z} = (0.01\pm 0.00) + (0.92\pm 0.02)\times B^{SP}_{Z} $  &  
$ 0.07\pm 0.01 $ & $ 0.99\pm 0.00 $ & $ 0.98\pm 0.01 $ \\
 & & & \\
Strong B Plage
 & $ B^{S2H}_{X} = (-0.04\pm 0.01) + (0.54\pm 0.13)\times B^{SP}_{X} $  &  
$ 0.25\pm 0.04 $ & $ 0.61\pm 0.12 $ & $ 0.39\pm 0.13 $ \\
 & $ B^{S2H}_{Y} = (0.01\pm 0.03) + (0.38\pm 0.08)\times B^{SP}_{Y} $  &  
$ 0.28\pm 0.01 $ & $ 0.44\pm 0.09 $ & $ 0.20\pm 0.08 $ \\
 & $ B^{S2H}_{Z} = (-0.01\pm 0.01) + (0.54\pm 0.05)\times B^{SP}_{Z} $  &  
$ 0.25\pm 0.01 $ & $ 0.76\pm 0.01 $ & $ 0.58\pm 0.02 $ \\
 & & & \\
Weak B Plage
 & $ B^{S2H}_{X} = (-0.03\pm 0.01) + (0.27\pm 0.03)\times B^{SP}_{X} $  &  
$ 0.11\pm 0.00 $ & $ 0.38\pm 0.04 $ & $ 0.15\pm 0.03 $ \\
 & $ B^{S2H}_{Y} = (0.00\pm 0.01) + (0.21\pm 0.03)\times B^{SP}_{Y} $  &  
$ 0.12\pm 0.01 $ & $ 0.30\pm 0.04 $ & $ 0.09\pm 0.02 $ \\
 & $ B^{S2H}_{Z} = (-0.00\pm 0.00) + (0.28\pm 0.01)\times B^{SP}_{Z} $  &  
$ 0.10\pm 0.01 $ & $ 0.61\pm 0.02 $ & $ 0.37\pm 0.03 $ \\
 & & & \\
 $\mathbf{Solar\ Feature}$ &  \vectorb{S2HFF1}{}  vs \vectorb{SPFF1}{} & $\mathbf{<s_e>}$ & $\mathbf{<r>}$ & $\mathbf{<r^{2}>}$ \\
\hline
Umbra
 & $ B^{S2H}_{X} = (0.01\pm 0.03) + (0.98\pm 0.04)\times B^{SPFF1}_{X} $  &  
$ 0.10\pm 0.05 $ & $ 0.99\pm 0.01 $ & $ 0.98\pm 0.02 $ \\
 & $ B^{S2H}_{Y} = (-0.00\pm 0.01) + (0.98\pm 0.01)\times B^{SPFF1}_{Y} $  &  
$ 0.05\pm 0.01 $ & $ 1.00\pm 0.00 $ & $ 0.99\pm 0.00 $ \\
 & $ B^{S2H}_{Z} = (-0.04\pm 0.10) + (1.00\pm 0.05)\times B^{SPFF1}_{Z} $  &  
$ 0.07\pm 0.01 $ & $ 0.99\pm 0.00 $ & $ 0.97\pm 0.01 $ \\
 & & & \\
Penumbra
 & $ B^{S2H}_{X} = (0.00\pm 0.00) + (1.01\pm 0.01)\times B^{SPFF1}_{X} $  &  
$ 0.09\pm 0.02 $ & $ 0.99\pm 0.00 $ & $ 0.98\pm 0.01 $ \\
 & $ B^{S2H}_{Y} = (-0.01\pm 0.01) + (1.00\pm 0.01)\times B^{SPFF1}_{Y} $  &  
$ 0.04\pm 0.00 $ & $ 1.00\pm 0.00 $ & $ 1.00\pm 0.00 $ \\
 & $ B^{S2H}_{Z} = (-0.00\pm 0.00) + (0.97\pm 0.00)\times B^{SPFF1}_{Z} $  &  
$ 0.03\pm 0.00 $ & $ 1.00\pm 0.00 $ & $ 1.00\pm 0.00 $ \\
 & & & \\
Strong B Plage
 & $ B^{S2H}_{X} = (-0.00\pm 0.01) + (0.78\pm 0.15)\times B^{SPFF1}_{X} $  &  
$ 0.19\pm 0.06 $ & $ 0.78\pm 0.15 $ & $ 0.63\pm 0.20 $ \\
 & $ B^{S2H}_{Y} = (-0.00\pm 0.02) + (0.59\pm 0.11)\times B^{SPFF1}_{Y} $  &  
$ 0.24\pm 0.03 $ & $ 0.60\pm 0.12 $ & $ 0.37\pm 0.14 $ \\
 & $ B^{S2H}_{Z} = (-0.00\pm 0.00) + (0.96\pm 0.02)\times B^{SPFF1}_{Z} $  &  
$ 0.03\pm 0.00 $ & $ 1.00\pm 0.00 $ & $ 0.99\pm 0.00 $ \\
 & & & \\
Weak B Plage
 & $ B^{S2H}_{X} = (-0.02\pm 0.00) + (0.55\pm 0.05)\times B^{SPFF1}_{X} $  &  
$ 0.09\pm 0.01 $ & $ 0.52\pm 0.05 $ & $ 0.28\pm 0.05 $ \\
 & $ B^{S2H}_{Y} = (0.00\pm 0.01) + (0.43\pm 0.06)\times B^{SPFF1}_{Y} $  &  
$ 0.10\pm 0.01 $ & $ 0.42\pm 0.06 $ & $ 0.18\pm 0.05 $ \\
 & $ B^{S2H}_{Z} = (-0.00\pm 0.00) + (0.97\pm 0.01)\times B^{SPFF1}_{Z} $  &  
$ 0.01\pm 0.00 $ & $ 1.00\pm 0.00 $ & $ 0.99\pm 0.00 $ \\
 & & & \\
\hline
\end{tabular}
\end{center}
\end{table*}
}

%% file: tablas_fieldincliazi_SPcgSPFF1S2HS2HFF1poly2d_sincg.tex
 
{
\begin{table*}[]
\scriptsize
\begin{center}
\caption{Comparison between the \newcorr{spherical components of the} vector magnetic field observed by SP and SP with FF=1, S2H and S2H with FF=1, S2H and SP, and S2H and SP both with FF=1.\label{tabla_compar_spher_SP_S2H}}
\begin{tabular}{llccc} 
 $\mathbf{Solar\ Feature}$ &  \vectorb{SPFF1}{}  vs \vectorb{SP}{} & $\mathbf{<s_e>}$ & $\mathbf{<r>}$ & $\mathbf{<r^{2}>}$ \\
\hline
Umbra
 & $ |B|^{SPFF1} = (-0.04\pm 0.12) + (1.00\pm 0.05)\times |B|^{SP} $  &  
$ 0.02\pm 0.01 $ & $ 0.99\pm 0.00 $ & $ 0.99\pm 0.01 $ \\
 & $ \theta_{B}^{SPFF1} = (1.93\pm 0.65) + (0.98\pm 0.00)\times \theta_{B}^{SP} $  &  
$ 0.37\pm 0.12 $ & $ 1.00\pm 0.00 $ & $ 1.00\pm 0.00 $ \\
 & $ \phi_{B}^{SPFF1} = (1.49\pm 1.24) + (0.99\pm 0.01)\times \phi_{B}^{SP} $  &  
$ 7.29\pm 3.35 $ & $ 0.99\pm 0.01 $ & $ 0.97\pm 0.02 $ \\
 & & & \\
Penumbra
 & $ |B|^{SPFF1} = (-0.10\pm 0.01) + (1.02\pm 0.01)\times |B|^{SP} $  &  
$ 0.09\pm 0.01 $ & $ 0.98\pm 0.00 $ & $ 0.97\pm 0.01 $ \\
 & $ \theta_{B}^{SPFF1} = (3.38\pm 0.95) + (0.96\pm 0.01)\times \theta_{B}^{SP} $  &  
$ 1.95\pm 0.60 $ & $ 1.00\pm 0.00 $ & $ 0.99\pm 0.00 $ \\
 & $ \phi_{B}^{SPFF1} = (1.90\pm 0.96) + (0.98\pm 0.01)\times \phi_{B}^{SP} $  &  
$ 10.32\pm 2.85 $ & $ 0.98\pm 0.01 $ & $ 0.96\pm 0.02 $ \\
 & & & \\
Strong B Plage
 & $ |B|^{SPFF1} = (0.09\pm 0.01) + (0.31\pm 0.03)\times |B|^{SP} $  &  
$ 0.14\pm 0.01 $ & $ 0.60\pm 0.03 $ & $ 0.35\pm 0.03 $ \\
 & $ \theta_{B}^{SPFF1} = (26.56\pm 1.62) + (0.71\pm 0.02)\times \theta_{B}^{SP} $  &  
$ 9.14\pm 0.29 $ & $ 0.96\pm 0.01 $ & $ 0.91\pm 0.01 $ \\
 & $ \phi_{B}^{SPFF1} = (13.43\pm 1.94) + (0.85\pm 0.02)\times \phi_{B}^{SP} $  &  
$ 26.91\pm 0.97 $ & $ 0.84\pm 0.02 $ & $ 0.70\pm 0.04 $ \\
 & & & \\
Weak B Plage
 & $ |B|^{SPFF1} = (0.11\pm 0.00) + (0.04\pm 0.01)\times |B|^{SP} $  &  
$ 0.04\pm 0.00 $ & $ 0.24\pm 0.04 $ & $ 0.06\pm 0.02 $ \\
 & $ \theta_{B}^{SPFF1} = (29.34\pm 0.92) + (0.68\pm 0.01)\times \theta_{B}^{SP} $  &  
$ 12.92\pm 0.35 $ & $ 0.92\pm 0.01 $ & $ 0.85\pm 0.02 $ \\
 & $ \phi_{B}^{SPFF1} = (23.94\pm 2.50) + (0.74\pm 0.03)\times \phi_{B}^{SP} $  &  
$ 35.08\pm 1.08 $ & $ 0.72\pm 0.03 $ & $ 0.52\pm 0.04 $ \\
 & & & \\
 $\mathbf{Solar\ Feature}$ &  \vectorb{S2HFF1}{}  vs \vectorb{S2H}{} & $\mathbf{<s_e>}$ & $\mathbf{<r>}$ & $\mathbf{<r^{2}>}$ \\
\hline
Umbra
 & $ |B|^{S2HFF1} = (-0.05\pm 0.06) + (1.00\pm 0.02)\times |B|^{S2H} $  &  
$ 0.04\pm 0.01 $ & $ 0.99\pm 0.00 $ & $ 0.98\pm 0.01 $ \\
 & $ \theta_{B}^{S2HFF1} = (1.26\pm 1.02) + (0.99\pm 0.01)\times \theta_{B}^{S2H} $  &  
$ 0.97\pm 0.40 $ & $ 1.00\pm 0.00 $ & $ 1.00\pm 0.00 $ \\
 & $ \phi_{B}^{S2HFF1} = (3.97\pm 3.75) + (0.95\pm 0.03)\times \phi_{B}^{S2H} $  &  
$ 14.39\pm 6.02 $ & $ 0.95\pm 0.03 $ & $ 0.90\pm 0.05 $ \\
 & & & \\
Penumbra
 & $ |B|^{S2HFF1} = (-0.04\pm 0.00) + (1.02\pm 0.00)\times |B|^{S2H} $  &  
$ 0.05\pm 0.00 $ & $ 0.99\pm 0.00 $ & $ 0.99\pm 0.00 $ \\
 & $ \theta_{B}^{S2HFF1} = (1.22\pm 0.32) + (0.99\pm 0.00)\times \theta_{B}^{S2H} $  &  
$ 1.20\pm 0.16 $ & $ 1.00\pm 0.00 $ & $ 1.00\pm 0.00 $ \\
 & $ \phi_{B}^{S2HFF1} = (1.20\pm 0.86) + (0.99\pm 0.01)\times \phi_{B}^{S2H} $  &  
$ 7.09\pm 2.10 $ & $ 0.99\pm 0.01 $ & $ 0.98\pm 0.01 $ \\
 & & & \\
Strong B Plage
 & $ |B|^{S2HFF1} = (0.13\pm 0.01) + (0.37\pm 0.02)\times |B|^{S2H} $  &  
$ 0.13\pm 0.01 $ & $ 0.63\pm 0.04 $ & $ 0.40\pm 0.05 $ \\
 & $ \theta_{B}^{S2HFF1} = (20.12\pm 1.49) + (0.78\pm 0.01)\times \theta_{B}^{S2H} $  &  
$ 7.24\pm 0.72 $ & $ 0.97\pm 0.00 $ & $ 0.94\pm 0.00 $ \\
 & $ \phi_{B}^{S2HFF1} = (11.60\pm 5.52) + (0.87\pm 0.07)\times \phi_{B}^{S2H} $  &  
$ 24.50\pm 6.94 $ & $ 0.87\pm 0.06 $ & $ 0.77\pm 0.10 $ \\
 & & & \\
Weak B Plage
 & $ |B|^{S2HFF1} = (0.13\pm 0.00) + (0.04\pm 0.01)\times |B|^{S2H} $  &  
$ 0.04\pm 0.00 $ & $ 0.20\pm 0.03 $ & $ 0.04\pm 0.01 $ \\
 & $ \theta_{B}^{S2HFF1} = (21.12\pm 1.31) + (0.77\pm 0.02)\times \theta_{B}^{S2H} $  &  
$ 7.51\pm 0.80 $ & $ 0.96\pm 0.00 $ & $ 0.92\pm 0.01 $ \\
 & $ \phi_{B}^{S2HFF1} = (12.02\pm 3.05) + (0.87\pm 0.04)\times \phi_{B}^{S2H} $  &  
$ 25.21\pm 4.19 $ & $ 0.87\pm 0.03 $ & $ 0.75\pm 0.05 $ \\
 & & & \\
 $\mathbf{Solar\ Feature}$ &  \vectorb{S2H}{}  vs \vectorb{SP}{} & $\mathbf{<s_e>}$ & $\mathbf{<r>}$ & $\mathbf{<r^{2}>}$ \\
\hline
Umbra
 & $ |B|^{S2H} = (-0.32\pm 0.12) + (1.13\pm 0.05)\times |B|^{SP} $  &  
$ 0.06\pm 0.01 $ & $ 0.97\pm 0.01 $ & $ 0.95\pm 0.01 $ \\
 & $ \theta_{B}^{S2H} = (-0.23\pm 1.85) + (1.01\pm 0.01)\times \theta_{B}^{SP} $  &  
$ 1.33\pm 0.18 $ & $ 1.00\pm 0.00 $ & $ 0.99\pm 0.00 $ \\
 & $ \phi_{B}^{S2H} = (7.32\pm 2.01) + (0.91\pm 0.02)\times \phi_{B}^{SP} $  &  
$ 18.76\pm 4.18 $ & $ 0.91\pm 0.02 $ & $ 0.83\pm 0.04 $ \\
 & & & \\
Penumbra
 & $ |B|^{S2H} = (0.02\pm 0.01) + (0.94\pm 0.01)\times |B|^{SP} $  &  
$ 0.10\pm 0.01 $ & $ 0.98\pm 0.01 $ & $ 0.95\pm 0.01 $ \\
 & $ \theta_{B}^{S2H} = (3.52\pm 0.99) + (0.96\pm 0.01)\times \theta_{B}^{SP} $  &  
$ 2.51\pm 0.39 $ & $ 0.99\pm 0.00 $ & $ 0.99\pm 0.00 $ \\
 & $ \phi_{B}^{S2H} = (8.55\pm 2.12) + (0.91\pm 0.02)\times \phi_{B}^{SP} $  &  
$ 20.36\pm 2.01 $ & $ 0.91\pm 0.01 $ & $ 0.84\pm 0.03 $ \\
 & & & \\
Strong B Plage
 & $ |B|^{S2H} = (0.13\pm 0.03) + (0.49\pm 0.07)\times |B|^{SP} $  &  
$ 0.24\pm 0.01 $ & $ 0.55\pm 0.04 $ & $ 0.31\pm 0.04 $ \\
 & $ \theta_{B}^{S2H} = (18.66\pm 1.98) + (0.79\pm 0.02)\times \theta_{B}^{SP} $  &  
$ 9.81\pm 0.15 $ & $ 0.96\pm 0.01 $ & $ 0.92\pm 0.01 $ \\
 & $ \phi_{B}^{S2H} = (25.69\pm 6.84) + (0.72\pm 0.08)\times \phi_{B}^{SP} $  &  
$ 35.39\pm 6.18 $ & $ 0.72\pm 0.07 $ & $ 0.52\pm 0.09 $ \\
 & & & \\
Weak B Plage
 & $ |B|^{S2H} = (0.16\pm 0.01)  (-0.02\pm 0.02)\times |B|^{SP} $  &  
$ 0.09\pm 0.01 $ & $ -0.04\pm 0.05 $ & $ 0.00\pm 0.00 $ \\
 & $ \theta_{B}^{S2H} = (32.49\pm 1.47) + (0.63\pm 0.01)\times \theta_{B}^{SP} $  &  
$ 10.85\pm 0.51 $ & $ 0.93\pm 0.00 $ & $ 0.87\pm 0.01 $ \\
 & $ \phi_{B}^{S2H} = (39.78\pm 4.71) + (0.57\pm 0.05)\times \phi_{B}^{SP} $  &  
$ 41.89\pm 3.47 $ & $ 0.57\pm 0.04 $ & $ 0.33\pm 0.04 $ \\
 & & & \\
 $\mathbf{Solar\ Feature}$ &  \vectorb{S2HFF1}{}  vs \vectorb{SPFF1}{} & $\mathbf{<s_e>}$ & $\mathbf{<r>}$ & $\mathbf{<r^{2}>}$ \\
\hline
Umbra
 & $ |B|^{S2HFF1} = (-0.36\pm 0.08) + (1.15\pm 0.03)\times |B|^{SPFF1} $  &  
$ 0.05\pm 0.01 $ & $ 0.98\pm 0.01 $ & $ 0.96\pm 0.02 $ \\
 & $ \theta_{B}^{S2HFF1} = (-1.33\pm 3.54) + (1.02\pm 0.02)\times \theta_{B}^{SPFF1} $  &  
$ 1.20\pm 0.25 $ & $ 1.00\pm 0.00 $ & $ 0.99\pm 0.00 $ \\
 & $ \phi_{B}^{S2HFF1} = (5.71\pm 1.88) + (0.92\pm 0.03)\times \phi_{B}^{SPFF1} $  &  
$ 16.74\pm 2.66 $ & $ 0.93\pm 0.02 $ & $ 0.86\pm 0.04 $ \\
 & & & \\
Penumbra
 & $ |B|^{S2HFF1} = (0.06\pm 0.01) + (0.95\pm 0.01)\times |B|^{SPFF1} $  &  
$ 0.04\pm 0.00 $ & $ 1.00\pm 0.00 $ & $ 0.99\pm 0.00 $ \\
 & $ \theta_{B}^{S2HFF1} = (1.31\pm 0.39) + (0.98\pm 0.00)\times \theta_{B}^{SPFF1} $  &  
$ 1.84\pm 0.42 $ & $ 1.00\pm 0.00 $ & $ 0.99\pm 0.00 $ \\
 & $ \phi_{B}^{S2HFF1} = (8.22\pm 2.41) + (0.91\pm 0.02)\times \phi_{B}^{SPFF1} $  &  
$ 19.61\pm 1.99 $ & $ 0.92\pm 0.01 $ & $ 0.85\pm 0.02 $ \\
 & & & \\
Strong B Plage
 & $ |B|^{S2HFF1} = (0.02\pm 0.00) + (0.93\pm 0.02)\times |B|^{SPFF1} $  &  
$ 0.03\pm 0.00 $ & $ 0.98\pm 0.00 $ & $ 0.96\pm 0.00 $ \\
 & $ \theta_{B}^{S2HFF1} = (3.16\pm 1.68) + (0.97\pm 0.02)\times \theta_{B}^{SPFF1} $  &  
$ 4.62\pm 0.32 $ & $ 0.99\pm 0.00 $ & $ 0.97\pm 0.01 $ \\
 & $ \phi_{B}^{S2HFF1} = (18.68\pm 7.29) + (0.80\pm 0.09)\times \phi_{B}^{SPFF1} $  &  
$ 30.61\pm 7.48 $ & $ 0.80\pm 0.08 $ & $ 0.64\pm 0.12 $ \\
 & & & \\
Weak B Plage
 & $ |B|^{S2HFF1} = (0.05\pm 0.00) + (0.70\pm 0.02)\times |B|^{SPFF1} $  &  
$ 0.03\pm 0.00 $ & $ 0.74\pm 0.02 $ & $ 0.55\pm 0.03 $ \\
 & $ \theta_{B}^{S2HFF1} = (22.55\pm 2.01) + (0.74\pm 0.02)\times \theta_{B}^{SPFF1} $  &  
$ 10.55\pm 0.46 $ & $ 0.92\pm 0.01 $ & $ 0.84\pm 0.02 $ \\
 & $ \phi_{B}^{S2HFF1} = (32.65\pm 3.32) + (0.65\pm 0.03)\times \phi_{B}^{SPFF1} $  &  
$ 38.81\pm 3.37 $ & $ 0.64\pm 0.02 $ & $ 0.42\pm 0.03 $ \\
 & & & \\
\hline
\end{tabular}
\end{center}
\end{table*}
}

%% file: tabla_unsigFlux_HMIvsSPcgSPFF1S2HS2HFF1poly2d_okmx.tex
 
{
\begin{table*}[]
\scriptsize
\begin{center}
\caption{Comparison between the apparent longitudinal magnetic flux density ($B_{app}$), the total magnetic flux ($\Phi$), and the total unsigned magnetic flux ($\widehat{\Phi}$) for HMI and SP, SP with FF=1, S2H, and S2H with FF=1. $B_{app}$ is given in $Mx/cm^2$. $\Phi$ and $\widehat{\Phi}$ are given in $10^{21} Mx$.\label{tabla_compar_flux_SPHMIS2H}}
\begin{tabular}{llccc} 
 $\mathbf{Solar\ Feature}$ &  $B_{app}^{HMI}$  vs $B_{app}^{SP/SPFF1/S2H/S2HFF1}$ & $\mathbf{<s_{e}>}$ & $\mathbf{<r>}$ & $\mathbf{<r^{2}>}$ \\
\hline
 & & & \\
Umbra  & $ B^{HMI}_{app} = (-0.26\pm 0.07) + (0.84\pm 0.04)\times B^{SP}_{app} $  &  
$ 0.09\pm 0.03 $ & $ 0.97\pm 0.02 $ & $ 0.93\pm 0.04 $ \\
 & $ B^{HMI}_{app} = (-0.20\pm 0.11) + (0.85\pm 0.06)\times B^{SPFF1}_{app} $  &  
$ 0.10\pm 0.03 $ & $ 0.96\pm 0.02 $ & $ 0.93\pm 0.04 $ \\
 & $ B^{HMI}_{app} = (-0.21\pm 0.06) + (0.83\pm 0.04)\times B^{S2H}_{app} $  &  
$ 0.10\pm 0.03 $ & $ 0.97\pm 0.02 $ & $ 0.93\pm 0.04 $ \\
 & $ B^{HMI}_{app} = (-0.18\pm 0.06) + (0.84\pm 0.03)\times B^{S2HFF1}_{app} $  &  
$ 0.09\pm 0.03 $ & $ 0.97\pm 0.02 $ & $ 0.94\pm 0.04 $ \\
 & & & \\
Penumbra  & $ B^{HMI}_{app} = (-0.02\pm 0.01) + (0.95\pm 0.02)\times B^{SP}_{app} $  &  
$ 0.09\pm 0.01 $ & $ 0.98\pm 0.01 $ & $ 0.96\pm 0.01 $ \\
 & $ B^{HMI}_{app} = (-0.03\pm 0.01) + (0.90\pm 0.02)\times B^{SPFF1}_{app} $  &  
$ 0.09\pm 0.01 $ & $ 0.98\pm 0.01 $ & $ 0.96\pm 0.01 $ \\
 & $ B^{HMI}_{app} = (-0.03\pm 0.01) + (0.94\pm 0.02)\times B^{S2H}_{app} $  &  
$ 0.09\pm 0.01 $ & $ 0.98\pm 0.01 $ & $ 0.96\pm 0.01 $ \\
 & $ B^{HMI}_{app} = (-0.03\pm 0.01) + (0.93\pm 0.02)\times B^{S2HFF1}_{app} $  &  
$ 0.09\pm 0.01 $ & $ 0.98\pm 0.01 $ & $ 0.96\pm 0.01 $ \\
 & & & \\
Strong B Plage  & $ B^{HMI}_{app} = (-0.02\pm 0.01) + (0.32\pm 0.06)\times B^{SP}_{app} $  &  
$ 0.09\pm 0.01 $ & $ 0.82\pm 0.02 $ & $ 0.67\pm 0.03 $ \\
 & $ B^{HMI}_{app} = (-0.02\pm 0.01) + (0.39\pm 0.05)\times B^{SPFF1}_{app} $  &  
$ 0.09\pm 0.01 $ & $ 0.84\pm 0.02 $ & $ 0.71\pm 0.04 $ \\
 & $ B^{HMI}_{app} = (-0.02\pm 0.01) + (0.35\pm 0.05)\times B^{S2H}_{app} $  &  
$ 0.09\pm 0.01 $ & $ 0.82\pm 0.02 $ & $ 0.67\pm 0.04 $ \\
 & $ B^{HMI}_{app} = (-0.02\pm 0.01) + (0.40\pm 0.04)\times B^{S2HFF1}_{app} $  &  
$ 0.10\pm 0.01 $ & $ 0.84\pm 0.02 $ & $ 0.70\pm 0.04 $ \\
 & & & \\
Weak B Plage  & $ B^{HMI}_{app} = (-0.00\pm 0.00) + (0.32\pm 0.08)\times B^{SP}_{app} $  &  
$ 0.05\pm 0.00 $ & $ 0.72\pm 0.03 $ & $ 0.51\pm 0.04 $ \\
 & $ B^{HMI}_{app} = (-0.00\pm 0.00) + (0.41\pm 0.08)\times B^{SPFF1}_{app} $  &  
$ 0.05\pm 0.00 $ & $ 0.76\pm 0.03 $ & $ 0.58\pm 0.05 $ \\
 & $ B^{HMI}_{app} = (-0.00\pm 0.00) + (0.37\pm 0.07)\times B^{S2H}_{app} $  &  
$ 0.05\pm 0.00 $ & $ 0.72\pm 0.03 $ & $ 0.52\pm 0.04 $ \\
 & $ B^{HMI}_{app} = (-0.00\pm 0.00) + (0.43\pm 0.07)\times B^{S2HFF1}_{app} $  &  
$ 0.05\pm 0.00 $ & $ 0.76\pm 0.03 $ & $ 0.57\pm 0.05 $ \\
 & & & \\
 $\mathbf{Solar\ Feature}$ &  $\Phi^{HMI}$  vs $\Phi^{SP/SPFF1/S2H/S2HFF1}$ & $\mathbf{s_{e}}$ & $\mathbf{r}$ & $\mathbf{r^{2}}$ \\
\hline
 & & & \\
Umbra  & $ \Phi^{HMI} = (-0.07\pm 0.02) + (0.93\pm 0.01)\times \Phi^{SP} $  &  
$ 0.01\pm 0.01 $ & $ 1.00\pm 0.01 $ & $ 1.00\pm 0.01 $ \\
 & $ \Phi^{HMI} = (-0.05\pm 0.02) + (0.92\pm 0.01)\times \Phi^{SPFF1} $  &  
$ 0.01\pm 0.01 $ & $ 1.00\pm 0.01 $ & $ 1.00\pm 0.01 $ \\
 & $ \Phi^{HMI} = (-0.00\pm 0.02) + (0.94\pm 0.01)\times \Phi^{S2H} $  &  
$ 0.01\pm 0.01 $ & $ 1.00\pm 0.01 $ & $ 1.00\pm 0.01 $ \\
 & $ \Phi^{HMI} = (0.03\pm 0.02) + (0.94\pm 0.01)\times \Phi^{S2HFF1} $  &  
$ 0.01\pm 0.01 $ & $ 1.00\pm 0.01 $ & $ 1.00\pm 0.01 $ \\
 & & & \\
Penumbra  & $ \Phi^{HMI} = (0.20\pm 0.07) + (1.14\pm 0.05)\times \Phi^{SP} $  &  
$ 0.04\pm 0.01 $ & $ 0.99\pm 0.01 $ & $ 0.98\pm 0.01 $ \\
 & $ \Phi^{HMI} = (0.15\pm 0.06) + (1.07\pm 0.04)\times \Phi^{SPFF1} $  &  
$ 0.04\pm 0.01 $ & $ 0.99\pm 0.01 $ & $ 0.98\pm 0.01 $ \\
 & $ \Phi^{HMI} = (0.14\pm 0.07) + (1.11\pm 0.05)\times \Phi^{S2H} $  &  
$ 0.04\pm 0.01 $ & $ 0.99\pm 0.01 $ & $ 0.98\pm 0.01 $ \\
 & $ \Phi^{HMI} = (0.19\pm 0.06) + (1.12\pm 0.04)\times \Phi^{S2HFF1} $  &  
$ 0.04\pm 0.01 $ & $ 0.99\pm 0.01 $ & $ 0.98\pm 0.01 $ \\
 & & & \\
Strong B Plage  & $ \Phi^{HMI} = (-0.04\pm 0.00) + (0.27\pm 0.08)\times \Phi^{SP} $  &  
$ 0.02\pm 0.01 $ & $ 0.69\pm 0.01 $ & $ 0.47\pm 0.01 $ \\
 & $ \Phi^{HMI} = (-0.03\pm 0.00) + (0.35\pm 0.10)\times \Phi^{SPFF1} $  &  
$ 0.02\pm 0.01 $ & $ 0.71\pm 0.01 $ & $ 0.50\pm 0.01 $ \\
 & $ \Phi^{HMI} = (-0.04\pm 0.00) + (0.34\pm 0.10)\times \Phi^{S2H} $  &  
$ 0.02\pm 0.01 $ & $ 0.70\pm 0.01 $ & $ 0.50\pm 0.01 $ \\
 & $ \Phi^{HMI} = (-0.04\pm 0.01) + (0.21\pm 0.08)\times \Phi^{S2HFF1} $  &  
$ 0.02\pm 0.01 $ & $ 0.62\pm 0.01 $ & $ 0.38\pm 0.01 $ \\
 & & & \\
Weak B Plage  & $ \Phi^{HMI} = (-0.01\pm 0.00) + (0.32\pm 0.03)\times \Phi^{SP} $  &  
$ 0.01\pm 0.01 $ & $ 0.96\pm 0.01 $ & $ 0.91\pm 0.01 $ \\
 & $ \Phi^{HMI} = (-0.01\pm 0.00) + (0.47\pm 0.04)\times \Phi^{SPFF1} $  &  
$ 0.01\pm 0.01 $ & $ 0.95\pm 0.01 $ & $ 0.90\pm 0.01 $ \\
 & $ \Phi^{HMI} = (-0.01\pm 0.00) + (0.43\pm 0.04)\times \Phi^{S2H} $  &  
$ 0.01\pm 0.01 $ & $ 0.95\pm 0.01 $ & $ 0.89\pm 0.01 $ \\
 & $ \Phi^{HMI} = (-0.01\pm 0.00) + (0.28\pm 0.03)\times \Phi^{S2HFF1} $  &  
$ 0.01\pm 0.01 $ & $ 0.93\pm 0.01 $ & $ 0.87\pm 0.01 $ \\
 & & & \\
 $\mathbf{Solar\ Feature}$ &  $\widehat{\Phi}^{HMI}$  vs $\widehat{\Phi}^{SP/SPFF1/S2H/S2HFF1}$ & $\mathbf{s_{e}}$ & $\mathbf{r}$ & $\mathbf{r^{2}}$ \\
\hline
 & & & \\
Umbra  & $ \widehat{\Phi}^{HMI} = (0.07\pm 0.02) + (0.93\pm 0.01)\times \widehat{\Phi}^{SP} $  &  
$ 0.01\pm 0.01 $ & $ 1.00\pm 0.01 $ & $ 1.00\pm 0.01 $ \\
 & $ \widehat{\Phi}^{HMI} = (0.05\pm 0.02) + (0.92\pm 0.01)\times \widehat{\Phi}^{SPFF1} $  &  
$ 0.01\pm 0.01 $ & $ 1.00\pm 0.01 $ & $ 1.00\pm 0.01 $ \\
 & $ \widehat{\Phi}^{HMI} = (0.00\pm 0.02) + (0.94\pm 0.01)\times \widehat{\Phi}^{S2H} $  &  
$ 0.01\pm 0.01 $ & $ 1.00\pm 0.01 $ & $ 1.00\pm 0.01 $ \\
 & $ \widehat{\Phi}^{HMI} = (-0.03\pm 0.02) + (0.94\pm 0.01)\times \widehat{\Phi}^{S2HFF1} $  &  
$ 0.01\pm 0.01 $ & $ 1.00\pm 0.01 $ & $ 1.00\pm 0.01 $ \\
 & & & \\
Penumbra  & $ \widehat{\Phi}^{HMI} = (-0.23\pm 0.08) + (1.13\pm 0.05)\times \widehat{\Phi}^{SP} $  &  
$ 0.03\pm 0.01 $ & $ 0.99\pm 0.01 $ & $ 0.98\pm 0.01 $ \\
 & $ \widehat{\Phi}^{HMI} = (-0.24\pm 0.08) + (1.09\pm 0.05)\times \widehat{\Phi}^{SPFF1} $  &  
$ 0.03\pm 0.01 $ & $ 0.99\pm 0.01 $ & $ 0.98\pm 0.01 $ \\
 & $ \widehat{\Phi}^{HMI} = (-0.31\pm 0.08) + (1.17\pm 0.05)\times \widehat{\Phi}^{S2H} $  &  
$ 0.03\pm 0.01 $ & $ 0.99\pm 0.01 $ & $ 0.98\pm 0.01 $ \\
 & $ \widehat{\Phi}^{HMI} = (-0.43\pm 0.08) + (1.22\pm 0.05)\times \widehat{\Phi}^{S2HFF1} $  &  
$ 0.03\pm 0.01 $ & $ 0.99\pm 0.01 $ & $ 0.98\pm 0.01 $ \\
 & & & \\
Strong B Plage  & $ \widehat{\Phi}^{HMI} = (0.32\pm 0.04) + (0.08\pm 0.05)\times \widehat{\Phi}^{SP} $  &  
$ 0.02\pm 0.01 $ & $ 0.45\pm 0.01 $ & $ 0.20\pm 0.01 $ \\
 & $ \widehat{\Phi}^{HMI} = (0.32\pm 0.04) + (0.11\pm 0.06)\times \widehat{\Phi}^{SPFF1} $  &  
$ 0.02\pm 0.01 $ & $ 0.46\pm 0.01 $ & $ 0.21\pm 0.01 $ \\
 & $ \widehat{\Phi}^{HMI} = (0.34\pm 0.03) + (0.07\pm 0.05)\times \widehat{\Phi}^{S2H} $  &  
$ 0.02\pm 0.01 $ & $ 0.39\pm 0.01 $ & $ 0.15\pm 0.01 $ \\
 & $ \widehat{\Phi}^{HMI} = (0.32\pm 0.04) + (0.08\pm 0.05)\times \widehat{\Phi}^{S2HFF1} $  &  
$ 0.02\pm 0.01 $ & $ 0.43\pm 0.01 $ & $ 0.18\pm 0.01 $ \\
 & & & \\
Weak B Plage  & $ \widehat{\Phi}^{HMI} = (0.08\pm 0.01) + (0.15\pm 0.02)\times \widehat{\Phi}^{SP} $  &  
$ 0.01\pm 0.01 $ & $ 0.89\pm 0.01 $ & $ 0.79\pm 0.01 $ \\
 & $ \widehat{\Phi}^{HMI} = (0.05\pm 0.01) + (0.28\pm 0.03)\times \widehat{\Phi}^{SPFF1} $  &  
$ 0.01\pm 0.01 $ & $ 0.93\pm 0.01 $ & $ 0.86\pm 0.01 $ \\
 & $ \widehat{\Phi}^{HMI} = (0.07\pm 0.01) + (0.22\pm 0.03)\times \widehat{\Phi}^{S2H} $  &  
$ 0.01\pm 0.01 $ & $ 0.91\pm 0.01 $ & $ 0.83\pm 0.01 $ \\
 & $ \widehat{\Phi}^{HMI} = (0.07\pm 0.01) + (0.13\pm 0.02)\times \widehat{\Phi}^{S2HFF1} $  &  
$ 0.02\pm 0.01 $ & $ 0.88\pm 0.01 $ & $ 0.78\pm 0.01 $ \\
 & & & \\
\hline
\end{tabular}
\end{center}
\end{table*}
}

%% file: tabla_unsigFlux_SP_S2H_ori_sincg_okmx.tex
 
{
\begin{table*}[]
\scriptsize
\begin{center}
\caption{Comparison between the apparent longitudinal magnetic flux density ($B_{app}$), the total magnetic flux ($\Phi$), and the total unsigned magnetic flux ($\widehat{\Phi}$) for SP, SP with FF=1, S2H, and S2H with FF=1. $B_{app}$ is given in $Mx/cm^2$. $\Phi$ and $\widehat{\Phi}$ are given in $10^{21} Mx$.\label{tabla_compar_flux_SP_S2H}}
\begin{tabular}{llccc} 
 $\mathbf{Solar\ Feature}$ &  $B_{app}^{SPFF1/S2HFF1}$  vs $B_{app}^{SP/S2H/SP/SPFF1}$ & $\mathbf{<s_{e}>}$ & $\mathbf{<r>}$ & $\mathbf{<r^{2}>}$ \\
\hline
 & & & \\
Umbra  & $ B^{SPFF1}_{app} = (-0.07\pm 0.05) + (0.98\pm 0.02)\times B^{SP}_{app} $  &  
$ 0.03\pm 0.01 $ & $ 1.00\pm 0.00 $ & $ 0.99\pm 0.00 $ \\
 & $ B^{S2HFF1}_{app} = (-0.05\pm 0.03) + (0.98\pm 0.02)\times B^{S2H}_{app} $  &  
$ 0.04\pm 0.01 $ & $ 1.00\pm 0.00 $ & $ 0.99\pm 0.01 $ \\
 & $ B^{S2HFF1}_{app} = (-0.10\pm 0.05) + (0.98\pm 0.02)\times B^{SP}_{app} $  &  
$ 0.06\pm 0.01 $ & $ 0.99\pm 0.00 $ & $ 0.98\pm 0.01 $ \\
 & $ B^{S2HFF1}_{app} = (-0.04\pm 0.11) + (0.99\pm 0.05)\times B^{SPFF1}_{app} $  &  
$ 0.07\pm 0.01 $ & $ 0.99\pm 0.00 $ & $ 0.97\pm 0.01 $ \\
 & & & \\
Penumbra  & $ B^{SPFF1}_{app} = (0.01\pm 0.00) + (1.05\pm 0.01)\times B^{SP}_{app} $  &  
$ 0.03\pm 0.01 $ & $ 1.00\pm 0.00 $ & $ 1.00\pm 0.00 $ \\
 & $ B^{S2HFF1}_{app} = (0.00\pm 0.00) + (1.01\pm 0.00)\times B^{S2H}_{app} $  &  
$ 0.01\pm 0.00 $ & $ 1.00\pm 0.00 $ & $ 1.00\pm 0.00 $ \\
 & $ B^{S2HFF1}_{app} = (0.01\pm 0.00) + (1.02\pm 0.01)\times B^{SP}_{app} $  &  
$ 0.04\pm 0.00 $ & $ 1.00\pm 0.00 $ & $ 0.99\pm 0.00 $ \\
 & $ B^{S2HFF1}_{app} = (-0.00\pm 0.00) + (0.97\pm 0.00)\times B^{SPFF1}_{app} $  &  
$ 0.03\pm 0.00 $ & $ 1.00\pm 0.00 $ & $ 1.00\pm 0.00 $ \\
 & & & \\
Strong B Plage  & $ B^{SPFF1}_{app} = (-0.01\pm 0.00) + (0.82\pm 0.04)\times B^{SP}_{app} $  &  
$ 0.07\pm 0.01 $ & $ 0.98\pm 0.00 $ & $ 0.96\pm 0.00 $ \\
 & $ B^{S2HFF1}_{app} = (-0.00\pm 0.00) + (0.88\pm 0.02)\times B^{S2H}_{app} $  &  
$ 0.06\pm 0.00 $ & $ 0.98\pm 0.00 $ & $ 0.97\pm 0.00 $ \\
 & $ B^{S2HFF1}_{app} = (-0.01\pm 0.00) + (0.79\pm 0.06)\times B^{SP}_{app} $  &  
$ 0.07\pm 0.01 $ & $ 0.97\pm 0.00 $ & $ 0.95\pm 0.00 $ \\
 & $ B^{S2HFF1}_{app} = (-0.00\pm 0.00) + (0.96\pm 0.03)\times B^{SPFF1}_{app} $  &  
$ 0.03\pm 0.00 $ & $ 1.00\pm 0.00 $ & $ 0.99\pm 0.00 $ \\
 & & & \\
Weak B Plage  & $ B^{SPFF1}_{app} = (-0.00\pm 0.00) + (0.74\pm 0.07)\times B^{SP}_{app} $  &  
$ 0.03\pm 0.00 $ & $ 0.95\pm 0.01 $ & $ 0.90\pm 0.01 $ \\
 & $ B^{S2HFF1}_{app} = (-0.00\pm 0.00) + (0.87\pm 0.05)\times B^{S2H}_{app} $  &  
$ 0.02\pm 0.00 $ & $ 0.97\pm 0.01 $ & $ 0.95\pm 0.02 $ \\
 & $ B^{S2HFF1}_{app} = (-0.00\pm 0.00) + (0.72\pm 0.06)\times B^{SP}_{app} $  &  
$ 0.02\pm 0.00 $ & $ 0.95\pm 0.01 $ & $ 0.90\pm 0.01 $ \\
 & $ B^{S2HFF1}_{app} = (-0.00\pm 0.00) + (0.97\pm 0.01)\times B^{SPFF1}_{app} $  &  
$ 0.01\pm 0.00 $ & $ 1.00\pm 0.00 $ & $ 0.99\pm 0.00 $ \\
 & & & \\
 $\mathbf{Solar\ Feature}$ &  $\widehat{\Phi}^{SPFF1/S2HFF1}$  vs $\widehat{\Phi}^{SP/S2H/SP/SPFF1}$ & $\mathbf{s_{e}}$ & $\mathbf{r}$ & $\mathbf{r^{2}}$ \\
\hline
 & & & \\
Umbra  & $ \Phi^{SPFF1} = (-0.02\pm 0.00) + (1.00\pm 0.00)\times \Phi^{SP} $  &  
$ 0.00\pm 0.01 $ & $ 1.00\pm 0.01 $ & $ 1.00\pm 0.01 $ \\
 & $ \Phi^{S2HFF1} = (-0.01\pm 0.00) + (1.00\pm 0.00)\times \Phi^{S2H} $  &  
$ 0.00\pm 0.01 $ & $ 1.00\pm 0.01 $ & $ 1.00\pm 0.01 $ \\
 & $ \Phi^{S2HFF1} = (-0.07\pm 0.00) + (0.99\pm 0.00)\times \Phi^{SP} $  &  
$ 0.00\pm 0.01 $ & $ 1.00\pm 0.01 $ & $ 1.00\pm 0.01 $ \\
 & $ \Phi^{S2HFF1} = (-0.05\pm 0.01) + (0.98\pm 0.00)\times \Phi^{SPFF1} $  &  
$ 0.00\pm 0.01 $ & $ 1.00\pm 0.01 $ & $ 1.00\pm 0.01 $ \\
 & & & \\
Penumbra  & $ \Phi^{SPFF1} = (0.05\pm 0.01) + (1.06\pm 0.00)\times \Phi^{SP} $  &  
$ 0.00\pm 0.01 $ & $ 1.00\pm 0.01 $ & $ 1.00\pm 0.01 $ \\
 & $ \Phi^{S2HFF1} = (-0.02\pm 0.00) + (1.00\pm 0.00)\times \Phi^{S2H} $  &  
$ 0.00\pm 0.01 $ & $ 1.00\pm 0.01 $ & $ 1.00\pm 0.01 $ \\
 & $ \Phi^{S2HFF1} = (0.03\pm 0.01) + (1.02\pm 0.01)\times \Phi^{SP} $  &  
$ 0.01\pm 0.01 $ & $ 1.00\pm 0.01 $ & $ 1.00\pm 0.01 $ \\
 & $ \Phi^{S2HFF1} = (-0.01\pm 0.01) + (0.96\pm 0.00)\times \Phi^{SPFF1} $  &  
$ 0.00\pm 0.01 $ & $ 1.00\pm 0.01 $ & $ 1.00\pm 0.01 $ \\
 & & & \\
Strong B Plage  & $ \Phi^{SPFF1} = (-0.02\pm 0.00) + (0.66\pm 0.04)\times \Phi^{SP} $  &  
$ 0.01\pm 0.01 $ & $ 0.98\pm 0.01 $ & $ 0.96\pm 0.01 $ \\
 & $ \Phi^{S2HFF1} = (-0.01\pm 0.00) + (0.86\pm 0.03)\times \Phi^{S2H} $  &  
$ 0.00\pm 0.01 $ & $ 0.99\pm 0.01 $ & $ 0.99\pm 0.01 $ \\
 & $ \Phi^{S2HFF1} = (-0.03\pm 0.00) + (0.54\pm 0.05)\times \Phi^{SP} $  &  
$ 0.01\pm 0.01 $ & $ 0.95\pm 0.01 $ & $ 0.90\pm 0.01 $ \\
 & $ \Phi^{S2HFF1} = (-0.01\pm 0.00) + (0.83\pm 0.04)\times \Phi^{SPFF1} $  &  
$ 0.01\pm 0.01 $ & $ 0.99\pm 0.01 $ & $ 0.97\pm 0.01 $ \\
 & & & \\
Weak B Plage  & $ \Phi^{SPFF1} = (-0.00\pm 0.00) + (0.61\pm 0.04)\times \Phi^{SP} $  &  
$ 0.00\pm 0.01 $ & $ 0.98\pm 0.01 $ & $ 0.96\pm 0.01 $ \\
 & $ \Phi^{S2HFF1} = (-0.00\pm 0.00) + (0.87\pm 0.03)\times \Phi^{S2H} $  &  
$ 0.00\pm 0.01 $ & $ 0.99\pm 0.01 $ & $ 0.98\pm 0.01 $ \\
 & $ \Phi^{S2HFF1} = (-0.01\pm 0.00) + (0.64\pm 0.05)\times \Phi^{SP} $  &  
$ 0.00\pm 0.01 $ & $ 0.97\pm 0.01 $ & $ 0.94\pm 0.01 $ \\
 & $ \Phi^{S2HFF1} = (-0.00\pm 0.00) + (1.04\pm 0.03)\times \Phi^{SPFF1} $  &  
$ 0.00\pm 0.01 $ & $ 1.00\pm 0.01 $ & $ 0.99\pm 0.01 $ \\
 & & & \\
 $\mathbf{Solar\ Feature}$ &  $\widehat{\Phi}^{SPFF1/S2HFF1}$  vs $\widehat{\Phi}^{SP/S2H/SP/SPFF1}$ & $\mathbf{s_{e}}$ & $\mathbf{r}$ & $\mathbf{r^{2}}$ \\
\hline
 & & & \\
Umbra  & $ \widehat{\Phi}^{SPFF1} = (0.02\pm 0.00) + (1.00\pm 0.00)\times \widehat{\Phi}^{SP} $  &  
$ 0.00\pm 0.01 $ & $ 1.00\pm 0.01 $ & $ 1.00\pm 0.01 $ \\
 & $ \widehat{\Phi}^{S2HFF1} = (0.01\pm 0.00) + (1.00\pm 0.00)\times \widehat{\Phi}^{S2H} $  &  
$ 0.00\pm 0.01 $ & $ 1.00\pm 0.01 $ & $ 1.00\pm 0.01 $ \\
 & $ \widehat{\Phi}^{S2HFF1} = (0.07\pm 0.00) + (0.99\pm 0.00)\times \widehat{\Phi}^{SP} $  &  
$ 0.00\pm 0.01 $ & $ 1.00\pm 0.01 $ & $ 1.00\pm 0.01 $ \\
 & $ \widehat{\Phi}^{S2HFF1} = (0.05\pm 0.01) + (0.98\pm 0.00)\times \widehat{\Phi}^{SPFF1} $  &  
$ 0.00\pm 0.01 $ & $ 1.00\pm 0.01 $ & $ 1.00\pm 0.01 $ \\
 & & & \\
Penumbra  & $ \widehat{\Phi}^{SPFF1} = (0.01\pm 0.01) + (1.03\pm 0.00)\times \widehat{\Phi}^{SP} $  &  
$ 0.00\pm 0.01 $ & $ 1.00\pm 0.01 $ & $ 1.00\pm 0.01 $ \\
 & $ \widehat{\Phi}^{S2HFF1} = (0.04\pm 0.01) + (0.99\pm 0.00)\times \widehat{\Phi}^{S2H} $  &  
$ 0.00\pm 0.01 $ & $ 1.00\pm 0.01 $ & $ 1.00\pm 0.01 $ \\
 & $ \widehat{\Phi}^{S2HFF1} = (0.10\pm 0.02) + (0.95\pm 0.01)\times \widehat{\Phi}^{SP} $  &  
$ 0.01\pm 0.01 $ & $ 1.00\pm 0.01 $ & $ 1.00\pm 0.01 $ \\
 & $ \widehat{\Phi}^{S2HFF1} = (0.09\pm 0.01) + (0.92\pm 0.01)\times \widehat{\Phi}^{SPFF1} $  &  
$ 0.00\pm 0.01 $ & $ 1.00\pm 0.01 $ & $ 1.00\pm 0.01 $ \\
 & & & \\
Strong B Plage  & $ \widehat{\Phi}^{SPFF1} = (0.18\pm 0.02) + (0.60\pm 0.02)\times \widehat{\Phi}^{SP} $  &  
$ 0.01\pm 0.01 $ & $ 0.99\pm 0.01 $ & $ 0.99\pm 0.01 $ \\
 & $ \widehat{\Phi}^{S2HFF1} = (0.07\pm 0.01) + (0.79\pm 0.01)\times \widehat{\Phi}^{S2H} $  &  
$ 0.01\pm 0.01 $ & $ 1.00\pm 0.01 $ & $ 1.00\pm 0.01 $ \\
 & $ \widehat{\Phi}^{S2HFF1} = (0.13\pm 0.01) + (0.55\pm 0.02)\times \widehat{\Phi}^{SP} $  &  
$ 0.01\pm 0.01 $ & $ 1.00\pm 0.01 $ & $ 0.99\pm 0.01 $ \\
 & $ \widehat{\Phi}^{S2HFF1} = (-0.03\pm 0.01) + (0.91\pm 0.02)\times \widehat{\Phi}^{SPFF1} $  &  
$ 0.01\pm 0.01 $ & $ 1.00\pm 0.01 $ & $ 1.00\pm 0.01 $ \\
 & & & \\
Weak B Plage  & $ \widehat{\Phi}^{SPFF1} = (0.10\pm 0.01) + (0.44\pm 0.04)\times \widehat{\Phi}^{SP} $  &  
$ 0.01\pm 0.01 $ & $ 0.95\pm 0.01 $ & $ 0.89\pm 0.01 $ \\
 & $ \widehat{\Phi}^{S2HFF1} = (0.05\pm 0.01) + (0.69\pm 0.03)\times \widehat{\Phi}^{S2H} $  &  
$ 0.00\pm 0.01 $ & $ 0.99\pm 0.01 $ & $ 0.98\pm 0.01 $ \\
 & $ \widehat{\Phi}^{S2HFF1} = (0.08\pm 0.01) + (0.50\pm 0.03)\times \widehat{\Phi}^{SP} $  &  
$ 0.01\pm 0.01 $ & $ 0.98\pm 0.01 $ & $ 0.96\pm 0.01 $ \\
 & $ \widehat{\Phi}^{S2HFF1} = (-0.03\pm 0.01) + (1.08\pm 0.05)\times \widehat{\Phi}^{SPFF1} $  &  
$ 0.01\pm 0.01 $ & $ 0.99\pm 0.01 $ & $ 0.98\pm 0.01 $ \\
 & & & \\
\hline
\end{tabular}
\end{center}
\end{table*}
}

%% file: tablaHMIvsSP_fieldaziincli_dis_ff1_newcutpoly2d.tex
 
{
\begin{table*}[]
\scriptsize
\begin{center}
\caption{Comparison between the \newcorr{spherical components of the} vector magnetic field observed by HMI and SP, SP  disambiguated with AZAM2, and  with AMBIG.\label{tabla_compar_SPcgHMIdiscut}}
\begin{tabular}{llccc} 
 $\mathbf{Solar Feature}$ &  \vectorb{HMI}{}  vs \vectorb{SP}{} & $\mathbf{<s_{e}>}$ & $\mathbf{<r>}$ & $\mathbf{<r^{2}>}$ \\
\hline
Umbra
 & $ |B|^{HMI} = (-0.11\pm 0.31) + (0.99\pm 0.12)\times |B|^{SP} $  &  
$ 0.10\pm 0.03 $ & $ 0.92\pm 0.05 $ & $ 0.84\pm 0.09 $ \\
 & $ \theta_{B}^{HMI} = (12.60\pm 1.89) + (0.91\pm 0.01)\times \theta_{B}^{SP} $  &  
$ 1.89\pm 0.57 $ & $ 0.99\pm 0.01 $ & $ 0.98\pm 0.02 $ \\
 & $ \phi_{B}^{HMI} = (14.77\pm 6.60) + (0.82\pm 0.08)\times \phi_{B}^{SP} $  &  
$ 29.04\pm 5.67 $ & $ 0.77\pm 0.12 $ & $ 0.61\pm 0.18 $ \\
 & $ \phi_{B}^{HMI\ DIS} = (5.48\pm 10.49) + (0.97\pm 0.05)\times \phi_{B}^{SP\ AZAM} $  &  
$ 19.11\pm 3.85 $ & $ 0.91\pm 0.04 $ & $ 0.83\pm 0.08 $ \\
 & $ \phi_{B}^{HMI\ DIS} = (6.65\pm 21.73) + (0.96\pm 0.10)\times \phi_{B}^{SP\ AZAM2} $  &  
$ 20.18\pm 6.44 $ & $ 0.90\pm 0.07 $ & $ 0.82\pm 0.11 $ \\
 & $ \phi_{B}^{HMI\ DIS} = (13.06\pm 14.68) + (0.93\pm 0.07)\times \phi_{B}^{SP\ AMBIG} $  &  
$ 20.88\pm 3.99 $ & $ 0.89\pm 0.06 $ & $ 0.80\pm 0.10 $ \\
 & & & \\
Penumbra
 & $ |B|^{HMI} = (0.01\pm 0.01) + (0.93\pm 0.02)\times |B|^{SP} $  &  
$ 0.13\pm 0.02 $ & $ 0.96\pm 0.01 $ & $ 0.91\pm 0.02 $ \\
 & $ \theta_{B}^{HMI} = (14.37\pm 2.01) + (0.86\pm 0.02)\times \theta_{B}^{SP} $  &  
$ 4.98\pm 0.40 $ & $ 0.97\pm 0.02 $ & $ 0.94\pm 0.03 $ \\
 & $ \phi_{B}^{HMI} = (13.00\pm 4.82) + (0.85\pm 0.04)\times \phi_{B}^{SP} $  &  
$ 24.95\pm 2.49 $ & $ 0.87\pm 0.03 $ & $ 0.75\pm 0.05 $ \\
 & $ \phi_{B}^{HMI\ DIS} = (13.11\pm 6.93) + (0.92\pm 0.03)\times \phi_{B}^{SP\ AZAM} $  &  
$ 36.28\pm 6.33 $ & $ 0.92\pm 0.03 $ & $ 0.85\pm 0.05 $ \\
 & $ \phi_{B}^{HMI\ DIS} = (13.44\pm 7.04) + (0.92\pm 0.03)\times \phi_{B}^{SP\ AZAM2} $  &  
$ 36.88\pm 6.73 $ & $ 0.92\pm 0.03 $ & $ 0.84\pm 0.05 $ \\
 & $ \phi_{B}^{HMI\ DIS} = (13.58\pm 6.84) + (0.92\pm 0.03)\times \phi_{B}^{SP\ AMBIG} $  &  
$ 37.17\pm 6.59 $ & $ 0.91\pm 0.03 $ & $ 0.84\pm 0.05 $ \\
 & & & \\
Strong B Plage
 & $ |B|^{HMI} = (0.19\pm 0.01) + (0.07\pm 0.03)\times |B|^{SP} $  &  
$ 0.12\pm 0.01 $ & $ 0.16\pm 0.08 $ & $ 0.03\pm 0.03 $ \\
 & $ \theta_{B}^{HMI} = (56.19\pm 1.83) + (0.39\pm 0.02)\times \theta_{B}^{SP} $  &  
$ 11.06\pm 0.77 $ & $ 0.83\pm 0.01 $ & $ 0.68\pm 0.02 $ \\
 & $ \phi_{B}^{HMI} = (52.10\pm 7.52) + (0.43\pm 0.09)\times \phi_{B}^{SP} $  &  
$ 42.11\pm 2.15 $ & $ 0.46\pm 0.06 $ & $ 0.21\pm 0.06 $ \\
 & $ \phi_{B}^{HMI\ DIS} = (122.92\pm 7.75) + (0.45\pm 0.03)\times \phi_{B}^{SP\ AZAM} $  &  
$ 82.86\pm 5.05 $ & $ 0.47\pm 0.04 $ & $ 0.22\pm 0.04 $ \\
 & $ \phi_{B}^{HMI\ DIS} = (125.45\pm 9.46) + (0.43\pm 0.03)\times \phi_{B}^{SP\ AZAM2} $  &  
$ 83.71\pm 4.63 $ & $ 0.45\pm 0.04 $ & $ 0.21\pm 0.03 $ \\
 & $ \phi_{B}^{HMI\ DIS} = (112.48\pm 10.89) + (0.49\pm 0.04)\times \phi_{B}^{SP\ AMBIG} $  &  
$ 84.97\pm 4.77 $ & $ 0.43\pm 0.04 $ & $ 0.18\pm 0.03 $ \\
 & & & \\
Weak B Plage
 & $ |B|^{HMI} = (0.13\pm 0.01) + (0.00\pm 0.01)\times |B|^{SP} $  &  
$ 0.04\pm 0.00 $ & $ 0.01\pm 0.04 $ & $ 0.00\pm 0.00 $ \\
 & $ \theta_{B}^{HMI} = (62.72\pm 2.41) + (0.30\pm 0.02)\times \theta_{B}^{SP} $  &  
$ 11.17\pm 0.97 $ & $ 0.78\pm 0.02 $ & $ 0.61\pm 0.03 $ \\
 & $ \phi_{B}^{HMI} = (72.95\pm 4.70) + (0.19\pm 0.03)\times \phi_{B}^{SP} $  &  
$ 44.44\pm 0.88 $ & $ 0.20\pm 0.02 $ & $ 0.04\pm 0.01 $ \\
 & $ \phi_{B}^{HMI\ DIS} = (174.89\pm 9.38) + (0.23\pm 0.02)\times \phi_{B}^{SP\ AZAM} $  &  
$ 85.83\pm 2.88 $ & $ 0.26\pm 0.02 $ & $ 0.07\pm 0.01 $ \\
 & $ \phi_{B}^{HMI\ DIS} = (179.99\pm 10.53) + (0.19\pm 0.03)\times \phi_{B}^{SP\ AZAM2} $  &  
$ 86.63\pm 2.84 $ & $ 0.22\pm 0.03 $ & $ 0.05\pm 0.01 $ \\
 & $ \phi_{B}^{HMI\ DIS} = (195.01\pm 11.89) + (0.10\pm 0.05)\times \phi_{B}^{SP\ AMBIG} $  &  
$ 88.48\pm 3.01 $ & $ 0.08\pm 0.04 $ & $ 0.01\pm 0.01 $ \\
 & & & \\
 $\mathbf{Solar Feature}$ &  \vectorb{HMI}{}  vs \vectorb{SPFF1}{} & $\mathbf{<s_{e}>}$ & $\mathbf{<r>}$ & $\mathbf{<r^{2}>}$ \\
\hline
Umbra
 & $ |B|^{HMI} = (-0.07\pm 0.21) + (0.98\pm 0.09)\times |B|^{SPFF1} $  &  
$ 0.09\pm 0.03 $ & $ 0.92\pm 0.05 $ & $ 0.85\pm 0.08 $ \\
 & $ \theta_{B}^{HMI} = (10.79\pm 2.17) + (0.92\pm 0.01)\times \theta_{B}^{SPFF1} $  &  
$ 1.84\pm 0.55 $ & $ 0.99\pm 0.01 $ & $ 0.98\pm 0.02 $ \\
 & $ \phi_{B}^{HMI} = (14.98\pm 6.84) + (0.81\pm 0.08)\times \phi_{B}^{SPFF1} $  &  
$ 29.06\pm 5.83 $ & $ 0.77\pm 0.12 $ & $ 0.60\pm 0.19 $ \\
 & $ \phi_{B}^{HMI\ DIS} = (5.45\pm 9.98) + (0.97\pm 0.05)\times \phi_{B}^{SPFF1\ AZAM} $  &  
$ 19.10\pm 3.83 $ & $ 0.91\pm 0.04 $ & $ 0.84\pm 0.07 $ \\
 & $ \phi_{B}^{HMI\ DIS} = (11.46\pm 18.91) + (0.94\pm 0.08)\times \phi_{B}^{SPFF1\ AZAM2} $  &  
$ 21.31\pm 5.65 $ & $ 0.89\pm 0.06 $ & $ 0.80\pm 0.10 $ \\
 & $ \phi_{B}^{HMI\ DIS} = (13.06\pm 14.68) + (0.93\pm 0.07)\times \phi_{B}^{SPFF1\ AMBIG} $  &  
$ 20.88\pm 3.99 $ & $ 0.89\pm 0.06 $ & $ 0.80\pm 0.10 $ \\
 & & & \\
Penumbra
 & $ |B|^{HMI} = (0.10\pm 0.02) + (0.92\pm 0.01)\times |B|^{SPFF1} $  &  
$ 0.09\pm 0.01 $ & $ 0.98\pm 0.01 $ & $ 0.96\pm 0.01 $ \\
 & $ \theta_{B}^{HMI} = (11.14\pm 2.27) + (0.90\pm 0.02)\times \theta_{B}^{SPFF1} $  &  
$ 4.56\pm 0.41 $ & $ 0.97\pm 0.01 $ & $ 0.95\pm 0.03 $ \\
 & $ \phi_{B}^{HMI} = (12.70\pm 4.82) + (0.86\pm 0.04)\times \phi_{B}^{SPFF1} $  &  
$ 24.57\pm 2.76 $ & $ 0.87\pm 0.03 $ & $ 0.76\pm 0.05 $ \\
 & $ \phi_{B}^{HMI\ DIS} = (13.50\pm 6.67) + (0.92\pm 0.03)\times \phi_{B}^{SPFF1\ AZAM} $  &  
$ 36.83\pm 6.26 $ & $ 0.92\pm 0.03 $ & $ 0.84\pm 0.05 $ \\
 & $ \phi_{B}^{HMI\ DIS} = (13.82\pm 6.89) + (0.92\pm 0.03)\times \phi_{B}^{SPFF1\ AZAM2} $  &  
$ 37.38\pm 6.56 $ & $ 0.91\pm 0.03 $ & $ 0.84\pm 0.05 $ \\
 & $ \phi_{B}^{HMI\ DIS} = (13.58\pm 6.84) + (0.92\pm 0.03)\times \phi_{B}^{SPFF1\ AMBIG} $  &  
$ 37.17\pm 6.59 $ & $ 0.91\pm 0.03 $ & $ 0.84\pm 0.05 $ \\
 & & & \\
Strong B Plage
 & $ |B|^{HMI} = (0.17\pm 0.01) + (0.35\pm 0.06)\times |B|^{SPFF1} $  &  
$ 0.11\pm 0.01 $ & $ 0.49\pm 0.07 $ & $ 0.25\pm 0.07 $ \\
 & $ \theta_{B}^{HMI} = (32.38\pm 3.30) + (0.65\pm 0.04)\times \theta_{B}^{SPFF1} $  &  
$ 9.14\pm 1.01 $ & $ 0.91\pm 0.01 $ & $ 0.82\pm 0.03 $ \\
 & $ \phi_{B}^{HMI} = (35.82\pm 8.28) + (0.60\pm 0.09)\times \phi_{B}^{SPFF1} $  &  
$ 37.83\pm 3.97 $ & $ 0.63\pm 0.07 $ & $ 0.40\pm 0.09 $ \\
 & $ \phi_{B}^{HMI\ DIS} = (75.07\pm 11.34) + (0.63\pm 0.06)\times \phi_{B}^{SPFF1\ AZAM} $  &  
$ 74.13\pm 7.92 $ & $ 0.64\pm 0.07 $ & $ 0.41\pm 0.09 $ \\
 & $ \phi_{B}^{HMI\ DIS} = (78.54\pm 8.88) + (0.61\pm 0.03)\times \phi_{B}^{SPFF1\ AZAM2} $  &  
$ 75.19\pm 5.68 $ & $ 0.63\pm 0.03 $ & $ 0.39\pm 0.04 $ \\
 & $ \phi_{B}^{HMI\ DIS} = (76.08\pm 9.66) + (0.62\pm 0.05)\times \phi_{B}^{SPFF1\ AMBIG} $  &  
$ 77.70\pm 6.27 $ & $ 0.59\pm 0.04 $ & $ 0.35\pm 0.05 $ \\
 & & & \\
Weak B Plage
 & $ |B|^{HMI} = (0.12\pm 0.01) + (0.10\pm 0.01)\times |B|^{SPFF1} $  &  
$ 0.05\pm 0.00 $ & $ 0.19\pm 0.03 $ & $ 0.04\pm 0.01 $ \\
 & $ \theta_{B}^{HMI} = (51.79\pm 2.84) + (0.42\pm 0.03)\times \theta_{B}^{SPFF1} $  &  
$ 11.29\pm 0.98 $ & $ 0.79\pm 0.02 $ & $ 0.62\pm 0.03 $ \\
 & $ \phi_{B}^{HMI} = (70.08\pm 5.61) + (0.22\pm 0.04)\times \phi_{B}^{SPFF1} $  &  
$ 44.30\pm 1.11 $ & $ 0.24\pm 0.03 $ & $ 0.06\pm 0.01 $ \\
 & $ \phi_{B}^{HMI\ DIS} = (171.25\pm 9.31) + (0.24\pm 0.02)\times \phi_{B}^{SPFF1\ AZAM} $  &  
$ 86.60\pm 3.05 $ & $ 0.26\pm 0.02 $ & $ 0.07\pm 0.01 $ \\
 & $ \phi_{B}^{HMI\ DIS} = (175.75\pm 9.34) + (0.20\pm 0.02)\times \phi_{B}^{SPFF1\ AZAM2} $  &  
$ 87.30\pm 2.99 $ & $ 0.23\pm 0.02 $ & $ 0.05\pm 0.01 $ \\
 & $ \phi_{B}^{HMI\ DIS} = (189.29\pm 11.56) + (0.12\pm 0.05)\times \phi_{B}^{SPFF1\ AMBIG} $  &  
$ 89.24\pm 3.13 $ & $ 0.10\pm 0.04 $ & $ 0.01\pm 0.01 $ \\
 & & & \\
 \hline
\end{tabular}
\end{center}
\end{table*}
}

%% file: tablaHMIvsS2H_fieldaziincli_dis_ff1_newcutpoly2d.tex
 
{
\begin{table*}[]
\scriptsize
\begin{center}
\caption{Comparison between the \newcorr{spherical components of the} vector magnetic field observed by HMI and S2H, HMI and S2H disambiguated by DIS, and the corresponding comparisons for S2H with FF=1  (S2HFF1).\label{tabla_compar_S2HcgHMIdiscut}}
\begin{tabular}{llccc} 
 $\mathbf{Solar Feature}$ &  \vectorb{HMI}{}  vs \vectorb{S2H}{} & $\mathbf{<s_{e}>}$ & $\mathbf{<r>}$ & $\mathbf{<r^{2}>}$ \\
\hline
Umbra
 & $ |B|^{HMI} = (0.23\pm 0.18) + (0.85\pm 0.07)\times |B|^{S2H} $  &  
$ 0.10\pm 0.02 $ & $ 0.91\pm 0.04 $ & $ 0.84\pm 0.07 $ \\
 & $ \theta_{B}^{HMI} = (13.57\pm 2.28) + (0.89\pm 0.02)\times \theta_{B}^{S2H} $  &  
$ 2.03\pm 0.52 $ & $ 0.99\pm 0.01 $ & $ 0.98\pm 0.02 $ \\
 & $ \phi_{B}^{HMI} = (16.30\pm 6.50) + (0.79\pm 0.08)\times \phi_{B}^{S2H} $  &  
$ 30.40\pm 5.28 $ & $ 0.75\pm 0.12 $ & $ 0.57\pm 0.19 $ \\
 & $ \phi_{B}^{HMI\ DIS} = (67.98\pm 50.21) + (0.67\pm 0.20)\times \phi_{B}^{S2H\ DIS} $  &  
$ 25.27\pm 3.31 $ & $ 0.84\pm 0.07 $ & $ 0.71\pm 0.11 $ \\
 & & & \\
Penumbra
 & $ |B|^{HMI} = (0.01\pm 0.02) + (0.98\pm 0.02)\times |B|^{S2H} $  &  
$ 0.11\pm 0.01 $ & $ 0.97\pm 0.01 $ & $ 0.95\pm 0.01 $ \\
 & $ \theta_{B}^{HMI} = (11.39\pm 2.56) + (0.90\pm 0.02)\times \theta_{B}^{S2H} $  &  
$ 4.62\pm 0.52 $ & $ 0.97\pm 0.02 $ & $ 0.95\pm 0.03 $ \\
 & $ \phi_{B}^{HMI} = (10.78\pm 4.47) + (0.88\pm 0.04)\times \phi_{B}^{S2H} $  &  
$ 23.36\pm 3.69 $ & $ 0.88\pm 0.04 $ & $ 0.78\pm 0.06 $ \\
 & $ \phi_{B}^{HMI\ DIS} = (8.39\pm 4.99) + (0.95\pm 0.02)\times \phi_{B}^{S2H\ DIS} $  &  
$ 25.11\pm 3.98 $ & $ 0.96\pm 0.01 $ & $ 0.93\pm 0.02 $ \\
 & & & \\
Strong B Plage
 & $ |B|^{HMI} = (0.19\pm 0.01) + (0.13\pm 0.02)\times |B|^{S2H} $  &  
$ 0.12\pm 0.01 $ & $ 0.28\pm 0.05 $ & $ 0.08\pm 0.03 $ \\
 & $ \theta_{B}^{HMI} = (47.78\pm 2.67) + (0.48\pm 0.03)\times \theta_{B}^{S2H} $  &  
$ 10.89\pm 0.82 $ & $ 0.84\pm 0.01 $ & $ 0.71\pm 0.02 $ \\
 & $ \phi_{B}^{HMI} = (47.55\pm 8.57) + (0.47\pm 0.09)\times \phi_{B}^{S2H} $  &  
$ 41.44\pm 2.67 $ & $ 0.50\pm 0.07 $ & $ 0.25\pm 0.07 $ \\
 & $ \phi_{B}^{HMI\ DIS} = (37.81\pm 7.52) + (0.81\pm 0.03)\times \phi_{B}^{S2H\ DIS} $  &  
$ 50.63\pm 5.02 $ & $ 0.85\pm 0.02 $ & $ 0.72\pm 0.04 $ \\
 & & & \\
Weak B Plage
 & $ |B|^{HMI} = (0.14\pm 0.01)  (-0.00\pm 0.01)\times |B|^{S2H} $  &  
$ 0.05\pm 0.00 $ & $ -0.01\pm 0.03 $ & $ 0.00\pm 0.00 $ \\
 & $ \theta_{B}^{HMI} = (54.02\pm 2.97) + (0.40\pm 0.03)\times \theta_{B}^{S2H} $  &  
$ 11.34\pm 1.00 $ & $ 0.78\pm 0.02 $ & $ 0.61\pm 0.03 $ \\
 & $ \phi_{B}^{HMI} = (72.72\pm 5.15) + (0.18\pm 0.03)\times \phi_{B}^{S2H} $  &  
$ 44.61\pm 1.04 $ & $ 0.20\pm 0.02 $ & $ 0.04\pm 0.01 $ \\
 & $ \phi_{B}^{HMI\ DIS} = (66.67\pm 9.49) + (0.69\pm 0.03)\times \phi_{B}^{S2H\ DIS} $  &  
$ 59.83\pm 3.12 $ & $ 0.74\pm 0.02 $ & $ 0.55\pm 0.04 $ \\
 & & & \\
 $\mathbf{Solar Feature}$ &  \vectorb{HMI}{}  vs \vectorb{S2HFF1}{} & $\mathbf{<s_{e}>}$ & $\mathbf{<r>}$ & $\mathbf{<r^{2}>}$ \\
\hline
Umbra
 & $ |B|^{HMI} = (0.25\pm 0.15) + (0.85\pm 0.06)\times |B|^{S2HFF1} $  &  
$ 0.08\pm 0.02 $ & $ 0.94\pm 0.04 $ & $ 0.88\pm 0.06 $ \\
 & $ \theta_{B}^{HMI} = (12.69\pm 2.93) + (0.90\pm 0.02)\times \theta_{B}^{S2HFF1} $  &  
$ 1.99\pm 0.56 $ & $ 0.99\pm 0.01 $ & $ 0.98\pm 0.02 $ \\
 & $ \phi_{B}^{HMI} = (16.15\pm 6.75) + (0.80\pm 0.08)\times \phi_{B}^{S2HFF1} $  &  
$ 30.24\pm 5.78 $ & $ 0.75\pm 0.12 $ & $ 0.58\pm 0.18 $ \\
 & $ \phi_{B}^{HMI\ DIS} = (68.01\pm 48.61) + (0.67\pm 0.19)\times \phi_{B}^{S2HFF1\ DIS} $  &  
$ 25.41\pm 4.22 $ & $ 0.84\pm 0.07 $ & $ 0.71\pm 0.11 $ \\
 & & & \\
Penumbra
 & $ |B|^{HMI} = (0.04\pm 0.01) + (0.96\pm 0.01)\times |B|^{S2HFF1} $  &  
$ 0.09\pm 0.01 $ & $ 0.98\pm 0.00 $ & $ 0.97\pm 0.01 $ \\
 & $ \theta_{B}^{HMI} = (10.31\pm 2.51) + (0.91\pm 0.02)\times \theta_{B}^{S2HFF1} $  &  
$ 4.50\pm 0.47 $ & $ 0.97\pm 0.01 $ & $ 0.95\pm 0.03 $ \\
 & $ \phi_{B}^{HMI} = (10.46\pm 4.22) + (0.88\pm 0.04)\times \phi_{B}^{S2HFF1} $  &  
$ 23.22\pm 3.69 $ & $ 0.88\pm 0.04 $ & $ 0.78\pm 0.06 $ \\
 & & & \\
 & $ \phi_{B}^{HMI\ DIS} = (8.23\pm 4.56) + (0.95\pm 0.02)\times \phi_{B}^{S2HFF1\ DIS} $  &  
$ 25.00\pm 3.95 $ & $ 0.96\pm 0.01 $ & $ 0.93\pm 0.02 $ \\
Strong B Plage
 & $ |B|^{HMI} = (0.15\pm 0.01) + (0.41\pm 0.05)\times |B|^{S2HFF1} $  &  
$ 0.10\pm 0.01 $ & $ 0.53\pm 0.06 $ & $ 0.28\pm 0.06 $ \\
 & $ \theta_{B}^{HMI} = (30.68\pm 4.33) + (0.67\pm 0.05)\times \theta_{B}^{S2HFF1} $  &  
$ 9.38\pm 1.07 $ & $ 0.90\pm 0.01 $ & $ 0.81\pm 0.03 $ \\
 & $ \phi_{B}^{HMI} = (39.12\pm 9.12) + (0.56\pm 0.09)\times \phi_{B}^{S2HFF1} $  &  
$ 39.62\pm 3.77 $ & $ 0.58\pm 0.08 $ & $ 0.34\pm 0.09 $ \\
 & $ \phi_{B}^{HMI\ DIS} = (29.07\pm 6.71) + (0.85\pm 0.03)\times \phi_{B}^{S2HFF1\ DIS} $  &  
$ 46.70\pm 5.81 $ & $ 0.87\pm 0.02 $ & $ 0.76\pm 0.04 $ \\
 & & & \\
Weak B Plage
 & $ |B|^{HMI} = (0.12\pm 0.01) + (0.09\pm 0.01)\times |B|^{S2HFF1} $  &  
$ 0.05\pm 0.00 $ & $ 0.16\pm 0.04 $ & $ 0.03\pm 0.01 $ \\
 & $ \theta_{B}^{HMI} = (43.64\pm 3.57) + (0.52\pm 0.04)\times \theta_{B}^{S2HFF1} $  &  
$ 10.72\pm 1.10 $ & $ 0.81\pm 0.02 $ & $ 0.66\pm 0.03 $ \\
 & $ \phi_{B}^{HMI} = (72.06\pm 5.52) + (0.19\pm 0.04)\times \phi_{B}^{S2HFF1} $  &  
$ 44.65\pm 1.11 $ & $ 0.21\pm 0.03 $ & $ 0.04\pm 0.01 $ \\
 & $ \phi_{B}^{HMI\ DIS} = (65.55\pm 8.87) + (0.69\pm 0.03)\times \phi_{B}^{S2HFF1\ DIS} $  &  
$ 59.79\pm 3.07 $ & $ 0.74\pm 0.02 $ & $ 0.55\pm 0.03 $ \\
 & & & \\
\hline 
\end{tabular}
\end{center}
\end{table*}
}

%% file: tablaHMIvsSP_SPAZAM2_Leka_poly2d.tex
 
{
\begin{table*}[]
\scriptsize
\begin{center}
\caption{Comparison between the \newcorr{Cartesian components of the} vector magnetic field observed by HMI and SP, and SP disambiguated with the AZAM2, and with AMBIG.\label{tabla_hmi_sp_disamb}}
\begin{tabular}{llccc} 
 $\mathbf{Solar\ Feature}$ &  \vectorb{HMI}{}  vs \vectorb{SP}{} & $\mathbf{<s_e>}$ & $\mathbf{<r>}$ & $\mathbf{<r^{2}>}$ \\
\hline
Umbra
 & $ B^{HMI}_{X} = (+0.05\pm 0.10) + (+0.99\pm 0.02)\times B^{SP}_{X} $  &  
$ 0.09\pm 0.02 $ & $ 0.99\pm 0.00 $ & $ 0.99\pm 0.01 $ \\
 & $ B^{HMI}_{Y} = (-0.00\pm 0.04) + (+0.98\pm 0.02)\times B^{SP}_{Y} $  &  
$ 0.09\pm 0.02 $ & $ 0.99\pm 0.00 $ & $ 0.98\pm 0.01 $ \\
 & $ B^{HMI}_{Z} = (-0.23\pm 0.10) + (+0.82\pm 0.06)\times B^{SP}_{Z} $  &  
$ 0.10\pm 0.03 $ & $ 0.96\pm 0.02 $ & $ 0.93\pm 0.04 $ \\
 & & & \\
Penumbra
 & $ B^{HMI}_{X} = (+0.00\pm 0.01) + (+0.97\pm 0.02)\times B^{SP}_{X} $  &  
$ 0.10\pm 0.01 $ & $ 0.99\pm 0.00 $ & $ 0.98\pm 0.01 $ \\
 & $ B^{HMI}_{Y} = (+0.01\pm 0.01) + (+0.95\pm 0.02)\times B^{SP}_{Y} $  &  
$ 0.11\pm 0.02 $ & $ 0.99\pm 0.00 $ & $ 0.97\pm 0.01 $ \\
 & $ B^{HMI}_{Z} = (-0.02\pm 0.01) + (+0.84\pm 0.05)\times B^{SP}_{Z} $  &  
$ 0.11\pm 0.01 $ & $ 0.97\pm 0.01 $ & $ 0.94\pm 0.01 $ \\
 & & & \\
Strong B Plage
 & $ B^{HMI}_{X} = (-0.03\pm 0.01) + (+0.40\pm 0.11)\times B^{SP}_{X} $  &  
$ 0.16\pm 0.02 $ & $ 0.65\pm 0.12 $ & $ 0.44\pm 0.14 $ \\
 & $ B^{HMI}_{Y} = (+0.02\pm 0.01) + (+0.33\pm 0.08)\times B^{SP}_{Y} $  &  
$ 0.16\pm 0.01 $ & $ 0.59\pm 0.10 $ & $ 0.35\pm 0.12 $ \\
 & $ B^{HMI}_{Z} = (-0.01\pm 0.01) + (+0.15\pm 0.01)\times B^{SP}_{Z} $  &  
$ 0.07\pm 0.01 $ & $ 0.72\pm 0.02 $ & $ 0.52\pm 0.04 $ \\
 & & & \\
Weak B Plage
 & $ B^{HMI}_{X} = (-0.02\pm 0.01) + (+0.21\pm 0.04)\times B^{SP}_{X} $  &  
$ 0.10\pm 0.01 $ & $ 0.40\pm 0.05 $ & $ 0.17\pm 0.04 $ \\
 & $ B^{HMI}_{Y} = (+0.01\pm 0.01) + (+0.17\pm 0.04)\times B^{SP}_{Y} $  &  
$ 0.10\pm 0.00 $ & $ 0.36\pm 0.05 $ & $ 0.14\pm 0.04 $ \\
 & $ B^{HMI}_{Z} = (-0.01\pm 0.00) + (+0.10\pm 0.02)\times B^{SP}_{Z} $  &  
$ 0.05\pm 0.00 $ & $ 0.63\pm 0.03 $ & $ 0.40\pm 0.04 $ \\
 & & & \\
 $\mathbf{Solar\ Feature}$ &  \vectorb{HMI}{}  vs \vectorb{SP\ AZAM2}{} & $\mathbf{<s_e>}$ & $\mathbf{<r>}$ & $\mathbf{<r^{2}>}$ \\
\hline
Umbra
 & $ B^{HMI}_{X} = (+0.05\pm 0.10) + (+0.99\pm 0.02)\times B^{SP}_{X} $  &  
$ 0.09\pm 0.02 $ & $ 0.99\pm 0.00 $ & $ 0.99\pm 0.01 $ \\
 & $ B^{HMI}_{Y} = (-0.00\pm 0.04) + (+0.97\pm 0.02)\times B^{SP}_{Y} $  &  
$ 0.10\pm 0.03 $ & $ 0.99\pm 0.00 $ & $ 0.98\pm 0.01 $ \\
 & $ B^{HMI}_{Z} = (-0.23\pm 0.10) + (+0.82\pm 0.06)\times B^{SP}_{Z} $  &  
$ 0.10\pm 0.03 $ & $ 0.96\pm 0.02 $ & $ 0.93\pm 0.04 $ \\
 & & & \\
Penumbra
 & $ B^{HMI}_{X} = (+0.00\pm 0.01) + (+0.97\pm 0.02)\times B^{SP}_{X} $  &  
$ 0.11\pm 0.02 $ & $ 0.99\pm 0.00 $ & $ 0.98\pm 0.01 $ \\
 & $ B^{HMI}_{Y} = (+0.01\pm 0.01) + (+0.95\pm 0.02)\times B^{SP}_{Y} $  &  
$ 0.11\pm 0.02 $ & $ 0.99\pm 0.00 $ & $ 0.97\pm 0.01 $ \\
 & $ B^{HMI}_{Z} = (-0.02\pm 0.01) + (+0.84\pm 0.05)\times B^{SP}_{Z} $  &  
$ 0.11\pm 0.01 $ & $ 0.97\pm 0.01 $ & $ 0.94\pm 0.01 $ \\
 & & & \\
Strong B Plage
 & $ B^{HMI}_{X} = (-0.03\pm 0.02) + (+0.39\pm 0.11)\times B^{SP}_{X} $  &  
$ 0.17\pm 0.02 $ & $ 0.63\pm 0.11 $ & $ 0.41\pm 0.14 $ \\
 & $ B^{HMI}_{Y} = (+0.02\pm 0.02) + (+0.32\pm 0.07)\times B^{SP}_{Y} $  &  
$ 0.16\pm 0.01 $ & $ 0.58\pm 0.09 $ & $ 0.34\pm 0.10 $ \\
 & $ B^{HMI}_{Z} = (-0.01\pm 0.01) + (+0.15\pm 0.01)\times B^{SP}_{Z} $  &  
$ 0.07\pm 0.01 $ & $ 0.72\pm 0.02 $ & $ 0.52\pm 0.04 $ \\
 & & & \\
Weak B Plage
 & $ B^{HMI}_{X} = (-0.02\pm 0.01) + (+0.18\pm 0.04)\times B^{SP}_{X} $  &  
$ 0.10\pm 0.01 $ & $ 0.36\pm 0.04 $ & $ 0.13\pm 0.03 $ \\
 & $ B^{HMI}_{Y} = (+0.01\pm 0.01) + (+0.16\pm 0.03)\times B^{SP}_{Y} $  &  
$ 0.10\pm 0.00 $ & $ 0.35\pm 0.04 $ & $ 0.12\pm 0.03 $ \\
 & $ B^{HMI}_{Z} = (-0.01\pm 0.00) + (+0.10\pm 0.02)\times B^{SP}_{Z} $  &  
$ 0.05\pm 0.00 $ & $ 0.63\pm 0.03 $ & $ 0.40\pm 0.04 $ \\
 & & & \\
 $\mathbf{Solar\ Feature}$ &  \vectorb{HMI}{}  vs \vectorb{SP\ AMBIG}{} & $\mathbf{<s_e>}$ & $\mathbf{<r>}$ & $\mathbf{<r^{2}>}$ \\
\hline
Umbra
 & $ B^{HMI}_{X} = (+0.05\pm 0.10) + (+0.99\pm 0.02)\times B^{SP}_{X} $  &  
$ 0.09\pm 0.02 $ & $ 0.99\pm 0.00 $ & $ 0.99\pm 0.01 $ \\
 & $ B^{HMI}_{Y} = (-0.00\pm 0.04) + (+0.98\pm 0.02)\times B^{SP}_{Y} $  &  
$ 0.09\pm 0.02 $ & $ 0.99\pm 0.00 $ & $ 0.98\pm 0.01 $ \\
 & $ B^{HMI}_{Z} = (-0.23\pm 0.10) + (+0.82\pm 0.06)\times B^{SP}_{Z} $  &  
$ 0.10\pm 0.03 $ & $ 0.96\pm 0.02 $ & $ 0.93\pm 0.04 $ \\
 & & & \\
Penumbra
 & $ B^{HMI}_{X} = (+0.00\pm 0.01) + (+0.97\pm 0.02)\times B^{SP}_{X} $  &  
$ 0.10\pm 0.01 $ & $ 0.99\pm 0.00 $ & $ 0.98\pm 0.01 $ \\
 & $ B^{HMI}_{Y} = (+0.01\pm 0.01) + (+0.95\pm 0.02)\times B^{SP}_{Y} $  &  
$ 0.11\pm 0.02 $ & $ 0.99\pm 0.00 $ & $ 0.97\pm 0.01 $ \\
 & $ B^{HMI}_{Z} = (-0.02\pm 0.01) + (+0.84\pm 0.05)\times B^{SP}_{Z} $  &  
$ 0.11\pm 0.01 $ & $ 0.97\pm 0.01 $ & $ 0.94\pm 0.01 $ \\
 & & & \\
Strong B Plage
 & $ B^{HMI}_{X} = (-0.02\pm 0.01) + (+0.39\pm 0.08)\times B^{SP}_{X} $  &  
$ 0.17\pm 0.01 $ & $ 0.65\pm 0.06 $ & $ 0.42\pm 0.07 $ \\
 & $ B^{HMI}_{Y} = (-0.01\pm 0.01) + (+0.36\pm 0.07)\times B^{SP}_{Y} $  &  
$ 0.16\pm 0.00 $ & $ 0.61\pm 0.07 $ & $ 0.38\pm 0.09 $ \\
 & $ B^{HMI}_{Z} = (-0.01\pm 0.01) + (+0.15\pm 0.01)\times B^{SP}_{Z} $  &  
$ 0.07\pm 0.01 $ & $ 0.72\pm 0.02 $ & $ 0.52\pm 0.04 $ \\
 & & & \\
Weak B Plage
 & $ B^{HMI}_{X} = (-0.02\pm 0.01) + (+0.14\pm 0.02)\times B^{SP}_{X} $  &  
$ 0.10\pm 0.01 $ & $ 0.28\pm 0.04 $ & $ 0.08\pm 0.02 $ \\
 & $ B^{HMI}_{Y} = (+0.00\pm 0.01) + (+0.14\pm 0.03)\times B^{SP}_{Y} $  &  
$ 0.10\pm 0.00 $ & $ 0.29\pm 0.03 $ & $ 0.08\pm 0.02 $ \\
 & $ B^{HMI}_{Z} = (-0.01\pm 0.00) + (+0.10\pm 0.02)\times B^{SP}_{Z} $  &  
$ 0.05\pm 0.00 $ & $ 0.63\pm 0.03 $ & $ 0.40\pm 0.04 $ \\
 & & & \\
 \hline
\end{tabular}
\end{center}
\end{table*}
}

%% file: tablaHMIvsSPnewcutposneg.tex
 
{
\begin{table*}[]
\scriptsize
\begin{center}
\caption{Comparison between the \newcorr{Cartesian components of the} vector magnetic field observed by HMI and SP taking into account the sign between the horizontal components.\label{tabla_HMISP_posneg}}
\begin{tabular}{lccccc} 
 $\mathbf{Solar Feature}$ &  \vectorb{HMI}{}  vs \vectorb{SP}{}\ Original & $\mathbf{<\sigma>}$ & $\mathbf{<r>}$ & $\mathbf{<r^{2}>}$  & \% Eval. \\
\hline
Strong B Plage
 & $ B^{HMI}_{X} = (-0.03\pm 0.01) + (+0.40\pm 0.11)\times B^{SP}_{X} $  &  
$ 0.16\pm 0.02 $ & $ +0.65\pm 0.12 $ & $ 0.44\pm 0.14 $ & $ 4\pm 0 $ \\
 & $ B^{HMI}_{Y} = (+0.02\pm 0.01) + (+0.33\pm 0.08)\times B^{SP}_{Y} $  &  
$ 0.16\pm 0.01 $ & $ +0.59\pm 0.10 $ & $ 0.35\pm 0.12 $ & $ 4\pm 0 $ \\
 & $ B^{HMI}_{Z} = (-0.01\pm 0.01) + (+0.15\pm 0.01)\times B^{SP}_{Z} $  &  
$ 0.07\pm 0.01 $ & $ +0.72\pm 0.02 $ & $ 0.52\pm 0.04 $ & $ 4\pm 0 $ \\
 & & & & \\
Weak B Plage
 & $ B^{HMI}_{X} = (-0.02\pm 0.01) + (+0.21\pm 0.04)\times B^{SP}_{X} $  &  
$ 0.10\pm 0.01 $ & $ +0.40\pm 0.05 $ & $ 0.17\pm 0.04 $ & $ 6\pm 0 $ \\
 & $ B^{HMI}_{Y} = (+0.01\pm 0.01) + (+0.17\pm 0.04)\times B^{SP}_{Y} $  &  
$ 0.10\pm 0.00 $ & $ +0.36\pm 0.05 $ & $ 0.14\pm 0.04 $ & $ 6\pm 1 $ \\
 & $ B^{HMI}_{Z} = (-0.01\pm 0.00) + (+0.10\pm 0.02)\times B^{SP}_{Z} $  &  
$ 0.05\pm 0.00 $ & $ +0.63\pm 0.03 $ & $ 0.40\pm 0.04 $ & $ 9\pm 1 $ \\
 & & & & \\
 $\mathbf{Solar Feature}$ &  \vectorb{HMI}{}  vs \vectorb{SP}{}\ Same Sign in XY (Pos.) & $\mathbf{<\sigma>}$ & $\mathbf{<r>}$ & $\mathbf{<r^{2}>}$  & \% Pos. \\
\hline
Strong B Plage
 & $ B^{HMI}_{X} = (-0.01\pm 0.01) + (+0.55\pm 0.08)\times B^{SP}_{X} $  &  
$ 0.12\pm 0.01 $ & $ +0.86\pm 0.03 $ & $ 0.74\pm 0.05 $ & $ 80\pm 4 $ \\
 & $ B^{HMI}_{Y} = (+0.01\pm 0.01) + (+0.50\pm 0.05)\times B^{SP}_{Y} $  &  
$ 0.12\pm 0.00 $ & $ +0.84\pm 0.03 $ & $ 0.71\pm 0.05 $ & $ 77\pm 5 $ \\
 & $ B^{HMI}_{Z} = (-0.01\pm 0.01) + (+0.15\pm 0.01)\times B^{SP}_{Z} $  &  
$ 0.07\pm 0.01 $ & $ +0.74\pm 0.02 $ & $ 0.54\pm 0.03 $ & $ 95\pm 1 $ \\
 & & & & \\
Weak B Plage
 & $ B^{HMI}_{X} = (-0.00\pm 0.00) + (+0.41\pm 0.06)\times B^{SP}_{X} $  &  
$ 0.08\pm 0.01 $ & $ +0.76\pm 0.03 $ & $ 0.58\pm 0.05 $ & $ 67\pm 3 $ \\
 & $ B^{HMI}_{Y} = (+0.01\pm 0.01) + (+0.37\pm 0.04)\times B^{SP}_{Y} $  &  
$ 0.07\pm 0.00 $ & $ +0.75\pm 0.02 $ & $ 0.56\pm 0.04 $ & $ 67\pm 2 $ \\
 & $ B^{HMI}_{Z} = (-0.01\pm 0.00) + (+0.11\pm 0.02)\times B^{SP}_{Z} $  &  
$ 0.05\pm 0.00 $ & $ +0.65\pm 0.03 $ & $ 0.42\pm 0.04 $ & $ 94\pm 1 $ \\
 & & & & \\
 $\mathbf{Solar Feature}$ &  \vectorb{HMI}{}  vs \vectorb{SP}{}\ Opposite Sign in XY (Neg.) & $\mathbf{<\sigma>}$ & $\mathbf{<r>}$ & $\mathbf{<r^{2}>}$  & \% Neg. \\
\hline
Strong B Plage
 & $ B^{HMI}_{X} = (+0.00\pm 0.02) + (-0.29\pm 0.04)\times B^{SP}_{X} $  &  
$ 0.08\pm 0.01 $ & $ -0.73\pm 0.11 $ & $ 0.54\pm 0.14 $ & $ 20\pm 4 $ \\
 & $ B^{HMI}_{Y} = (-0.01\pm 0.01) + (-0.28\pm 0.03)\times B^{SP}_{Y} $  &  
$ 0.08\pm 0.01 $ & $ -0.78\pm 0.03 $ & $ 0.61\pm 0.05 $ & $ 23\pm 5 $ \\
 & $ B^{HMI}_{Z} = (-0.00\pm 0.00) + (-0.02\pm 0.01)\times B^{SP}_{Z} $  &  
$ 0.02\pm 0.01 $ & $ -0.48\pm 0.10 $ & $ 0.24\pm 0.09 $ & $ 5\pm 1 $ \\
 & & & & \\
Weak B Plage
 & $ B^{HMI}_{X} = (-0.02\pm 0.01) + (-0.26\pm 0.04)\times B^{SP}_{X} $  &  
$ 0.07\pm 0.00 $ & $ -0.59\pm 0.05 $ & $ 0.35\pm 0.06 $ & $ 33\pm 3 $ \\
 & $ B^{HMI}_{Y} = (+0.00\pm 0.01) + (-0.27\pm 0.04)\times B^{SP}_{Y} $  &  
$ 0.06\pm 0.00 $ & $ -0.70\pm 0.03 $ & $ 0.48\pm 0.04 $ & $ 33\pm 2 $ \\
 & $ B^{HMI}_{Z} = (-0.00\pm 0.00) + (-0.03\pm 0.01)\times B^{SP}_{Z} $  &  
$ 0.02\pm 0.00 $ & $ -0.30\pm 0.05 $ & $ 0.09\pm 0.03 $ & $ 6\pm 1 $ \\
 & & & & \\
 \hline
\end{tabular}
\end{center}
\end{table*}
}

%% file: tablaHMIvsSPFF1newcutposneg.tex
 
{
\begin{table*}[]
\scriptsize
\begin{center}
\caption{Comparison between the \newcorr{Cartesian components of the} vector magnetic field observed by HMI and SPFF1 taking into account the sign between the horizontal components.\label{tabla_compar_SPFF1cgHMInewcutposneg}}
\begin{tabular}{lccccc} 
 $\mathbf{Solar Feature}$ &  \vectorb{HMI}{}  vs \vectorb{SPFF1}{}\ Original & $\mathbf{<s_e>}$ & $\mathbf{<r>}$ & $\mathbf{<r^{2}>}$  & \% Eval. \\
\hline
Strong B Plage
 & $ B^{HMI}_{X} = (-0.01\pm 0.01) + (+0.74\pm 0.17)\times B^{SPFF1}_{X} $  &  
$ 0.16\pm 0.04 $ & $ +0.78\pm 0.14 $ & $ 0.62\pm 0.19 $ & $ 2\pm 0 $ \\
 & $ B^{HMI}_{Y} = (+0.01\pm 0.02) + (+0.66\pm 0.12)\times B^{SPFF1}_{Y} $  &  
$ 0.17\pm 0.02 $ & $ +0.74\pm 0.10 $ & $ 0.55\pm 0.14 $ & $ 2\pm 0 $ \\
 & $ B^{HMI}_{Z} = (-0.02\pm 0.01) + (+0.39\pm 0.05)\times B^{SPFF1}_{Z} $  &  
$ 0.09\pm 0.01 $ & $ +0.84\pm 0.02 $ & $ 0.71\pm 0.04 $ & $ 1\pm 0 $ \\
 & & & & \\
Weak B Plage
 & $ B^{HMI}_{X} = (-0.02\pm 0.01) + (+0.50\pm 0.08)\times B^{SPFF1}_{X} $  &  
$ 0.10\pm 0.01 $ & $ +0.50\pm 0.06 $ & $ 0.26\pm 0.06 $ & $ 6\pm 0 $ \\
 & $ B^{HMI}_{Y} = (+0.01\pm 0.01) + (+0.42\pm 0.09)\times B^{SPFF1}_{Y} $  &  
$ 0.10\pm 0.00 $ & $ +0.47\pm 0.06 $ & $ 0.23\pm 0.06 $ & $ 6\pm 1 $ \\
 & $ B^{HMI}_{Z} = (-0.00\pm 0.00) + (+0.41\pm 0.08)\times B^{SPFF1}_{Z} $  &  
$ 0.05\pm 0.00 $ & $ +0.76\pm 0.03 $ & $ 0.58\pm 0.05 $ & $ 9\pm 1 $ \\
 & & & & \\
 $\mathbf{Solar Feature}$ &  \vectorb{HMI}{}  vs \vectorb{SPFF1}{}\ Same Sign in XY (Pos.) & $\mathbf{<s_e>}$ & $\mathbf{<r>}$ & $\mathbf{<r^{2}>}$  & \% Pos. \\
\hline
Strong B Plage
 & $ B^{HMI}_{X} = (-0.01\pm 0.01) + (+0.89\pm 0.08)\times B^{SPFF1}_{X} $  &  
$ 0.10\pm 0.02 $ & $ +0.93\pm 0.03 $ & $ 0.87\pm 0.05 $ & $ 90\pm 6 $ \\
 & $ B^{HMI}_{Y} = (+0.01\pm 0.01) + (+0.83\pm 0.06)\times B^{SPFF1}_{Y} $  &  
$ 0.10\pm 0.01 $ & $ +0.92\pm 0.02 $ & $ 0.85\pm 0.04 $ & $ 87\pm 5 $ \\
 & $ B^{HMI}_{Z} = (-0.02\pm 0.01) + (+0.39\pm 0.05)\times B^{SPFF1}_{Z} $  &  
$ 0.09\pm 0.01 $ & $ +0.84\pm 0.02 $ & $ 0.71\pm 0.04 $ & $ 99\pm 1 $ \\
 & & & & \\
Weak B Plage
 & $ B^{HMI}_{X} = (-0.01\pm 0.00) + (+0.84\pm 0.07)\times B^{SPFF1}_{X} $  &  
$ 0.07\pm 0.01 $ & $ +0.84\pm 0.02 $ & $ 0.70\pm 0.04 $ & $ 67\pm 3 $ \\
 & $ B^{HMI}_{Y} = (+0.01\pm 0.00) + (+0.76\pm 0.08)\times B^{SPFF1}_{Y} $  &  
$ 0.07\pm 0.00 $ & $ +0.82\pm 0.02 $ & $ 0.67\pm 0.03 $ & $ 67\pm 2 $ \\
 & $ B^{HMI}_{Z} = (-0.00\pm 0.00) + (+0.42\pm 0.08)\times B^{SPFF1}_{Z} $  &  
$ 0.05\pm 0.00 $ & $ +0.77\pm 0.03 $ & $ 0.60\pm 0.05 $ & $ 94\pm 1 $ \\
 & & & & \\
 $\mathbf{Solar Feature}$ &  \vectorb{HMI}{}  vs \vectorb{SPFF1}{}\ Opposite Sign in XY (Neg.) & $\mathbf{<s_e>}$ & $\mathbf{<r>}$ & $\mathbf{<r^{2}>}$  & \% Neg. \\
\hline
Strong B Plage
 & $ B^{HMI}_{X} = (+0.02\pm 0.02) + (-0.63\pm 0.06)\times B^{SPFF1}_{X} $  &  
$ 0.13\pm 0.01 $ & $ -0.79\pm 0.06 $ & $ 0.63\pm 0.09 $ & $ 10\pm 6 $ \\
 & $ B^{HMI}_{Y} = (-0.02\pm 0.02) + (-0.54\pm 0.10)\times B^{SPFF1}_{Y} $  &  
$ 0.11\pm 0.01 $ & $ -0.81\pm 0.04 $ & $ 0.65\pm 0.07 $ & $ 13\pm 5 $ \\
 & $ B^{HMI}_{Z} = (-0.01\pm 0.02) + (+0.09\pm 0.22)\times B^{SPFF1}_{Z} $  &  
$ 0.04\pm 0.05 $ & $ -0.25\pm 0.76 $ & $ 0.61\pm 0.30 $ & $ 1\pm 1 $ \\
 & & & & \\
Weak B Plage
 & $ B^{HMI}_{X} = (-0.02\pm 0.00) + (-0.62\pm 0.08)\times B^{SPFF1}_{X} $  &  
$ 0.07\pm 0.01 $ & $ -0.65\pm 0.05 $ & $ 0.43\pm 0.06 $ & $ 33\pm 3 $ \\
 & $ B^{HMI}_{Y} = (+0.00\pm 0.01) + (-0.57\pm 0.09)\times B^{SPFF1}_{Y} $  &  
$ 0.06\pm 0.00 $ & $ -0.70\pm 0.03 $ & $ 0.50\pm 0.04 $ & $ 33\pm 2 $ \\
 & $ B^{HMI}_{Z} = (-0.00\pm 0.00) + (-0.17\pm 0.05)\times B^{SPFF1}_{Z} $  &  
$ 0.02\pm 0.00 $ & $ -0.42\pm 0.07 $ & $ 0.18\pm 0.05 $ & $ 6\pm 1 $ \\
 & & & & \\
 \hline
\end{tabular}
\end{center}
\end{table*}
}

%% file: tablaSPcgvsSPcglekadisnewcutposneg.tex
 
{
\begin{table*}[]
\scriptsize
\begin{center}
\caption{Comparison between the \newcorr{Cartesian components of the} vector magnetic field observed by SP and SP disambiguated by AMBIG taking into account the sign between the horizontal components.\label{tabla_SPSPLeka_posneg}}
\begin{tabular}{lccccc} 
 $\mathbf{Solar Feature}$ &  \vectorb{SP\ AMBIG}{}  vs \vectorb{SP}{}\ Original & $\mathbf{<s_e>}$ & $\mathbf{<r>}$ & $\mathbf{<r^{2}>}$  & \% Eval. \\
\hline
Strong B Plage
 & $ B^{SP\ AMBIG}_{X} = (-0.02\pm 0.04) + (+0.54\pm 0.14)\times B^{SP}_{X} $  &  
$ 0.31\pm 0.05 $ & $ +0.54\pm 0.14 $ & $ 0.31\pm 0.11 $ & $ 3\pm 0 $ \\
 & $ B^{SP\ AMBIG}_{Y} = (+0.09\pm 0.01) + (+0.46\pm 0.13)\times B^{SP}_{Y} $  &  
$ 0.31\pm 0.03 $ & $ +0.48\pm 0.15 $ & $ 0.25\pm 0.14 $ & $ 3\pm 0 $ \\
 & $ B^{SP\ AMBIG}_{Z} = (-0.00\pm 0.00) + (+1.00\pm 0.00)\times B^{SP}_{Z} $  &  
$ 0.00\pm 0.00 $ & $ +1.00\pm 0.00 $ & $ 1.00\pm 0.00 $ & $ 4\pm 0 $ \\
 & & & & \\
Weak B Plage
 & $ B^{SP\ AMBIG}_{X} = (-0.01\pm 0.02) + (+0.30\pm 0.07)\times B^{SP}_{X} $  &  
$ 0.11\pm 0.00 $ & $ +0.30\pm 0.07 $ & $ 0.10\pm 0.05 $ & $ 4\pm 1 $ \\
 & $ B^{SP\ AMBIG}_{Y} = (+0.06\pm 0.00) + (+0.26\pm 0.06)\times B^{SP}_{Y} $  &  
$ 0.09\pm 0.00 $ & $ +0.31\pm 0.06 $ & $ 0.10\pm 0.03 $ & $ 3\pm 0 $ \\
 & $ B^{SP\ AMBIG}_{Z} = (+0.00\pm 0.00) + (+1.00\pm 0.00)\times B^{SP}_{Z} $  &  
$ 0.00\pm 0.00 $ & $ +1.00\pm 0.00 $ & $ 1.00\pm 0.00 $ & $ 5\pm 0 $ \\
 & & & & \\
 $\mathbf{Solar Feature}$ &  \vectorb{SP\ AMBIG}{}  vs \vectorb{SP}{}\ Same Sign in XY (Pos.) & $\mathbf{<s_e>}$ & $\mathbf{<r>}$ & $\mathbf{<r^{2}>}$  & \% Pos. \\
\hline
Strong B Plage
 & $ B^{SP\ AMBIG}_{X} = (+0.00\pm 0.00) + (+1.00\pm 0.00)\times B^{SP}_{X} $  &  
$ 0.00\pm 0.00 $ & $ +1.00\pm 0.00 $ & $ 1.00\pm 0.00 $ & $ 75\pm 5 $ \\
 & $ B^{SP\ AMBIG}_{Y} = (+0.00\pm 0.00) + (+1.00\pm 0.00)\times B^{SP}_{Y} $  &  
$ 0.00\pm 0.00 $ & $ +1.00\pm 0.00 $ & $ 1.00\pm 0.00 $ & $ 73\pm 5 $ \\
 & $ B^{SP\ AMBIG}_{Z} = (-0.00\pm 0.00) + (+1.00\pm 0.00)\times B^{SP}_{Z} $  &  
$ 0.00\pm 0.00 $ & $ +1.00\pm 0.00 $ & $ 1.00\pm 0.00 $ & $ 100\pm 0 $ \\
 & & & & \\
Weak B Plage
 & $ B^{SP\ AMBIG}_{X} = (+0.00\pm 0.00) + (+1.00\pm 0.00)\times B^{SP}_{X} $  &  
$ 0.00\pm 0.00 $ & $ +1.00\pm 0.00 $ & $ 1.00\pm 0.00 $ & $ 61\pm 3 $ \\
 & $ B^{SP\ AMBIG}_{Y} = (+0.00\pm 0.00) + (+1.00\pm 0.00)\times B^{SP}_{Y} $  &  
$ 0.00\pm 0.00 $ & $ +1.00\pm 0.00 $ & $ 1.00\pm 0.00 $ & $ 62\pm 3 $ \\
 & $ B^{SP\ AMBIG}_{Z} = (+0.00\pm 0.00) + (+1.00\pm 0.00)\times B^{SP}_{Z} $  &  
$ 0.00\pm 0.00 $ & $ +1.00\pm 0.00 $ & $ 1.00\pm 0.00 $ & $ 100\pm 0 $ \\
 & & & & \\
 $\mathbf{Solar Feature}$ &  \vectorb{SP\ AMBIG}{}  vs \vectorb{SP}{}\ Opposite Sign in XY (Neg.) & $\mathbf{<s_e>}$ & $\mathbf{<r>}$ & $\mathbf{<r^{2}>}$  & \% Neg. \\
\hline
Strong B Plage
 & $ B^{SP\ AMBIG}_{X} = (+0.00\pm 0.00) + (-1.00\pm 0.00)\times B^{SP}_{X} $  &  
$ 0.00\pm 0.00 $ & $ -1.00\pm 0.00 $ & $ 1.00\pm 0.00 $ & $ 25\pm 5 $ \\
 & $ B^{SP\ AMBIG}_{Y} = (+0.00\pm 0.00) + (-1.00\pm 0.00)\times B^{SP}_{Y} $  &  
$ 0.00\pm 0.00 $ & $ -1.00\pm 0.00 $ & $ 1.00\pm 0.00 $ & $ 27\pm 5 $ \\
 & $ B^{SP\ AMBIG}_{Z} = (-0.00\pm 0.00) + (+1.00\pm 0.00)\times B^{SP}_{Z} $  &  
$ 0.00\pm 0.00 $ & $ +1.00\pm 0.00 $ & $ 1.00\pm 0.00 $ & $ 0\pm 0 $ \\
 & & & & \\
Weak B Plage
 & $ B^{SP\ AMBIG}_{X} = (+0.00\pm 0.00) + (-1.00\pm 0.00)\times B^{SP}_{X} $  &  
$ 0.00\pm 0.00 $ & $ -1.00\pm 0.00 $ & $ 1.00\pm 0.00 $ & $ 39\pm 3 $ \\
 & $ B^{SP\ AMBIG}_{Y} = (+0.00\pm 0.00) + (-1.00\pm 0.00)\times B^{SP}_{Y} $  &  
$ 0.00\pm 0.00 $ & $ -1.00\pm 0.00 $ & $ 1.00\pm 0.00 $ & $ 38\pm 3 $ \\
 & $ B^{SP\ AMBIG}_{Z} = (+0.00\pm 0.00) + (+1.00\pm 0.00)\times B^{SP}_{Z} $  &  
$ 0.00\pm 0.00 $ & $ +1.00\pm 0.00 $ & $ 1.00\pm 0.00 $ & $ 0\pm 0 $ \\
 & & & & \\
 \hline
\end{tabular}
\end{center}
\end{table*}
}